\documentclass[twoside,12pt]{article}

\usepackage{epsfig,amsmath,amssymb,color}

\newcommand{\be}{\begin{equation}}

\newcommand{\ee}{\end{equation}}

\newcommand{\bea}{\begin{eqnarray}}

\newcommand{\eea}{\end{eqnarray}}

\newcommand{\nn}{\nonumber}

\newcommand \tie {{\it i.e.}}

\newcommand \f {\not\!}

\newcommand \ra  {\rightarrow}

\newcommand \vk {\vec{k}}
\newcommand \vl {\vec{l}}

\newcommand \vp {\vec{p}}

\newcommand \A {\alpha}
\newcommand \B {\beta}
\newcommand \g {\gamma}

\newcommand \kd  {\delta}

\newcommand \e {\epsilon}

\newcommand \h {\theta}

\newcommand \sg {\sigma}

\newcommand \w  {\omega}
\newcommand \ld {\lambda}

\newcommand \Op {{\mathcal O}}

\newcommand \x {\cdot}
\newcommand \hf {\frac{1}{2}}

\newcommand \lc {\langle}
\newcommand \rc {\rangle}

\newcommand \lt {\left}
\newcommand \rt {\right}
\newcommand \prt {\partial}
\newcommand \nt {\noindent}

\newcommand \gmn {g^{\mu \nu}}

\newcommand \bvec{\left( \begin{array}{c} }
\newcommand \evec{\end{array} \right)}
\newcommand \tr {\mbox{{\bf Tr}}}
\newcommand \epem {$e^+ e^-$}
\newcommand \mbx {\mbox{}}

\newcommand \psibar {\bar{\psi}}
\newcommand \ata {& \times &}

\newcommand \slm {\sum\limits}
\newcommand{\pt}{\ensuremath{p_{T}}}
\newcommand{\GeVc}{GeV/$c$}
\newcommand{\dphi}{\ensuremath{\Delta\phi}}
\newcommand{\rtsnn}{\ensuremath{\sqrt{s_{NN}}}}
\newcommand{\am}{}
\newcommand{\amnew} {}
\newcommand{\mvl}{}
\newcommand{\mvlnew}{}

\begin{document}
\title{ The theory and phenomenology of \\ perturbative QCD based jet quenching}

\author{A. Majumder \\  Department of Physics, The Ohio-State University, Columbus, OH 43210, USA \\ 
 M. van Leeuwen, \\  Department of Physics, Utrecht University, Utrecht, The Netherlands.}
\maketitle

\begin{abstract}
The study of the structure of strongly interacting dense matter via hard jets is reviewed. 
High momentum partons produced in hard collisions produce a shower of 
gluons prior to undergoing the non-perturbative process of hadronization. 
In the presence of a dense medium this shower is modified due to scattering of the 
various partons off the constituents in the medium. The modified 
pattern of the final detected hadrons is then a probe of the structure 
of the medium as perceived by the jet. Starting from the factorization paradigm 
developed for the case of particle collisions, we review the basic underlying theory 
of medium induced gluon radiation based on perturbative~Quantum Chromo Dynamics (pQCD)
and current experimental results from
 Deep Inelastic Scattering on large nuclei and high energy
heavy-ion collisions, emphasizing how these results constrain our understanding of energy loss.
\am{This review contains introductions to the theory of radiative energy loss, elastic energy loss, 
and the corresponding experimental observables and issues.}
We close with a discussion of important calculations and measurements that need to 
be carried out to complete the description of jet modification at high energies 
\am{at future high energy colliders.} 
\end{abstract}

\tableofcontents

%%%%%%%%%%%%%%%%%%%%%%%%%%%%%%%%%%%%%%%
%%%%%%%%%%%%%%%%%%%%%%%%%%%%%%%%%%%%%%%
%%%%%%%%%%%%%%%%%%%%%%%%%%%%%%%%%%%%%%%

\section{Introduction} \label{intro:main}

%%%%%%%%%%%%%%%%%%%%%%%%%%%%%%%%%%%%%%%
%%%%%%%%%%%%%%%%%%%%%%%%%%%%%%%%%%%%%%%
%%%%%%%%%%%%%%%%%%%%%%%%%%%%%%%%%%%%%%%

During the past nine years, high energy heavy-ion collisions have been studied at
the the Relativistic Heavy-Ion Collider (RHIC) at
Brookhaven National Laboratory (BNL). The three most striking findings
in heavy-ion collisions at RHIC are the observation of a large
azimuthal asymmetry in soft ($p_{T}< 2$ GeV) {hadron production}, denoted as elliptic flow; 
the observation of a large suppression of hadron yields at high transverse momentum ($p_{T} > 6$) GeV 
and the finding that the yields and the elliptic flow at intermediate \pt{} follow a scaling
behaviour that separates baryons and mesons by a factor of 3/2. Each of
these observations indicates that hot and dense strongly interacting
matter is formed in collisions at RHIC, possibly a deconfined
Quark-Gluon Plasma. Soon, the LHC will start colliding nuclei at much
larger energies, which will provide critical tests of the
current understanding of hot and dense QCD matter at RHIC.

In this review, we will concentrate solely on the suppression of high-\pt{}
hadron production and specifically its theoretical description within a perturbative QCD 
(pQCD) based formalism.  
In the collision of two heavy-ions at very high energies, most of the hard valence partons 
go through the collision un-deflected. The prevalent interaction is that between 
the softer ``sea'' partons. It is these soft interactions which lead to the formation of the 
hot and dense matter. Here ``soft'' means involving momentum scales of the order of 
$\Lambda_{QCD}$. Occasionally the hard partons {undergo} hard scattering (at scales 
much larger than $\Lambda_{QCD}$) 
leading to the formation of two back-to-back hard partons with a large \pt{}. 
High-\pt{} hadrons originate in the fragmentation of such high \pt{}
partons after their escape from the medium.
The hard partons lose energy through
interactions with the hot and dense medium. High $p_{T}$ signatures
are of special interest, because they {are expected to be described by} perturbative QCD due to the hard scale of the jets. At the LHC,
it will be possible to reach $p_{T} \sim 100-200$ GeV, i.e. a factor
of 10 increase in the hard scale of the jets compared to RHIC. Jet
physics based on perturbative QCD is therefore expected to play a
central role {in the study of hot QCD matter} at the LHC.

While the development of the theoretical framework to describe parton energy loss was 
started a little less than 20 years ago~\cite{Wang:1991xy_1}, the basic observation that
high-\pt{} hadrons are suppressed was only made with the first RHIC
data~\cite{Adler:2002xw,Adcox:2001jp}. Since then there has been significant progress
in our understanding of parton energy loss. On the
experimental side, the variety of measurements has greatly increased, now
extending to a variety of di-hadron and heavy flavour measurements. The precision and momentum
reach of these measurements is also much more extended compared to the first results from
RHIC. Phenomenological calculations have become much more detailed,
incorporating a variety of initial state effects such as shadowing, 
Cronin broadening~\cite{Antreasyan:1978cw} and more realistic models of the collision geometry 
based on hydrodynamical calculations constrained to describe the soft
hadron spectrum.
There are now four different theoretical schemes for energy loss
calculations, which incorporate different physical assumptions
regarding the scales involved and the microscopic structure of the
medium. These approaches can be compared and contrasted to provide
insight into the essential aspects of the dynamics of energy loss
in heavy ion collisions.

In this review, we aim to provide a survey of the current state of the
theoretical understanding of parton energy loss and the experimental tests
of this understanding. Our objective is not to provide an exhaustive
compilation of all experimental results and theoretical calculations
that have been performed, but rather to provide an overview of the
current understanding and methods which can serve as a `reading
guide' for the extensive literature on the subject. The theoretical
part of the review introduces the pQCD based formalism of jet
quenching as an extension of the standard factorized approach to hard
processes in vacuum and then discusses the similarities and
differences between the four different theoretical approaches to
parton energy loss in some detail. In the experimental part of the
review the emphasis is on a subset of observables which are, or can
be, rigorously calculated within pQCD and focus on what has been
learnt from these.
This field is under active development, both theoretically and
experimentally. On the theory side, the topic of most current interest is the
development of Monte Carlo routines to address some of the
shortcomings of the analytical approaches and to compute more
exclusive jet based observables. On the experimental side,
measurements have been extended to include photon-jet events and
multi-hadron observables, including first attempts to reconstruct jets
in heavy ion collisions at RHIC. In conclusion, we will discuss these developments
and how they are expected to address some of the main open question
about parton energy loss.

In the remainder of this chapter we provide a very brief introduction of how the
factorized formalism arises in pQCD. In Chapter 2,
we re-derive the multiple scattering induced single gluon emission
cross section within this language. This is most straightforwardly
achieved in a particular variant of what has come to be denoted as the higher twist approach.  After
this we relate our results to the derivations in other schemes. In
Chapter 3, we describe the inclusion of multiple emissions. This is as
yet a theoretically unsettled regime and we review the three different
means by which multiple emissions have been included. This is followed
by a direct comparison of the various formalisms, where the medium
modified fragmentation function is calculated in an identical
medium. In Chapter 4 we extend our formalism to include heavy flavor
modification along with elastic loss and diffusion. In Chapter 5 we describe
the phenomenological setup used to compute jet modification in a
static large nucleus and in a dynamically evolving quark gluon
plasma. In Chapter 6, we compare the results of recent calculations
in the literature on single hadron, dihadron, photon triggered and
heavy flavor observables to measurements in Deep-Inelastic Scattering (DIS) from HERA and those in
high-energy nuclear collisions from RHIC. In Chapter 7, we provide an outlook to the
LHC and forthcoming theory calculations.

\subsection{Background} \label{intro:background}

Quantum chromodynamics  (QCD) is now the accepted theory of strong interactions. 
In spite of the large diversity of hadronic states observed in nature, the Lagrangian of QCD is rather 
simple, involving only two types of fundamental fields: quarks and gluons, interacting via an $SU(N_c=3)$ 
gauge theory~\cite{Fritzsch:1973pi}
\bea
\mathcal{L}_{QCD}(x)= - \frac{1}{4} F^a_{\mu \nu}(x)  {F^a}^{\mu \nu}(x) 
+ \sum_{q=1}^{n_f} {\psibar^q}_i (x) \lt[  i \g^\mu {\mathfrak{D}_{\mu}}_{i,j}  + m_q \rt] {\psi^q}_j (x). \label{QCD_lag},
\eea 
where $i,j$ run from $1$ to $N_c$ representing the colors of the fundamental quarks of flavor $q$ 
and $a$ runs from $1$ to $N_c^2-1$ representing the colors of the adjoint gluon gauge field. 
For the purposes of this review $n_f$ will mostly be limited to 3, {for the light quarks $u$, $d$, and $s$, which will be treated as massless}. The heavy flavors charm and bottom will be discussed separately and 
will have non-zero mass terms. The top quark will not be included in this review (thus $n_f \leq 5$). 
The covariant 
derivative and the adjoint field strength have the usual definitions, 
\bea
\mathfrak{D^\mu}_{i,j} = \kd_{i,j} \prt^\mu  - ig t^a_{i,j} {A^a}^\mu  \,\,\,\,\,{\rm and }  \,\,\,\,\,
F^a_{\mu \nu} = \prt_\mu A^a_\nu - \prt_\nu A^a_\mu + g f^{abc} A^b_\mu A^c_\nu.
\eea
In the equation above, $A^a_\mu$ is the gluon four-vector potential with adjoint color $a$, $t^a_{i,j}$ are the Gell-Mann 
matrices, $f^{abc}$ is the completely antisymmetric tensor 
and $g = \sqrt{ 4 \pi \A_s}$ is the strong interaction coupling constant. In this review, we will often 
refer to the fine structure constant $\A_s$ also as the ``coupling''. Which is meant, will be 
denoted using the appropriate symbol.

Like all renormalizable quantum field theories, the coupling depends on the renormalization scale $\mu$,  
usually chosen as the relevant hard scale in the problem $Q^2$ to minimize the effect of higher order 
contributions.  Unlike the 
case of QED or weak interactions, QCD is asymptotically free, which implies that the coupling 
becomes weaker as the scale is raised~\cite{Gross:1973id,Politzer:1973fx}. Solving the (one loop) 
renormalization group equation, and comparing 
with experimental data to obtain the initial condition, one obtains,
\bea
\A_s(Q^2) = \frac{4 \pi  }{ \lt(\frac{11 N_c}{3} -  \frac{2 n_f}{3} \rt) \log\lt( \frac{Q^2}{\Lambda_{QCD}^2}\rt)  }.
\eea
The above equation implies that at scales far above $\Lambda_{QCD} \sim 200$MeV, QCD should be weakly 
coupled and a perturbative expansion in terms of $\A_{s}$ should become feasible. 
The applicability of this statement, of course, depends on the process in question. In the case of the 
total cross section in \epem~annihilation, single hadron inclusive \epem~annihilation, the total cross section in 
Deep Inelastic Scattering (DIS), single hadron inclusive DIS and the Drell-Yan effect, 
pQCD begins to become 
applicable beyond a $Q^{2} \gtrsim 2$ GeV$^{2}$~\cite{Martin:1999ww,Pumplin:2002vw,Albino:2005me}. The quantity $Q^{2}$ 
here refers to the relevant invariant 
mass that sets the hard scale in the problem. In all the three cases mentioned, this is the invariant mass
squared of the intermediate photon.  In the case of hadron-hadron scattering, one requires a specific 
hard momentum transfer process, 
such as the production of a high transverse momentum (high $p_T$) particle. The 
hard interaction in this case is mediated by the strong force (meaning the higher order corrections are 
of a different type) and such processes seem to need a minimum 
$p_{T} \gtrsim 2$ GeV ($Q^{2} \gtrsim 4$ GeV$^{2}$) for pQCD to be applicable~\cite{Jager:2002xm}. 
\amnew{In all cases mentioned above the applicability of pQCD refers to the agreement of calculations based on pQCD 
and experimental data. }

With the diminishing of momentum transfers, the strong coupling fine structure constant $\A_{s}(Q^{2})$ 
continues to grow and confines all particles which carry color charge within composite color singlet 
hadrons. While one may still construct an effective Lagrangian using the underlying symmetries of QCD and
carry out a perturbative analysis, such theories will not constitute any part of this review. 
All interactions 
with a momentum scale below a $Q^{2} \sim 2$ GeV$^{2}$ will be considered as non-perturbative and 
will usually be contained in a non-perturbative distribution such as a fragmentation function, a parton 
distribution function or an in-medium transport coefficient. 
The {separation} of the non-perturbative part from the hard perturbative 
part of the calculation will be discussed in the next subsection.

%%%%%%%%%%%%%%%%%%%%%%%%%%%%%%%%%%%%%%%
%%%%%%%%%%%%%%%%%%%%%%%%%%%%%%%%%%%%%%%
%%%%%%%%%%%%%%%%%%%%%%%%%%%%%%%%%%%%%%%

\subsection{pQCD and factorization} \label{intro:pQCD_factorization}

%%%%%%%%%%%%%%%%%%%%%%%%%%%%%%%%%%%%%%%
%%%%%%%%%%%%%%%%%%%%%%%%%%%%%%%%%%%%%%%
%%%%%%%%%%%%%%%%%%%%%%%%%%%%%%%%%%%%%%%

The ability to apply pQCD to describe a particular process should not be 
confused with the application of perturbative expansions in other gauge 
theories such as QED, the electro-weak theory or even in low energy 
effective theories of QCD such as chiral perturbation theory. In those theories, the 
incoming and outgoing asymptotic states in a scattering event carry the quantum numbers 
of the fields in the respective Lagrangian densities. 
While the QCD Lagrangian is cast in terms of quarks and gluons and 
by pQCD we do mean a diagrammatic expansion involving those 
fields, asymptotic states in strong interactions, due to confinement, 
are never quarks or gluons but composite hadrons. The ability to 
apply pQCD thus means simply the ability to isolate a section of the 
interaction which can be systematically described using a perturbative 
expansion in $\A_{s}$ involving quarks and gluons, from the remaining 
part of the process which is non-perturbative.

In all the reactions mentioned above where pQCD is applicable, there exist  
sub-processes over a range of energy scales up to the hard scale $Q^2$. 
Most of these cannot be described using pQCD and need to be separated 
from those sub-processes which involve hard scales. 
The technical machinery 
which demonstrates this separation order-by-order in the coupling constant is called ``factorization''~\cite{Collins:1983ju,Collins:1985ue,Collins:1988ig}.
The result of factorization is usually stated as a theorem with corrections power suppressed at 
very large $Q^2$. 
We illustrate an example of this theorem for the case of 
a high energy $p$-$p$ collision leading to the formation of a high 
$p_T$ hadron. Assume the collision is in the center of mass frame with each proton carrying  
a momentum $P(-P)$. The physical picture of this process is one where a 
hard parton in one of the incoming nucleons, carrying a light-cone momentum fraction $x_{a}$ ($p_{a} = x_{a}P^+$)
scatters of a hard parton in the other nucleon with light-cone momentum fraction $x_{b}$ ($p_{b} = - x_{b} P^- $) and 
produces two back-to-back partonic jets. 
Light-cone momenta are defined as $P^+ = (P^0 + P^3)/\sqrt{2}$ and $P^- = (P^0 - P^3)/\sqrt{2}$.
The hadronization products of one of these jets ($c$) will yield 
the high $p_{T}$ hadron $h$ carrying a momentum fraction $z$ of the original jet's momentum.
The factorization theorem for the differential cross section (in $p_{T}$ and rapidity $y$) 
states that the process can be expressed as~\cite{Collins:1985ue,Collins:1989gx}, 
\bea
\frac{d^2 \sg^h}{dy d^2 p_T} = \int dx_a d x_b \, G_a(x_a,\mu_{f}) G_b(x_b,\mu_{f}) 
\frac{d \sg_{ab \ra cX} (\mu_{R},\mu_{f},\mu_{f}^\prime,x_{a}P,x_{b}P, p_{T}/z )}{d \hat{t}} 
\frac{D_c^h(z,\mu_{f}^\prime)}{\pi z} +
\mathcal{O} \lt( \frac{\Lambda^{2}}{Q^{2}} \rt),  \label{fact_1}
\eea
where $G_{a}(x_{a})[G_{b}(x_{b})]$ is the parton distribution function (PDF) to find a hard parton with a momentum 
fraction $x_a$ ($x_b$) in the 
incoming proton, $d \sigma/d\hat{t}$ is the hard partonic cross section (with a partonic Mandelstam variable $\hat{t}$) 
and $D^{h}(z)$  is the fragmentation function (FF), the distribution of hadrons with a momentum 
fraction $z$ produced in the hadronization of the outgoing hard parton.  The 
second term on the $r.h.s.$ indicates corrections to the factorization theorem that are suppressed by 
powers of the hard scale $Q^{2}$. The quantity $\Lambda$ represents the scale of soft processes in the 
collision. This contribution becomes negligible as $Q^{2} \ra \infty$. Even in this asymptotic limit, 
the only term in the equation above that {can} be completely calculated within pQCD is the hard 
partonic cross section; the PDFs and the FF are non-perturbative objects which represent 
physics at softer scales.

The primary utility of factorization is based on the lack of interference between the hard and the 
soft scale, i.e., the PDFs and the FF (as well as the partonic cross section) are all squares 
of amplitudes. Eq.~\eqref{fact_1} contains only a convolution in two parameters ($x_a,x_b$) 
between these  
probabilities. To compute the hard cross section one proceeds order by order and calculates 
all the amplitudes that may contribute to the process of two incoming partons scattering to 
$n$ outgoing partons [$n=2$ for Leading Order (LO), $n=3$ for Next-to-Leading Order (NLO)].
These are then summed and squared to obtain $d\sg/d\hat{t}$.  The non-perturbative quantities 
$G(x,\mu)$ or $D(z,\mu)$ cannot be calculated from first principles, however they have well 
defined operator expressions on the light-cone, e.g., the PDF to find a quark with momentum 
fraction $x$, in a proton state $|p\rc$ traveling with a large light-cone momentum $p^+$, is defined as, 
\bea
G(x) &=& \int \frac{dy^-}{2\pi} e^{ - i x p^+  y^-} \lc p | \psibar(y^-) \g^+ \psi(0) | p \rc .
\eea

Given the operator expression, one may calculate higher order corrections. These tend to have 
collinear divergences which are absorbed into a renormalization of the operator product (or non-perturbative expectation values in the case of the FF).  
Such a redefinition introduces a scale dependence of the expectation. Divergent contributions up to 
this scale, denoted as the factorization scale $\mu_f$, are absorbed into the definition of the PDF. 
If $\mu_f$ is large compared to $\Lambda_{QCD}$, 
the change of the PDF {and the} FF can be calculated by means of the 
Dokshitzer-Gribov-Lipatov-Altarelli-Parisi (DGLAP) equation~\cite{Dokshitzer:1977sg,Gribov:1972ri,Altarelli:1977zs},
\bea
\frac{ \prt G(x,\mu)}{\prt \log \mu} &=& \frac{\A_s}{2 \pi} \int_{x}^1 \frac{d y}{y} P(y) G\lt( \frac{x}{y}, \mu^2 \rt).
\label{dglap_1}
\eea
In the equation above, $P(y)$ is the splitting function which represents the probability 
for a hard quark to radiate a gluon and still retain a momentum fraction $y$. The homogeneous 
DGLAP equation, as expressed in Eq.~\eqref{dglap_1} applies only to the non-singlet quark 
distribution [$G_{NS}(x) =   G_Q (x) - G_{\bar{Q}} (x)$]. 
The singlet distributions ($ G_{S}(x) =   [G_Q (x) + G_{\bar{Q}} (x)]/2 $) 
have couplings that mix the quark and gluon distributions. 

Equation~\eqref{dglap_1} is a differential equation and thus requires an initial condition for its solution.
This requires the experimental measurement of the non-perturbative distribution at one value of 
the scale. Herein lies the other advantage of factorization: Once the non-perturbative distributions  
are factorized from the hard cross section, they are independent of the process and become 
universal functions in the sense that they may be given an identical definition and measured in a completely 
different process. In the case of totally inclusive DIS, the factorization theorem yields,
\bea
\frac{d \sg}{d Q^{2}} = \int d x G(x,\mu^{2}) \frac{ d \hat{\sg} (\mu_{R},\mu,Q,xP) }{d Q^{2}} \label{fact_DIS}
\eea
The operator expression for the PDF as well as its evolution equation, derived in this case, is 
identical to that in Eq.~\eqref{fact_1}. Thus, the PDF measured in DIS~\eqref{fact_DIS}, 
may be directly substituted in Eq.~\eqref{fact_1} to compute single particle inclusive 
cross section in $p$-$p$ collisions.

To summarize, the applicability of pQCD to vacuum processes, such as in Eqs.~(\ref{fact_1},\ref{fact_DIS}), 
consists of the ability to calculate the hard partonic cross section order by order, as well as to compute 
the scale dependence of the non-perturbative distributions. While these non-perturbative distributions 
need to be measured at one scale in experiment, they have rigorous operator definitions which are 
identical in all processes where they can be factorized and thus are universal functions. Even 
though the discussion above was focused on the PDF, an almost identical factorization theorem 
and DGLAP evolution equation may be written down for the FF. Most of the remaining review will 
deal with the vacuum and in-medium modification of the FF.

\subsection{Hard jets in Semi-inclusive DIS on a nucleus and heavy-ion collisions}~\label{intro:SIDIS}

The preceding subsection briefly described the factorized formalism of high energy pQCD as applied 
to hard processes. The underlying assumption of factorization has been rigorously proven in 
all the examples mentioned above. In this subsection, we outline the extension of this 
factorized formalism to the modification of hard jets in a dense medium. Unlike the case of 
hard processes in vacuum, \amnew{the factorization theorems which are the underlying assumptions in 
these calculations, have not been proven to the same degree of rigor.} 
So far there have only been a handful of attempts in this direction both in pQCD~\cite{Qiu:1990xy,Doria:1980ak,DiLieto:1980dt} as well as in an effective-field theory 
approach~\cite{Idilbi:2008vm}. These factorization theorems will not be discussed further; 
we will assume their applicability. In this review we will focus on the application of the factorized theory
to jet propagation in dense extended media and the ensuing phenomenology.

\begin{figure}[tb]

\epsfysize=9.0cm

\begin{center}

\begin{minipage}[t]{8 cm}

\epsfig{file=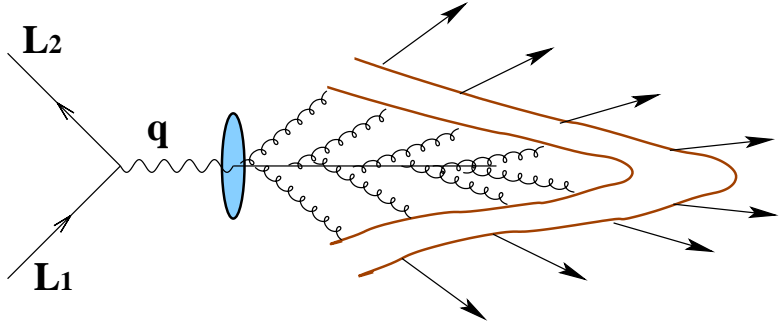,scale=0.5}

\end{minipage}

\begin{minipage}[t]{16.5 cm}

\caption{DIS on a nucleon leading to the formation of a quark jet which showers gluons 
in vacuum leading to the formation of a collimated jet of hadrons.\label{fig1}}

\end{minipage}

\end{center}

\end{figure}

Imagine the single-inclusive DIS of a hard lepton on a proton in the 
Breit frame. This is illustrated in Fig.~\ref{fig1}. The virtual photon with 
large negative light-cone momentum $q\equiv [Q^2/2q^-,q^-,0,0] $ strikes a hard quark which 
carries a momentum fraction $x_B$ of the Lorentz contracted proton with large 
positive light-cone momentum $P^+$. The quark is turned around and then 
exits the proton. Being considerably off-shell from the hard collision, it begins to shower gluons. 
The initial radiations have large 
transverse momentum $l_\perp$ (due to the large virtuality), and thus have shorter formation times:
\bea
\tau_f \sim \frac{ 2 q^- y (1-y)}{ l_\perp^2 }. \label{formation_time}
\eea 
Where $y$ is the momentum fraction of the parent parton carried away by the radiated gluon.
Later radiations (those with larger formation times) have smaller transverse momentum.
While not illustrated in the figure, the radiated gluons are also off-shell and tend to radiate 
gluons with smaller off-shellness. Eventually the local virtualities of the partons is so low that 
one cannot apply a perturbative partonic picture and hadronization begins to set in. 
As a result, the collection of hard collinear partons turns into a collimated jet of hadrons. 
The leading hadron in this picture is the one with the largest longitudinal momentum and,
given some form of localized momentum conservation, is the result of hadronization of the 
highest longitudinal momentum part or the largest formation time part of the jet. 

The picture in $p$-$p$ is somewhat similar except that one has at least 2 back-to-back jets being produced.
Both of these jets are produced considerably off-shell and tend to lose this virtuality by successive 
emissions. Unlike the case of DIS, jets which arise from the 
produced hard partons in large $Q^2$ {$p$-$p$ collisions} have a large momentum transverse 
to the incoming hard partons. 
We will always denote the momentum in this direction as $p_T$, differentiating it 
from the momentum of gluons transverse to the produced jet which will be indicated with a $\perp$, as in $l_\perp$. 
The factorized formula to obtain the cross section of single hadrons is given by the first term on the 
right hand side of Eq.~\eqref{fact_1}. At leading order, this formula simplifies to, 
\bea
\frac{d^2 \sg^h}{dy d^2 p_T} = \int dx_a d x_b \, G_a(x_a,Q) G_b(x_b,Q) 
\frac{d \sg_{ab \ra cX} (Q,x_{a}P,x_{b}P, p_{T}/z )}{d \hat{t}} 
\frac{D_c^h(z,Q)}{\pi z} .
\eea
The reader will note that we have chosen all factorization and renormalization scales to be $Q$ 
which is the hard scale in the problem. At leading order all hard scales that appear in the 
calculation are equivalent; in most cases one simply picks \amnew{$Q \simeq p_T$}.

\begin{figure}[tb]

\epsfysize=10.0cm

\begin{center}

\begin{minipage}[t]{8 cm}

\epsfig{file=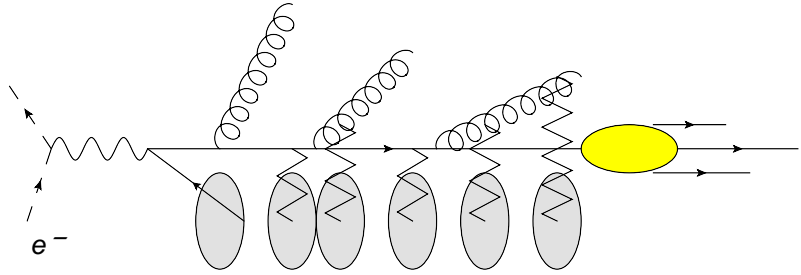,scale=1}

\end{minipage}

\begin{minipage}[t]{16.5 cm}

\caption{DIS on a large nucleus. A quark in a nucleon on the front side is struck and then 
propagates through the nucleons behind the struck nucleon. In the process both the 
quark and the ensuing shower gluons scatter off the soft glue field inside the nucleons. This 
modifies the shower pattern.\label{fig2}}

\end{minipage}

\end{center}

\end{figure}

We now consider the process of DIS on a large nucleus.  This will often be referred to as 
A-DIS (A referring to a nucleus). In the Breit frame, one may 
neglect the soft interactions between the various nucleons as these occur over a long 
time scale. The nucleus may then be modeled as a weakly interacting gas of nucleons 
traveling in the positive light-cone direction. The virtual photon strikes a hard quark in 
one of the nucleons and turns it backwards as in the case of the 
DIS on a single proton. 
The quark then
propagates through the nucleons directly behind the one that is
struck, as illustrated in Fig.~\ref{fig2}, where we draw the struck quark
as propagating outside the line of nucleons for clarity.
The quark is virtual at production 
and radiates a shower of gluons with progressively longer 
formation times, similar to the case of DIS on a proton. 
In this case however, both the quark and the radiated gluons tend to scatter 
off the soft gluon field in the nucleons. This is indicated by the zig-zag lines. Note, the 
zig-zag lines do not indicate a pomeron or double gluon exchange, but rather are
single-gluon exchanges which are distinct from the gluons in the
shower of the jet. 
Diffractive exchanges with the nucleons may be neglected in the 
case of very large nuclei. This will be justified in the next chapter.

The multiple 
scattering of the partons in the shower changes their momentum distributions and as a result the final hadronization pattern 
is modified. 
In the case of  a jet produced in a heavy ion collision, the picture is qualitatively similar except 
for the absence of nucleons. The different components of the jet now scatter off the 
quark-gluon substructure of the degrees  of freedom in the hot deconfined matter. 
The leading effect of the multiple scattering on the shower profile is a broadening of the 
distribution in transverse momentum (recently this has been experimentally {observed in cold nuclear matter}~\cite{Airapetian:2009jy}). 
Besides transferring transverse momentum, the medium also exchanges energy and longitudinal 
momentum with the jet (in the language of light-cone momentum, both $(+)$ and $(-)$ components are 
exchanged between the jet and the medium). While energy and momentum exchange (including transverse momentum) 
in the right proportion may cause minimal change in the virtuality of a given jet parton, arbitrary momentum 
exchanges may noticeably change the virtuality leading to induced radiation. 
In the case that the energy (and longitudinal momentum) of the jet partons far exceeds that of any constituent in the medium, 
such exchanges lead to a depletion of the light-cone momentum in the forward part of the jet. 
This is often referred to as radiative energy loss and results in a suppressed yield of 
leading 
particles~\cite{Wang:1991xy,Gyulassy:1993hr, Wang:1994fx,Baier:1994bd,Baier:1996kr,Baier:1996sk, Baier:1998yf}. 
This suppression has been quantitatively established in both DIS on 
a large nucleus~\cite{Airapetian:2007vu} and in jets in heavy ion collisions~\cite{Adler:2002xw,Adcox:2001jp}. The primary difference between the 
two cases is the fact that a heavy-ion collision is not a static environment and evolves 
rapidly with time and then finally disintegrates into a cascade of hadrons. 

The calculation of the modification of the hadronization pattern due to the multiple 
scattering induced radiation will be the subject of the next two chapters. 
Until recently, experiments have only been able to measure the modification 
of the yield of the leading hadron or the correlation between the leading and 
next-to-leading hadrons. The basic theoretical object required to describe this 
aspect of jet modification is called the medium modified fragmentation function. 
Along with the perturbative calculation of the evolution of the fragmentation 
function in vacuum, one computes the change in the fragmentation function 
due to the broadening and stimulated emission that occurs as the jet passes through 
matter. 
The medium 
modified fragmentation function is now a function of not just the momentum fraction 
and the scale $Q^2$, but also of the energy of the jet and the distance travelled 
in the medium. 
While it has become conventional to call this a ``medium modified'' fragmentation 
function, it should be pointed out that in all pQCD based calculations, the 
fragmentation is assumed to occur once the jet has escaped the medium.

We should point out that while this review will describe jet modification 
in dense matter as an extension of the factorized processes in vacuum pQCD 
[this is often referred to as the Higher-Twist (HT) scheme~\cite{Guo:2000nz,Majumder:2007hx,Majumder:2009ge}], there 
are other approaches to this problem. An entirely orthogonal approach is that based 
on finite temperature field theory based on the work of Arnold, Moore, Yaffe and 
collaborators~\cite{Arnold:2001ba,Arnold:2001ms,Arnold:2002ja,Turbide:2005fk,Qin:2007rn,Qin:2009bk}.
Referred to as the AMY scheme, this formalism considers the effective Hard Thermal 
Loop (HTL) effective theory of dense matter and considers the hard jet to have the 
same virtuality or mass scale as a thermal plasma particle but with energy $E \gg T$.
Following this, one identifies and resums the collinear enhanced contributions emanating 
from the scattering and induced radiation off the hard parton. The entire calculation is 
carried out at the scale of the temperature $T$ which is assumed to be large ($T \ra \infty$) 
so that the effective coupling $g(T)$ is small [$g(T) \ra 0 $]. Another approach developed 
by Armesto, Salgado, Wiedemann and collaborators models the medium as a series of 
Debye-screened, heavy, colored scattering centers. In this approach, referred to as the 
ASW scheme, the hard parton radiates a virtual 
gluon which is then progressively brought on shell by a large number of soft scatterings 
off these heavy 
centers~\cite{Wiedemann:2000ez,Wiedemann:2000za,Wiedemann:2000tf,Salgado:2003gb,Armesto:2003jh}. 
Yet another approach developed by Gyulassy, Levai, Vitev and their 
collaborators considers the same medium as ASW, however both the hard parton and the 
radiated gluon undergo a few but hard interactions with the centers leading to the emission of the
gluon~\cite{Gyulassy:1999zd,Gyulassy:2000fs,Gyulassy:2000er,Gyulassy:2001nm,Djordjevic:2003zk,Djordjevic:2004nq}. After the calculation of the single gluon emission kernel, the AMY scheme uses rate 
equations to incorporate multiple emissions whereas the GLV and the ASW use a Poisson 
emission \emph{Ansatz}.

The comparison of the yield of leading particles 
between the case with and without a medium {is} a measure 
of the properties of the medium as felt by the jet. In DIS, one 
measures the fragmentation function to produce a hadron with a 
momentum fraction $z$ of the original quark momentum immediately 
after being struck. For the case where the virtual photon carries a 
momentum, $q\equiv [-Q^2/(2q^-), q^- , {\bf 0}]$ and the incoming struck 
quark has $p \equiv [ Q^2/2q^-,0,{\bf 0}  ]$, the outgoing quark has a momentum $[0,q^-,{\bf 0}]$.
Thus the hadron momentum is $p_h \simeq z q^-$. The fragmentation 
function for DIS on a proton (nucleus $A$) and the ratio R are defined as,
\bea
D_{ p(A)}(z) = \frac{1}{\sg_{p(A)}}\frac{ d \sg_{p(A)} }{d z}, \,\,\,\, \mbox{and} \,\,\,\, R = \frac{ \sg_p }{\sg_A} \frac{ d \sg_A / d z }{ d \sg_p  / d z}.
\eea
In the equation above, {$\sg_{p(A)}$ is the differential DIS cross section off a proton $p$ (nucleus $A$)
integrated over a limited range of $Q^{2}$ and energy imparted from the electron $\nu = q^{-}$} . 
In order to compare with theoretical calculations, the fragmentation function in the 
case of DIS on a large nucleus $D_A (z)$ should be identified with the 
medium modified fragmentation function $\tilde{D}(z)$.

In the case of $p$-$p$ or heavy-ion collisions, the jets span a range of momentum 
depending on the momenta of the two partons that undergo the hard scattering. 
In this 
case one cannot isolate separate bins in the final state momentum fraction $z$ and thus 
cannot measure the fragmentation function directly. Instead one measures the 
ratio of the binary scaled differential yield to produce a high $p_T$ hadron in a heavy-ion collision to 
that in a $p$-$p$ collision referred to as the {nuclear modification factor} $R_{AA}$. The $R_{AA}$ may be measured 
both differentially as a function of angle with the reaction plane and {the transverse distance between the centers of the colliding nuclei, the impact parameter $b$}, or integrated.
The angle integrated $R_{AA}$ [also integrated over a small range of impact parameter ($b_{min}$ to $b_{max}$)],  is defined as
{
\begin{eqnarray}
R_{AA} &=& \frac{ \frac{d N^{AA}(b_{min}, b_{max})}{dy d^2 p_T} }
{ \lc N_{\rm bin} (b) \rc \frac{d N^{pp}(p_T,y)}{dy d^2 p_T}  }, \label{raa}
\end{eqnarray} 
\nt
where, $ \lc N_{\rm bin} (b) \rc$ is 
the mean number of binary nucleon nucleon 
encounters per ion-ion collision in the range of impact parameters chosen. 
The invariant yield in a heavy-ion collision and in a $p$-$p$ collision are 
related to the the invariant differential cross section by the relation, 
$d N = d \sg / \sg$, where the total cross section for a heavy-ion collision may be estimated as 
the geometrical cross section in that range of impact parameter selected. The 
relation between $N_{\rm bin}$ and the nuclear density profile will be discussed in Chapter~\ref{modeling}.
}

The calculation of the nuclear cross section to produce a hard hadron may be expressed 
as the convolution of the nuclear PDFs, the hard partonic cross sections and the medium 
modified fragmentation function. The 
jets may be produced at various locations in the hot matter and propagate in any direction in the 
transverse plane. The location of  the production point and the direction determine the extent 
and intensity of the medium as felt by the jet. One thus needs to integrate over all allowed production 
points and directions. This procedure will be described in Chapter~\ref{modeling}.

\subsection{Medium transport properties: what can be learnt from jets}~\label{intro:transport_coeff}

As described in the preceding subsections, a jet is essentially a collection of high momentum 
particles which are somewhat collimated in a given direction. In the partonic part of the jet, 
these are virtual partons. 
To make this explicit we ascribe a virtuality $\mu^{2} \ll Q^{2}$ to the hard jet, where $Q^{2}$ 
describes the scale of the hard interaction which produces the jet. In effective field 
theory approaches to hard jets such as that of Soft Collinear Effective Theory (SCET)~\cite{Bauer:2000yr,Bauer:2001ct}
one introduces a small parameter $\lambda$ with $\lambda  \ll 1$; collinear radiations 
have a transverse momentum of $l_{\perp} \sim \lambda Q$ and the virtuality of the jet may 
then be surmised as $\lambda^{2} Q^{2}$. Thus $\mu^{2} \sim (\lambda Q)^{2}$; the two 
terminologies will often be used interchangeably in this review.

\begin{figure}[tb]

\epsfysize=10.0cm

\begin{center}

\begin{minipage}[t]{8 cm}

\epsfig{file=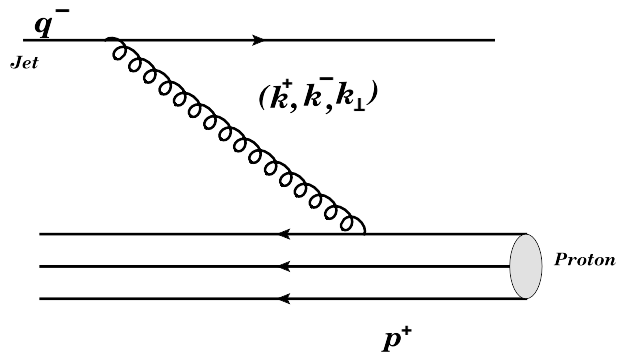,scale=0.7}

\end{minipage}

\begin{minipage}[t]{16.5 cm}

\caption{A single Coulomb (or Glauber) gluon interaction between a hard jet and a proton in the 
nucleus. 
\label{fig3}}

\end{minipage}

\end{center}

\end{figure}

As the jet passes through matter 
each of these particles will scatter off the various constituents that it encounters. In the Breit 
frame (or infinite momentum frame) the nucleus has a large boost in the direction opposed to that of the  
jet. As we pointed out earlier, one may consider each nucleon within the nucleus to be traveling 
in almost straight lines independent of each other. At sufficiently high energies, the ``large-$x$'' 
partons in the nucleons may also be considered {to be} traveling in straight lines independently 
of each other (see Fig.~\ref{fig3}). The hard virtual partons in the jet with virtuality $\mu^{2}$ will resolve this sub-structure
down to transverse sizes of order $1/\mu^{2} \sim (\lambda Q)^{-2}$. 
The hard partons in the jet will scatter of these 
partons by exchanging gluons with transverse momenta  $k_{\perp} \sim \mu \sim \lambda Q$. 
As a result, the hard partons in the jet will undergo a transverse diffusion as they propagate 
through the extended matter.

Along with the exchanged transverse momentum there may also be a certain amount of negative 
light-cone momentum ($k^{-}$) which may be exchanged between the partons in the jet and those 
from the medium. In this review, we will refer to this as ``elastic energy loss'' or ``drag'', even though 
the incoming parton from the medium is not on-shell, is not a quasi-particle of the medium and 
may not go back on shell after the scattering. This is primarily done to distinguish this 
type of light-cone momentum loss from the light-cone momentum loss that occurs due to 
radiation. This is somewhat different from what is traditionally referred to as elastic energy 
loss, which involves a certain energy and momentum transfer to an  in-medium quasi-particle which 
remains as one quasi-particle after the interaction, but with larger energy.  
The remaining component of the transferred momentum $k^{+}$ is completely determined 
by insisting on the criterion that the jet parton virtuality does not change too much after the momentum 
transfer. Thus if the original off-shellness is $2 q^{+} q^{-} \simeq \mu^{2 }$, then after the 
transfer, one obtains the virtuality as, 
\bea
2 (q^{+} + k^{+} ) ( q^{-} + k^{-}) - k_{\perp}^{2} = \mu^{2} + 2 \frac{\mu^{2}}{q^{-}} k^{-} + k^{+} (q^{-} + k^{-}) - k_{\perp}^{2}.
\eea
Thus, for $k^{-} \ll q^{-}$ we obtain that the virtuality of the parton will remain more or less unchanged  if 
$k^{+} = k_{\perp}^{2}/[2 ( q^{-}  + k^{-} ) ] $. For cases 
where $k^{+}$ exceeds this value, the hard parton from the jet is taken further off-shell 
leading to an induced radiation.

Treating the case of momentum change of the hard parton due to induced radiation separately, 
we obtain three separate components of 
momentum being exchanged between the jet and the medium: two components of transverse 
momentum and one component of negative light-cone momentum. 
At every exchange there is a distribution in transverse and light-cone momentum 
being imparted between the jet and the medium. Given the large number of 
exchanges, we invoke the Gaussian approximation, i.e., we approximate 
these distributions to be Gaussian and consider only the mean and the 
variance. The Gaussian approximation based on the central limit theorem is 
not completely unjustified. Each parton interacts multiple times with the medium and 
the shower distributions contain several hard partons. Along with this, one should 
also consider that except for the highest $p_{T}$ (or highest energy) hadrons, each 
bin in hadron momentum (or momentum fraction) contains several events that need 
to be summed over.  

Given cylindrical symmetry around the jet axis, on may further argue that the imparted 
${k_{x}}_{\perp}^{2} = {k_{y}}_{\perp}^{2}$, and the mean of the transverse momentum 
distribution is vanishing. Thus the first transport coefficient may be defined as the 
variance of the distribution of imparted transverse momentum per unit length traversed. 
This is referred to as $\hat{q}$, 
\bea
\hat{q} = \frac{ | {k_{x}}_{\perp} |_{L}^{2}  + | {k_{y}}_{\perp} |_{L}^{2}   }{ L}.
\eea
Where $|{k_{x,y}}_{\perp}|^{2}_{L}$  is the total transverse momentum gained 
in traversing a length $L$. 
In the case of negative light-cone momentum exchange, the mean of the 
distribution yields the drag per unit length ($\hat{e}$), whereas the variance is the fluctuation in light cone 
momentum transfer per unit length ($\hat{e}_{2}$):
\bea
\hat{e} = \frac{ k^{-} }{ L}, \, \, \, \mbox{and} \, \, \, \hat{e}_{2} = \frac{ (\Delta k^{-})^{2}}{L}
\eea
Given these two quantities, the Gaussian distribution of the negative light-cone momentum transfer is completely specified.

Thus the modification of hard jets will reveal at most these three quantities. Due to the large 
number of scatterings that need to be summed over in order to obtain measurements with 
small error bars, access to higher moments in these distributions is somewhat limited. 
These transport coefficients can in principle be calculated given a model of the medium, 
e.g., in the asymptotically high temperature limit, the system can be described in the 
weakly coupled quasi-particle picture afforded by leading order Hard Thermal Loop (HTL) effective 
theory. This derivation will be described in some detail in the remaining review. 
Even in this case, the value of the transport coefficients are unknown unless the value of 
the coupling is specified. {The relevant coupling} is the in-medium coupling between thermal quasi-particles which 
is \emph{a priori}  unknown and is therefore set by fitting calculations to at least one 
data point. It should be pointed out that if the medium is determined to not be weakly coupled, the HTL formulae cannot be applied and some other model will have to be used. 
{The value of the transport coefficient}, in principle, will have to be evolved up to the scale of the hard jet prior to 
use in a jet quenching calculation.

%%%%%%%%%%%%%%%%%%%%%%%%%%%%%%%%%%%%%%%
%%%%%%%%%%%%%%%%%%%%%%%%%%%%%%%%%%%%%%%

\subsection{{Other approaches and medium response observables}} \label{intro:other_approaches}

%%%%%%%%%%%%%%%%%%%%%%%%%%%%%%%%%%%%%%%%
%%%%%%%%%%%%%%%%%%%%%%%%%%%%%%%%%%%%%%%%

{ In this review, we will primarily address the question of how a hard jet can be 
used as a weakly coupled probe to study the properties of a dense QCD medium.
As such we will solely be interested in how different properties of the medium, 
codified as a set of transport coefficients will modify the shower pattern of the jet. 
\amnew{ As mentioned above, throughout this review, we will solely describe the 
the modification of the partonic portion of the jet shower, i.e., that part which may be computed 
using pQCD. This is often referred to in the literature as partonic energy loss to 
distinguish it from an alternate mechanism of hadronic energy loss. In this alternate 
hadronic energy loss scenario, one assumes that 
the jet hadronizes within a very short distance (of the order of $1$ fm) and 
produces a shower of hadrons. These hadrons then multiply scatter and lose forward momentum 
in the dense medium which is also assumed to be hadronic~\cite{Falter:2004uc,Gallmeister:2002us}.}

\amnew{ 
When applied to heavy-ion collisions, 
these calculations assume a minimal partonic phase, if any. While such theories have 
experienced a degree of success in the description of jets produced in DIS on a large 
nucleus~\cite{Airapetian:2007vu}, they have performed less favorably in comparison to RHIC data~\cite{Cassing:2004xr}. 
A slightly more successful variant has been the model based on pre-hadronic 
absorption~\cite{Cassing:2003sb,Accardi:2002tv,Accardi:2005jd}  which tend to model the 
absorption of the QCD string prior to hadronization. No doubt, all three mechanisms will be present 
in any description of jet modification where one attempts to capture 
the fate of all the hadrons which materialize from the shower, in media of arbitrary length. However, 
in sufficiently short media, and for considerably energetic jets, one may assume that the leading hadrons 
have hadronized outside the medium (these statements will be made more quantitative in the ensuing chapters). 
The yield of these leading hadrons may then be described by a factorized vacuum fragmentation function 
convoluted with the medium modified distribution of hard partons which have escaped the medium. 
The modification of the distribution of these few hard partons may then 
be calculated using pQCD with a minimal amount of non-perturbative input in the form of transport 
coefficients. The vacuum fragmentation functions are obtained from \epem experiments and are 
thus well known. Thus the study of this subset of observables involves only one set of unknowns: 
the transport coefficients of the medium. It is in this very restricted sense that pQCD based jet modification 
may be used as a probe of the medium. This review will be focused on this very 
restricted set of observables.}

As the medium modifies the jet, the jet in turn modifies the medium. There is a 
growing consensus that part of the energy radiated by a jet is deposited 
in the medium and this modifies the evolution of the soft medium at that location~\cite{Neufeld:2009ep,Qin:2009uh}.
Given the ``supersonic'' velocity of the hard jet, this may lead to the production 
of a Mach Cone~\cite{CasalderreySolana:2004qm} which trails the hard jet and 
may be observable as an excess in the hadronic yield at a large angle to the associated 
away side jet in the final soft hadronic spectra that is associated with a hard 
jet trigger~\cite{Adler:2005ee,:2008nda}. }

While the response of a soft medium to a hard jet may not be completely calculable in 
pQCD, in recent years an alternate theory based on the Anti-DeSitter space Conformal 
Field Theory  (AdS/CFT) conjecture~\cite{Maldacena:1997re} has been used to compute both the 
drag experienced by a heavy quark~\cite{Gubser:2006bz,Herzog:2006gh}
and the Mach cone left in a wake of of a hard jet~\cite{Gubser:2007xz,Chesler:2007an}.
Unlike calculations based on pQCD, these theories assume that the hard jet 
is strongly coupled with the dense medium. In the absence of a pQCD based picture 
it is not clear if the final hadronization process or the initial production process 
may be consistently factorized from the energy loss calculation. 
\amnew{
The three topics of (pre-)hadronic 
energy loss, 
medium response \mvlnew{to} a hard jet and the alternate theories of energy loss based on 
the AdS/CFT conjecture will not be covered in this review. For the reader interested in 
such topics we recommend these other excellent reviews: Refs.~\cite{Accardi:2009qv,CasalderreySolana:2007zz,Gubser:2009sn,dEnterria:2009am}}.

\section{Scattering induced single gluon radiation.}~\label{single_gluon}

In order to compute the energy loss of hard jets and the medium modified fragmentation 
function, one needs to compute a series of multiple scattering and multiple 
emission diagrams. In all formalisms, the means to achieve this is to first 
compute the single gluon emission kernel due to multiple scattering. 
This is then iterated to include the effect of multiple emissions. 
The methodology of iteration is somewhat related to the approximations 
made in the underlying calculation of the single gluon emission kernel.
In this chapter, we will describe this calculation in some detail. The 
iteration of the kernel will be dealt with in the next chapter. As stated 
in the introduction, our description of the single gluon emission kernel will 
follow that of the higher-twist approach. The differences with the other 
approaches will be pointed out at the end of this chapter.

We present the formalism in the Breit frame. 
In the case of DIS on a large nucleus this is characterized by the frame 
where the absolute magnitude of the ($+$) and ($-$) components of the 
photon momentum $q$ and the large ($+$) component of the initial  
momentum $p$ of the struck quark are equal:
\bea
q \equiv \left( -\frac{Q}{\sqrt{2}} , \frac{Q}{\sqrt{2}}, {\bf 0} \right)  
\,\,\, \mbox{and} \,\,\, p^+ = x_B P^+ = \frac{Q}{\sqrt{2}}, \, p^- \simeq|\vp_\perp|\ra 0.
\eea
$P^+$ is the momentum of a proton. The struck quark has a final 
($-$) momentum $\simeq q^- = Q/\sqrt{2}$ 
and travels in the negative $z$ direction. It now scatters off the gluons in the nucleons
which follow the struck nucleon. The picture is similar to that in Fig.~\ref{fig3}.

We will 
present the calculations in negative light-cone gauge $A^- = 0$. This makes the 
discussion of gluons emitted collinear to the outgoing quark particularly simple. 
The first step is to quantify the magnitude of the different components of the 
gluons being exchanged between the outgoing quark and the incoming nucleons. 
The final outgoing quark has a negative light-cone momentum of $q_f^- \sim Q$. 
Being close to on-shell, it has a virtuality $q_f^2 \sim \lambda^2 Q^2$, which 
is built up from some combination of a $q_f^+ \sim \lambda^2 Q$ and a small 
transverse momentum $q_{f \perp} \sim \lambda Q$.

\subsection{DIS on a proton and vacuum radiation}~\label{single_gluon:vacuum_radiation}

\begin{figure}[tb]

\epsfysize=9.0cm

\begin{center}

\begin{minipage}[t]{12 cm}

\epsfig{file=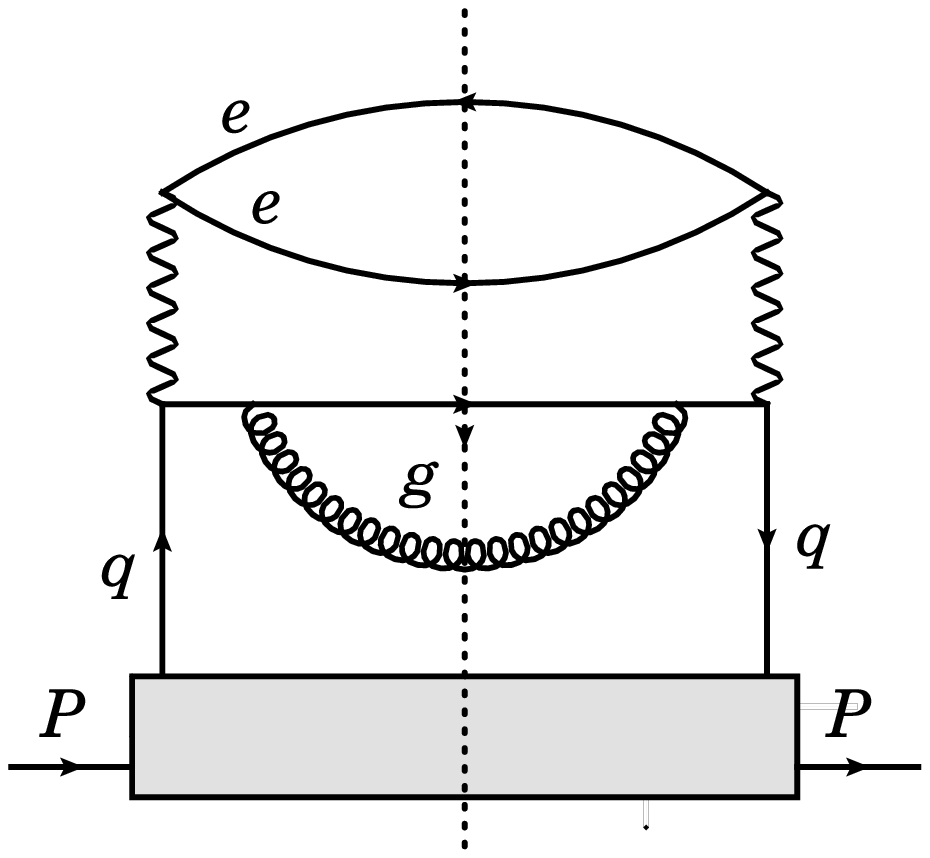,scale=0.6} \hfill
\epsfig{file=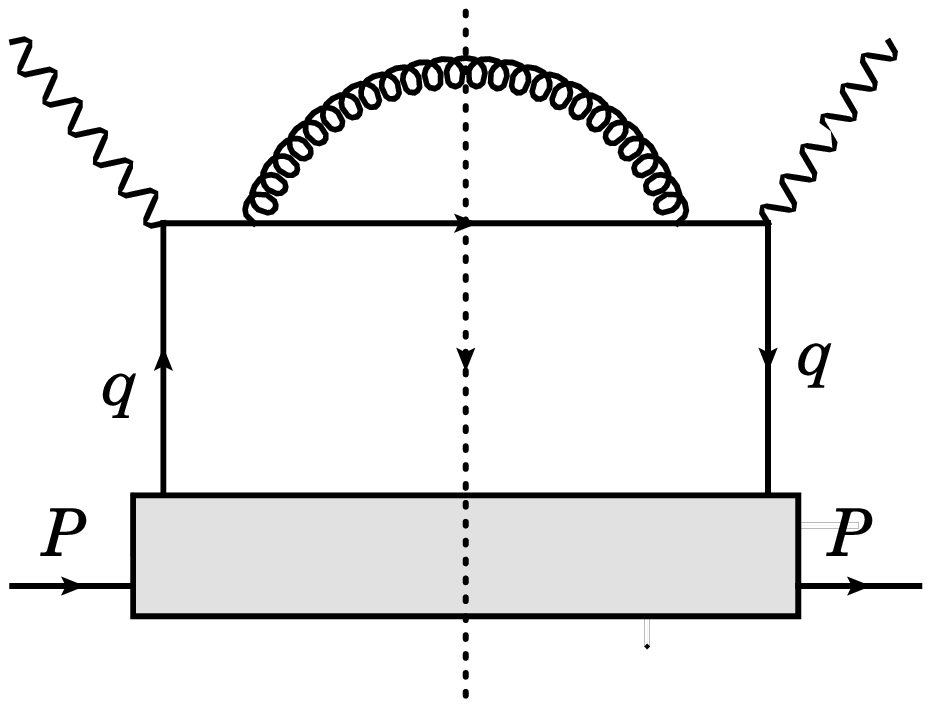,scale=0.6}

\end{minipage}

\begin{minipage}[t]{16.5 cm}

\caption{DIS on a nucleon leading to the formation of a quark and a radiated gluon.  Left panel 
represents the complete diagram whereas the right panel represents the hadronic tensor. The dangling 
photon lines do not represent photon propagators, but rather the scattering of the quark off a photon. 
The photon propagator is not contained in the hadronic tensor, see text for details.
\label{fig3b}}

\end{minipage}

\end{center}

\end{figure}

To familiarize the reader with the notation, we compute the simple process
of vacuum gluon radiation from a quark struck by a hard photon. This constitutes
the primary contribution to the scale evolution of fragmentation functions in 
vacuum. With this process we will also demonstrate an elementary example of 
factorization at leading order, where we factorize the parton distribution 
function (PDF) from the hard cross section. The basic quantity to be computed is the 
differential cross section for an electron with an initial momentum $L_1$ 
to scatter off a proton (with momentum $P^+$) 
with final electron momentum $L_2$ and producing an outgoing quark with momentum
$l_q = [ |{l_q}_\perp|^2/\{2q^-(1-y)\} , q^- (1 - y) , {l_q}_\perp ]$ and a gluon with momentum 
$l = [ |l_\perp|^2/(2 q^- y) , q^- y , l_\perp ]$:
\bea
{\mathcal L} (L_1) + p(P^+) \longrightarrow  \mathcal{L} (L_2) + Q( l_{q_\perp} ) + G(l_\perp) + X .
\label{chemical_eqn}
\eea
The complete cross section for the process is represented by Fig.~\ref{fig3b}. The rectangular blob at the 
bottom of the diagram represents the contents of the proton after the hard quark $q$ is struck by the 
photon and removed from the proton.

The matrix element that needs to be evaluated may be represented as 
\bea
\mathcal{M} = \lc X, l,l_{q} ; L_{2}| \mathcal{T} \exp \lt[ - i \int\limits_{-\infty}^{\infty} d t H_{I}(t) \rt] | P ; L_{1} \rc.
\eea
The state $| X \rc$ represents an arbitrary 
hadronic state. For the case of interest, $|L_{2}, l, l_{q}, X \rc$ represents a 
state with at least a quark and a gluon endowed with the requisite momenta and 
an outgoing electron with momentum $L_{2}$. At lowest 
order $| X \rc$ is simply the vacuum state. 
In the equation above, $H_{I}(t)$ represents the interaction Hamiltonian in the interaction 
picture for QCD and QED. The eventual evaluation of the exponent will involve two orders 
of the electro-magnetic (EM) interaction and all orders of the strong interaction. For the case of 
one gluon emission in this 
section we will only have to expand the exponent to one order of the strong interaction (and two 
orders of the EM interaction) i.e.,
\bea
\mathcal{M} &=& \lc X, l, l_{q}; L_{2} | \frac{1}{3!} 
\int_{-\infty}^{\infty} d^{4} y_e \psibar_{e} (y_e) e\g^{\mu} A_{\mu}(y_e) \psi_{e} (y_e) 
\int_{-\infty}^{\infty} d^{4} y_{0} \psibar (y_{0}) e\g^{\mu} A_{\mu} (y_{0}) \psi (y_{0}) \nn \\
\ata \int d^{4} y_{1} \psibar (y_{1}) g  \g^{\nu} t^{a} A^{a}_{\nu} (y_{1}) \psi (y_{1}) | P ; L_{1} \rc.
\eea
In the equation above, $\psi_{e}$ is the wave function of the lepton, while $\psi$
is that for a quark. 
The restriction to only two orders of the EM interaction, constrains the calculation 
to the one photon approximation. 
Squaring the matrix element one may calculate the cross section for this process. 
It may be straightforwardly demonstrated, in the one photon limit, that the process 
can be decomposed into a purely leptonic part and a partonic (or hadronic)  part connected 
by a single photon propagator. The leptonic part will involve the trace over the 
Dirac matrix structure of the electron momenta; we will simply write this down without proof. 
The hadronic part will be somewhat more 
complicated as it will contain both a perturbative component describing the scattering of a quark off 
the virtual photon followed by the emission of a hard gluon and a non-perturbative component which 
describes the probability to find a quark with a particular momentum fraction in the incoming nucleon.
This component will be described in some detail.

The quadruple differential cross section for the process in Eq.~\eqref{chemical_eqn}, in terms of 
$L_2,l_\perp,{l_q}_\perp,y$ can be decomposed, as stated above, into a purely leptonic part and a hadronic 
part in the one photon exchange approximation. This is symbolically represented as,  
\bea
\frac{E_{L_2} d \sigma } {d^3 L_2 d^2 l_{q_\perp}  d^2 l_\perp dy  } &=&
\frac{\A_{em} ^2}{2\pi s  Q^4}  L^{\A \B}   g_{ \A \mu} g_{\B \nu}
\frac{d W^{\mu \nu}}{d^2 l_{q_\perp} d^2 l_\perp dy}, \label{LO_cross}
\eea
where $L^{\A \B}$ and $W^{\mu \nu}$ are the leptonic and hadronic tensors 
respectively. The Mandelstam variable $s = (p+L_1)^2$.  Each of the factors of 
$g_{\A \mu}/Q^2$ and $g_{\B \nu}/Q^2$ represents a photon propagator, one from 
the amplitude and one from the complex conjugate.

The leptonic tensor has the straightforward definition, 
$L^{\A \B} = \tr [ \f L_1 \g^{\A} \f L_2 \g^{\B}]/2$, while the hadronic tensor, 
at leading order in the strong coupling, 
may be decomposed as 
\bea
{W_1}^{\mu \nu} 
% &\equiv & 2 Im \left[ \int d^4y e^{iq \x y} \lnuc  J^{\mu}(y) J^\nu (0)  \rnuc \right] \nn \\
%
=  \int d^4y_0 \lc P |  \psibar(y_0) \g^\mu  \widehat{\Op^{00}} \g^\nu \psi(0) | P \rc  
= \int d^4 y_0\tr [\frac{\g^-}{2} \g^\mu \frac{\g^+}{2} \g^\nu ] F(y_0) \Op^{00}(y_0) .  \label{w_mu_nu_twist=2}
\eea
This expression requires some explanation. The interaction terms in $H_{I}$ are those contained in the 
QCD [Eq.~\eqref{QCD_lag}] and QED Lagrangians. These are composed of solely quarks, gluons, leptons and 
photons. While we have, somewhat artificially, constrained the final state to be a quark and a gluon, the initial 
state is a proton. In the Breit frame, the high energy proton can be approximated 
as a weakly interacting gas of collinear quarks and gluons. The distribution of these partons 
in bins of light-cone momentum fraction $x$ depends on the transverse size of these partons (i.e., resolution 
of the probe). In this high energy limit, the projection of the product of $\psi |X \rc$ and its complex 
conjugate ($\lc X | \psibar$) along the large momentum direction 
may be viewed as an annihilation and creation of a near on-shell quark  in the proton's 
wave function. The  Fourier transformation of the expectation of this operator product 
is referred to as the parton distribution function (PDF),
\bea
f(x) = \int d y_0^- e^{-i x P^+ y_0^- } F(y_0^-) 
=  \int d y_0^- e^{-i x P^+ y_0^- } \lc P | \psibar(y_0^-,0) \frac{\g^+}{2} \psi(0,0) | P \rc .\label{F(y_0)}
\eea

Unlike the parton distribution function in Eq.~\eqref{F(y_0)} above, the function $F(y^0)$ that appears in Eq.~\eqref{w_mu_nu_twist=2} is not yet on the light-cone. Note that in the equation above only $y_{0}^{-}$ appears.
This happens after the invocation of the high energy or collinear approximation. 
The incoming parton is assumed to be endowed with very high forward momentum 
$({p_0}^+ = x_0 p^+, p_0^- \ra 0)$
with negligible transverse momentum ${p_0}_\perp \ll p_0^+$.  Within the 
kinematics chosen, the incoming virtual photon also has no transverse 
momentum. As a result, the produced final state parton also has a vanishingly small transverse momentum 
(\tie, with a distribution that may be approximated as $\kd^2(\vec{p}_\perp)$ up to corrections 
suppressed by powers of the hard scale). 
In this limit, the leading spin projection of  the pieces which represent the initial state 
and final state may be taken. The factors, 
\bea
\g^+ = \frac{\g^0 + \g^3}{ \sqrt{2} } \,\,\, ;\,\,\,\,\,    \g^- = \frac{\g^0 - \g^3}{\sqrt{2}}, 
\eea
are used to obtain  the spin projections along the leading momenta of the outgoing state and the incoming 
 state. See Refs.~\cite{Majumder:2009ge,Majumder:2007ne} for precise details of how this is done.

The other function in Eq.~\eqref{w_mu_nu_twist=2} is $\Op^{00}$ which represents 
the physics of the final state after the hard scattering with the photon.
The superscript on the operator $\Op^{00}$ implies that the 
quark undergoes no scattering in the final state. Taking the leading 
projection in Dirac matrix structure we obtain 
\bea 
\Op^{00} &=& \tr \left[ \frac{\g^-}{2} \widehat{\Op^{00}} \right]  
 =  \int \frac{d^4 l}{(2\pi)^4}   d^4 z d^4 z'
\frac{d^4 l_q}{(2\pi)^4}\frac{d^4 p_0}{(2\pi)^4} \frac{d^4 p'_0}{(2\pi)^4}
e^{i q \x y_0 }e^{ -i (p_0 + q) \x ( y_0 - z) } 
e^{-i l \x (z - z')} e^{-i l_q \x (z - z')} 
 e^{ -i (p_0'+ q) \x z }   \nn \\ 
\ata g^2 \tr \left[  \frac{\g^-}{2} \frac{-i (\f p_0 + \f q )}{ (p_0 + q )^2 - i \e }  
i\g^\A  \f l_q   2\pi \kd (l_q^2)  
 G_{\A \B} (l)   2\pi \kd (l^2)  (-i\g^{\B})
\frac{ i (\f p_0'+ \f q )}{ (p_0'+ q )^2 +  i \e }  \right]  \tr[ t^a t^b]. \label{O_00}
\eea
In the equation above, $G_{\A \B} (l)$ is the sum over the product of polarization vectors of the 
radiated gluon with momentum $l$. In light cone gauge $A \cdot n =  A^- = 0$ (with $n = (1,0,0,0)$) 
this is given as, 
\bea
G_{\A \B} (l) = - g_{\A \B} + \frac{ l_\A n\B  + l_\B n_\A }{ l.n }.
\eea

To evaluate the above Feynman integral one integrates over the internal 
locations $z,z'$, which will yield the momentum conserving delta functions. This will 
set $p_0 = p'_0$.  Then one approximates the numerators of the fermion propagators as 
$\g^+ q^-$. The denominators may be expressed as $2 P^+ q^- ( x_0 - x_B \pm i \e)$ where 
$x_0 = p^+/P^+$. The $\kd(l^2)$ sets $l^+ = l_\perp^2/2l^-$. 
The other $\kd$-function over ${l_q}_\perp$ sets $x_0 = x_B + x_L $ where $x_L =  l_\perp^2/[2P^+ q^- y(1-y)]$.
The remaining steps involve carrying out the simplifications associated 
with the $\g$ matrices and carrying out the integrations over the momenta that do not 
appear in the integrand. This yields 
\bea 
\Op^{00} &=& \kd(y_0^+) \kd^2(y_{0_\perp}) 
e^{-i (x_B + x_L)p^+ y_0^-}   \frac{\A_{s} C_{F}}{2\pi}  
\int \frac{dy dl_\perp^2}{l_\perp^2}  \frac{2 - 2y + y^2}{y}  . \label{O_00_3}
\eea
The resulting phase factors constrain $F(y^0)$ to the light cone and convert it into the 
PDF. The factor $(2 - 2y + y^2)/y$ is the splitting function $P(y)$ which expresses the 
probability for the hard quark to radiate a gluon. Re-incorporating this expression back 
into Eq.~\eqref{w_mu_nu_twist=2}, we obtain the differential hadronic tensor for 
a quark in a proton, struck by a hard virtual photon, to radiate a gluon with momentum $[q^-y,l_\perp]$ 
and itself have a momentum $[q^-(1-y),{l_q}_\perp]$:
\bea
\frac{ d W_1^{\mu \nu} }{d y d l_\perp^2 d^2 {l_q}_\perp }
= \sum_q 2 \pi Q_q^2 f_q (x_B+x_L) \frac{\A_s C_{F}}{2\pi l_\perp^2} P(y) \kd^2 (  \vec{l_q}_\perp + \vl_\perp) .
\label{W_mu_nu_gluon}
\eea
The above result should be compared with the simpler result of DIS without any radiation in the 
final state. The derivation is similar to that presented for the case of a produced radiation, we here
simply state the result (referring the interested reader to Ref.~\cite{Majumder:2007hx}). The 
differential hadronic tensor for a quark in a proton, struck by a hard photon, to have a momentum
$[q^-,{l_q}_\perp]$ is
\bea
\frac{d W_0^{\mu \nu}}{ d^2 {l_q}_\perp} = \sum_q 2 \pi Q_q^2 f_q (x_B) \kd^2 (\vec{l_q}_\perp).
\label{W_mu_nu}
\eea

In the equation above, $f_q(x)$ is the quark PDF defined in Eq.~\eqref{F(y_0)}. Note that the 
arguments $x$ in Eqs.(\ref{W_mu_nu_gluon},\ref{W_mu_nu}) are slightly different. 
They differ by the quantity $x_L = l_\perp^2/[ 2 P^+ q^- y (1-y) ]$. This should be 
understood as the ratio of the off-shellness of the outgoing quark to $Q^2/x_B = s$, i.e.,
\bea
x_L s = \frac{x_L}{x_B} Q^2 = x_L 2 P^+ q^- =  \frac{ l_\perp^2 }{y (1-y)}
= 2( l_q^+  +  l^+  ) (l_q^- + l^-) - \lt| \vl_\perp + \vec{l_q}_\perp \rt|^2 .
\eea
The momentum component $p^+ = x_L P^+$ is the extra ($+$)-component of the momentum 
that must be brought in by the incoming quark such that the quark after being 
struck should be off-shell enough to radiate the gluon with momentum [$l_\perp^2/(2q^-y), q^-y, \vl_\perp$].
We work in the limit where a collinear jet is produced in the hard interaction, {so that} the 
off-shellness is 
very small compared to the hard scale $Q^2$:
\bea
x_L 2 P^+ q^- = \frac{l_\perp^2}{2 P^+ q^-} \ll Q^2 = x_B 2 P^+ q^- \Longrightarrow x_L \ll x_B. 
\eea
As mentioned in the introduction, this ratio of the off-shellness of the produced quark to $Q^2$ is 
used to qualify the small parameter $\lambda$, i.e., $l_\perp \sim \lambda Q$. In this limit 
of transverse momenta, $x_L = \lambda^2 x_B (\lambda x_B)$ if $y \sim 1 (\lambda)$. In either 
case, this small correction to $x_B$ in Eq.~\eqref{W_mu_nu_gluon} can be neglected.  

In both equations ~(\ref{W_mu_nu_gluon},\ref{W_mu_nu}), the distribution of the final outgoing 
quark involves a $\kd$-function. This is meant to be a simplification. A better expression is to use a narrow, normalized Gaussian distribution. The width is of order $\sim \lambda Q$, i.e., small compared to the 
hard scale but still perturbatively large. The width of this outgoing quark distribution is related to the transverse 
momentum and off-shellness of the incoming quark; this controls the virtuality of outgoing quark and 
thus the virtualities involved in all final state processes. This  also controls the 
scale at which the coupling constant is evaluated in the case of final state gluon emissions.

In the limit of small $x_L$, the ratio of the cross sections for both processes, computed using Eq.~\eqref{LO_cross}, 
assumes a rather simple form.  
Integrating out the transverse momentum of the produced quark gives the differential number of 
gluons radiated, 
\bea
\frac{d N_g}{dy dl_\perp^2} = \frac{1}{ \sigma_{ [q + \g \ra q + X ] }  }
\frac{ d \sigma_{[ q + \g \ra q + g +X ] }}{ dy d l_\perp^2 }
= \frac{\A_s (l_\perp^2 ) C_{F}}{2\pi l_\perp^2} P(y). \label{vac_gluon_number}
\eea
In the subsequent section, we will derive the cross section for a hard quark to radiate a 
single gluon while undergoing multiple scattering in the medium. This will involve modifying 
a related version of Eqs.~(\ref{W_mu_nu_gluon},\ref{W_mu_nu},\ref{vac_gluon_number}).

We close with a comment regarding Eq.~\eqref{vac_gluon_number}. The reader will note that 
the expression for the differential number of gluons in $y,l_\perp$ possesses both an infra-red $y \ra 0$ 
[$P(y) \ra 2/y$ as $y \ra 0$] and 
a collinear singularity $l_\perp \ra 0$. 
\amnew{In the calculation of the final hadronic distribution, one convolutes this gluon distribution and the related 
final quark distribution with a fragmentation function $D$ as shown below in Eq.~\eqref{vac_dglap}. In this 
convolution one also has to include the effect of virtual corrections, i.e., instances where the leading 
parton radiated a gluon and then reabsorbed it. Such diagrams also contain an infrared and collinear 
divergence. In the convoluted expression to obtain the yield of hadrons, the infrared singularity cancels 
between real and virtual diagrams.
The real splitting function is replaced with the well known $(+)$-functions~\cite{Altarelli:1977zs}. The collinear divergence,
however remains even after the inclusion of virtual corrections.}
The source of this divergence are gluons with $l_\perp \ra 0$, 
and thus those with formation times $\tau_f \ra \infty$. In a factorized approach, 
such long distance effects are absorbed into a renormalization of the final fragmentation 
function to produce hadrons. In the case of single hadron inclusive DIS, absorbing gluons with transverse momenta up to the 
final factorization scale $l_\perp = \mu \ll Q$ yields the scale dependent fragmentation function \amnew{[$D(z,\mu^{2})$] }to produce a
hadron  with momentum fraction $z= p_h/q^-$. Including the effect of multiple emissions yields the 
DGLAP evolution equation for the fragmentation function:
\bea
\frac{ \prt D(z , \mu^2)}{ \prt \log (\mu^2) } = \frac{  \A_s C_{F}}{ 2 \pi} 
\int_{z}^1 \frac{dy}{y} P (y) D \lt( \frac{z}{y} , \mu^2\rt). \label{vac_dglap}
\eea
 
In the in-medium version of Eq.~\eqref{vac_gluon_number}, Landau-Pomeranchuck-Migdal (LPM) interference 
tends to cancel the collinear divergence making the differential yield of gluons finite at $l_\perp \ra 0$.  
As a result, one obtains a finite energy of gluons emitted on integrating over $l_\perp$. Two of the 
formalisms use this value to construct an iterative formalism, thus ignoring the fact that a fraction of 
the number of gluons are produced much later in time and should not be included in the calculation. 
This will be {further} discussed in Chapter~\ref{multiple_radiation}.

%%%%%%%%%%%%%%%%%%%%%%%%%%%%%%%%%%%%%%%%%%
%%%%%%%%%%%%%%%%%%%%%%%%%%%%%%%%%%%%%%%%%%
%%%%%%%%%%%%%%%%%%%%%%%%%%%%%%%%%%%%%%%%%%

\subsection{Multiple scattering induced single {gluon} radiation}~\label{single_gluon:multiple_scattering}

%%%%%%%%%%%%%%%%%%%%%%%%%%%%%%%%%%%%%%%%%%
%%%%%%%%%%%%%%%%%%%%%%%%%%%%%%%%%%%%%%%%%%
%%%%%%%%%%%%%%%%%%%%%%%%%%%%%%%%%%%%%%%%%%

In this section, we compute the cross section for single gluon emission induced by multiple scattering. 
Iterating this process and convoluting with the fragmentation function will produce the medium 
modified fragmentation function.
We will approximate  each scattering to 
only transfer a small transverse momentum $k_\perp \sim  \lambda Q$ and a 
($+$)-component $\sim \lambda^2 Q$  between the jet and the medium partons. There can also be 
an exchange of a small ($-$)-component; this leads to elastic energy loss and will be discussed 
separately in Chapter~\ref{heavy_flavor}.

%%%%%%%%%%%%%%%%%%%%%%%%%%%%%%%%%%%%%%%%%%
%%%%%%%%%%%%%%%%%%%%%%%%%%%%%%%%%%%%%%%%%%
%%%%%%%%%%%%%%%%%%%%%%%%%%%%%%%%%%%%%%%%%%

\subsubsection{{Multiple scattering without gluon radiation}}~\label{single_gluon:multiple_scattering:no_radiation}

%%%%%%%%%%%%%%%%%%%%%%%%%%%%%%%%%%%%%%%%%%
%%%%%%%%%%%%%%%%%%%%%%%%%%%%%%%%%%%%%%%%%%
%%%%%%%%%%%%%%%%%%%%%%%%%%%%%%%%%%%%%%%%%%

\begin{figure}[tb]
\epsfysize=9.0cm
\begin{center}
\begin{minipage}[t]{18 cm}
\epsfig{file=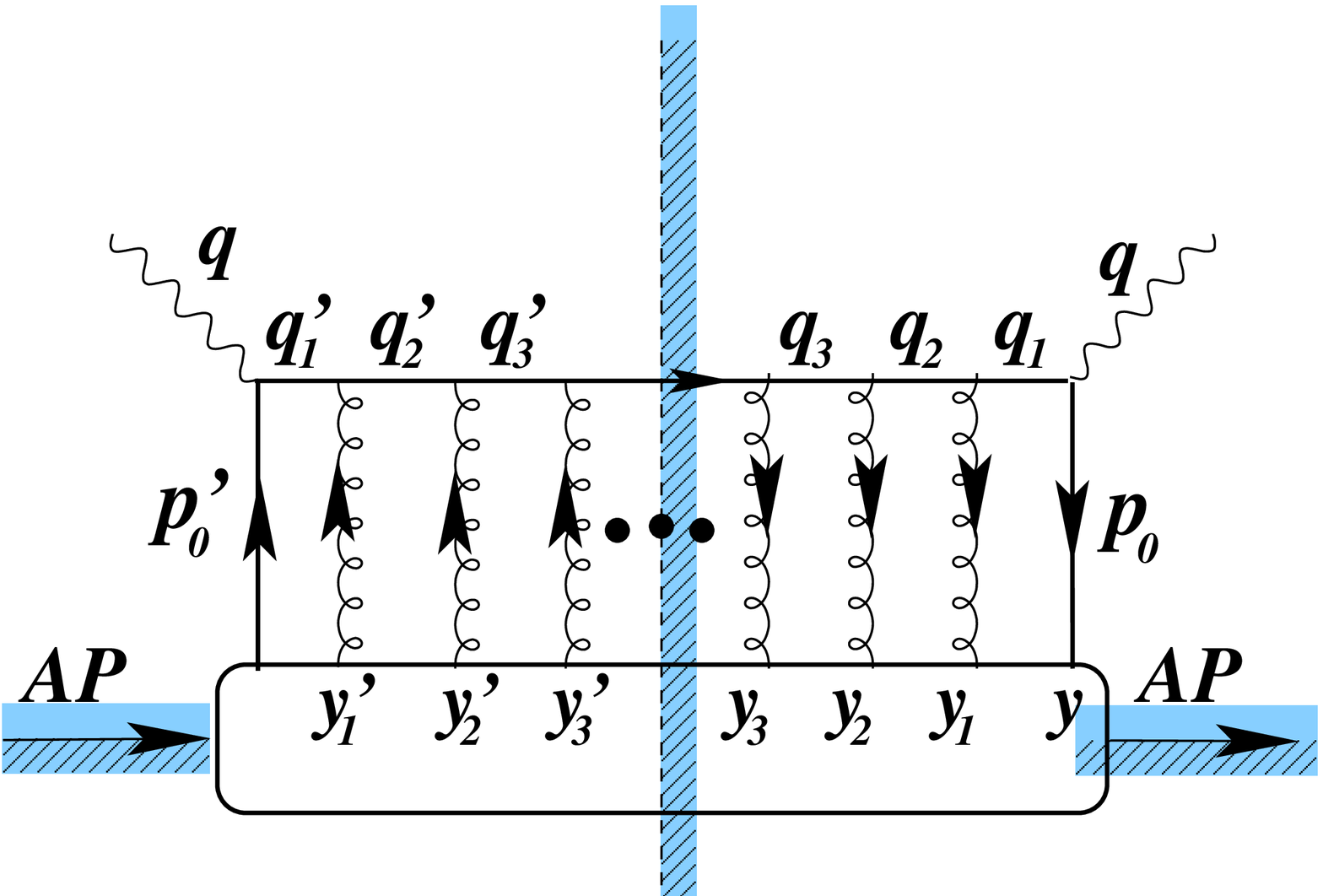,scale=0.5}\hfill
\epsfig{file=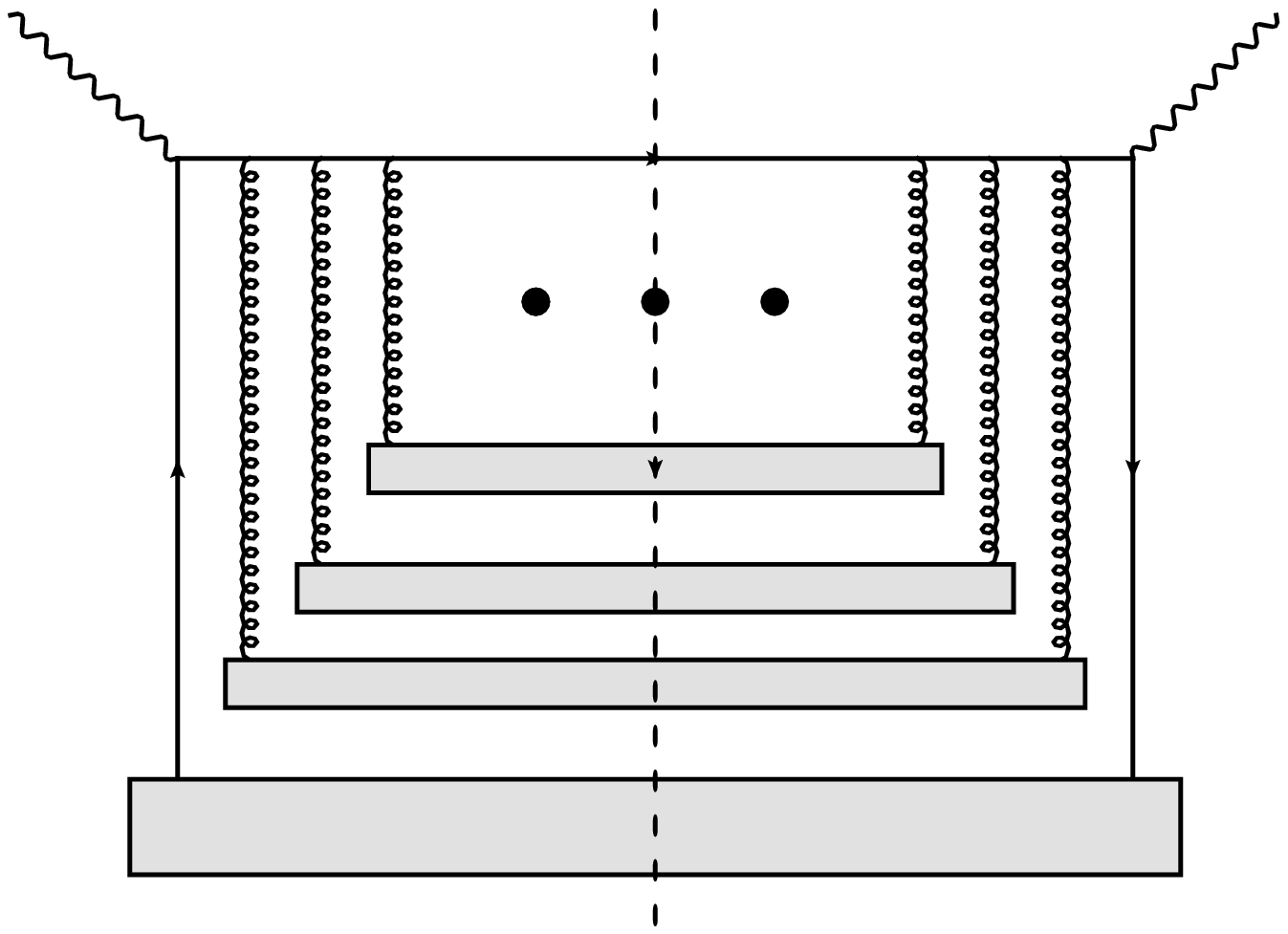,scale=0.5}
\end{minipage}
\begin{minipage}[t]{16.5 cm}
\caption{Left panel: DIS on a nucleus leading to the formation of a quark jet which 
which is constrained to propagate through the nucleus without radiating. 
Its only interactions are scatterings. Right panel: The dominant length enhancement 
arises from nested scattering diagrams, {where the blobs represent individual nucleons}, see text for details. \label{fig4}}
\end{minipage}
\end{center}
\end{figure}

We begin by considering the case of a hard quark produced in a 
DIS on a large nucleus, {which produces a hard quark propagating through the nucleus 
encountering multiple scattering without radiation}. 
The virtual photon strikes a hard quark in one of the nucleons and turns it back towards 
the nucleus. The quark then propagates through the nucleus without radiating. In this process the 
hard quark scatters $n$ times in the amplitude and in the complex conjugate. This is 
represented by the diagram in the left panel of Fig.~\ref{fig4}. The diagram represents a quark in 
one of the incoming nucleons with momentum $p_0'$, being struck with the photon with 
momentum $q = [-x_B P^+ , q^- , 0,0]$.  The outgoing quark has momentum $q_1' = p_0' + q$.
After encountering the $j^{\rm th}$ scattering, its momentum is 
$q_{j+1}'  = p_0 + q + \sum_{i=1}^j k_i'$. 

In the case of the higher-twist scheme, these gluon lines 
represent the gluon field at the point at which scattering takes place. Thus there is no 
meaning to crossing of gluon lines. The sources of the gluon lines are the nucleons in the 
nucleus (or rather the partons within these nucleons). 
The only assumption made regarding the gluon momenta are simple scaling 
relations regarding the momentum of the exchanged gluons and the magnitude of the 
corresponding components of the vector potentials. This is somewhat different in the 
other schemes. Under the assumption that the gluons transfer a transverse momentum 
of $k_\perp \sim \lambda Q$, and the incoming and outgoing quark lines remain close to 
on-shell, the $k^+ $ component is constrained with the leading contribution given by the equation, 
\bea
(q + k)^2 = 2q^- k^+ - k_\perp^2 = 0 \,\,\, \Longrightarrow \,\,\, k^+ = \frac{k_\perp^2}{2q^- } \sim \lambda^2 Q.
\eea
Ascribing the same criteria to the hard quark or gluon in the nucleons from which the 
exchanged gluon originates, implies that $k^- \simeq k_\perp^2/2p^+$ (where $p^+$ here is 
the momentum of some hard parton) also scales as $\sim \lambda^2 Q$. Exchanged 
gluons with transverse momenta much larger than their longitudinal components are 
referred to as Glauber gluons or Coulomb gluons. If the ($+$)-component were to 
become larger than $k_\perp^2/2q^-$,  it would drive the jet parton to go off-shell 
and radiate a hard gluon. If the $k^{-}$ component were to become larger, this 
will lead to energy loss from the jet parton but the parton in the nucleon will then 
go off shell and may radiate. As jet quenching measurements do not concern themselves 
with the fate of the target, the latter sort of momentum transfer is often referred to as 
elastic energy loss for the jet. 

Given these approximations we may now compute the scaling of the different 
components of the gauge field. To remind the reader we are calculating in the 
Breit frame where the hard quark moves in the $-z$ direction with a large 
light cone momentum $q^- \sim Q$ and the valence quarks inside the nucleon 
are moving in the $+z$ direction with a large $p^+ \sim Q$. 
We use the linear response formula to ascertain the power counting of the $A^+$ field.
Suppressing the color superscripts we obtain, 
\bea
A^\mu (x_1) &=& \int d^4 y_1 \mathcal{D}^{\mu \nu} (x_1-y_1) J_\nu (y_1) . \label{linear_resp}
\eea
In the equation above, $\mathcal{D}$ is the gluon propagator and at leading
order in the light cone gauge $A^- = 0$ is given as, 
\bea
\mathcal{D}^{\mu \nu} (x_1 - y_1 ) &=& \int \frac{d^4 k}{ (2 \pi)^4} 
\frac{ i \lt(  - \gmn +  \frac{k^\mu n^\nu + k^\nu n^\mu}{k\x n} \rt)  e^{-i k \x ( x_1 - y_1 )}  }{ k^2 + i \e}. 
\label{lightcone_gauge_gluon}
\eea
In Eq.~\eqref{linear_resp}, $J^{\nu} (y_1) = \psibar (y_1) \g^\nu \psi (y_1)$ is the current of partons in the target which generates the gluon field. 
The fermionic operator may be decomposed as,
\bea
\mbx\!\!\!\psi (y_1) \!\!&=&\!\! \int \frac{d p^+ d^2 p_\perp }{(2\pi)^3 \sqrt{p^+ + \frac{p_\perp^2}{2p^+}  }} 
\sum_s u^s(p) a_p^s e^{-ip\x y_1} + \ldots
\eea
The scaling of the fermionic operator depends on the range of momentum which are selected 
from the in-state by the annihilation operator. Note that this influences both the scaling of the 
annihilation operator $a_p$ as well as the bispinor $u(p)$. The power counting of the annihilation 
operator may be surmised from the standard anti-commutation relation, 
$
\{ a_p^r , {a_{p'}^s}^{\dag} \} = (2 \pi)^3 \kd^3 (\vp - \vp') \kd^{rs}.
$ 
The power counting of the bispinor can be obtained from the completeness relation,
$
\sum_s u_p^s \bar{u}_p^s = \f\! p = \g^- p^+ + \g^+ p^- - \g_\perp \x p_\perp
$. 
Substituting the equation for the current in Eq.~{\eqref{linear_resp}},
%\eqref{lightcone_gauge_gluon},
and integrating out 
$y$, we obtain, 

\bea
\mbx\!\!\!\!\!\!\!A^+ \!\!\!&\simeq&\! \!\!\!\int\!\! \frac{d^3 p d^3 q}{(2\pi)^6  \sqrt{p^+} \sqrt{q^+}}  
\frac{i \lt( - g^{+ -} + \frac{n^+ (p^- - q^-) }{(p^- - q^-)} \rt)  e^{-i(p-q)\x x_1} }{ (p-q)^2} a_q^\dag a_p 
\bar{u}(q) \g^+ u(q).
\eea
If the incoming and out going momenta $p$ and $q$ scale as collinear momenta in the ($+$)-direction, 
i.e., $p \sim Q(1,\lambda^2, \lambda)$, then we get, $\kd^3 ( \vp - \vp' ) \sim [ \lambda^2 Q^3 ]^{-1}$, as one 
of the momenta will involve the large ($+$)-component and the remaining are the small transverse 
components. Thus the annihilation (and creation) operator scales as $\lambda^{-1} Q^{-3/2}$. Also in the spin 
sum $\f\! p \sim Q$ and thus $u(p) \sim u(q) \sim Q^{1/2}$; one can check that the $\g^+$ 
projects out the large components in $u$ and  $\bar{u}$ in the expression $\bar{u}(q) \g^+ u(p)$. 
We also institute the Glauber condition that 
$p^+ - q^+ \sim \ld^2 Q$, $p^- - q^- \sim \ld^2 Q$ and $p_\perp - q_\perp \sim \ld Q$.

Using these scaling relations we correctly find that the bispinor scales as 
$\lambda Q^{3/2}$. However, to obtain the correct scaling of the gauge field $A^+$ 
one needs to institute the approximation that $q^+ = p^+ + k^+$ where $k^+ \sim \lambda^2 Q$.
This condition is introduced by insisting that the $(+)$ momentum of the incoming and outgoing 
state, which control the scaling of $a_q^\dag$ and $a_p$, are separated by $k^+ \sim \ld^2 Q$. This is used to 
shift the $dq^+ \ra d k^+$ and as a result we obtain the scaling of the $A^+$ field as $\ld^2 Q$.
Following a similar derivation, with the replacement $\g^+ \ra \g_\perp$ we obtain the scaling of the 
$A_\perp \sim \ld^3 Q$. Thus in the Breit frame and $A^-=0$ gauge, the $A^+$ field is dominant over 
the $A_\perp$ components which may be neglected. 

The evaluation of the right panel of Fig.~\ref{fig4} is straightforward within the approximations outlined above. 
The numerator 
of every propagator is replaced with $\g^+ q^-$ and every vertex with $\g^- A^+$. The denominators are 
simplified by contour integration,  e.g., for the simple case of the first propagator, we would get, 
\bea
\int\frac{ d k_1^+ }{2 \pi}\frac{ e^{- i k_1^+ (x_1^- - x_{0}^-)} }{ (q^- + k_1)^2   } 
\simeq \int \frac{d k_1^+}{ 2 \pi} \frac{ e^{- i k_1^+ (x_1^- - x_{0}^-)} }{ q^- k_1^+ - (k^1_\perp)^2 + i \e  } 
= - i \h (x_1^- - x_0^-)  e^{ - i \frac{ | k^1_\perp|^2 }{2 q^-} (x_1^- - x_0^- ) }.
\eea 
Thus carrying out the contour integrations, simply orders the locations of the scatterings and sets 
the ($+$) components of the momentum to be $k_\perp^2/2q^-$.
Carrying out these integrations on both sides of the cut we obtain an expression which is a modification 
of Eq.~\eqref{W_mu_nu}, 
\bea 
\frac{d W^{\mu \nu }}{d {l_q}_\perp^2} &=& \frac{ - g_\perp^{ \mu \nu}  g^{n + n'}  }{(2\pi)^2}  
\int \prod_{i=0}^n d y_i^- d^2 y_\perp^i 
\prod_{j=1}^{n'} d {y'}_j^- d^2 {y'}^j_\perp 
\int \prod_{i=0}^n \frac{ d^2 p^i_\perp}{(2\pi)^2} 
\prod_{j=0}^{n' - 1} \frac{d^2 {p'}^j_\perp} { (2\pi)^2} 
(2\pi)^2 \kd^2 ( \vec{l_q}_\perp - \sum_i\vec{k^i}_\perp ) \nn \\
\ata e^{-ix_B P^+ y^-} \prod_{i=0}^n e^{-ix_D^i P^+ y_i^- } 
e^{i p^i_\perp \x  y^i_\perp  } 
\prod_{j=0}^{n'} e^{i {x'}_D^j P^+ {y'}_j^- } 
e^{-i{p'}^j_\perp \x  {y'}^j_\perp  } 
\prod_{i=n}^1 \h ( y_i^-   - y_{i-1}^- ) 
\prod_{j = n'}^1 \h ( {y'}_j^-  -  {y'}_{j-1}^- ) \nn \\
\ata \lc A; p | \psibar(y^-,y_\perp) \frac{\g^+}{2} \psi(0) 
\tr \left[ \prod_{i=1}^{n} t^{a_i}  A_{a_i}^+ (y_i^-,y^i_\perp) 
\prod_{j=n'}^{1} t^{a_j} A_{a_j}^+ ( {y'}_j^-, {y'}^j_\perp ) \right]  | A;p \rc.
\label{W_mu_nu_simple}
\eea
In the equation above,  we use the short hand $ x_D^i = \lt[\slm_{k=0}^{i} 2p_\perp^i \cdot p_\perp^k + |p_\perp^i|^2 \rt]/2P^+q^-$ to save writing, and similarly for the complex conjugate.

Further evaluation 
requires two sets of approximations. The first arises from the large length limit of a large nucleus. 
Each of the $d y_i^-$ or $ d y_i'^-$ integrals may be extended up to the light-cone length of the 
nucleus $L^-$. The largest factor of $L^-$ is obtained in the limit that the gluon operators $A^+ (y_i)$ are 
distributed over the largest number of nucleons. The color singlet nature of nucleons requires that 
the extraction of a gluon in the amplitude from a particular nucleon be matched by an identical gluon 
being extracted from the same nucleon in the complex conjugate. This along with the strong ordering of 
positions of the scatterings restricts the calculation to the set of nested scattering diagrams where 
the locations $y_{i}$ and $y'_{i}$ are within the same nucleon. These diagrams are represented by the 
right panel of Fig.~\ref{fig4}. 
Diagrams with more gluons per nucleon 
have smaller length enhancement factors, this is the reason that pomeron like exchanges which 
include at least four gluons per nucleon are suppressed in the large nucleus limit.
Since the transverse size of a nucleon is 
much larger than the inverse transverse momentum carried by either exchanged gluon, one may integrate 
over the mean transverse location of the two insertions. 
This equates the transverse momentum that is exchanged between the jet and the nucleon, 
in the amplitude and the complex conjugate. For the $i^{\rm th}$ scattering, this may be expressed as
\bea
\mbx & & 
\int d y_{i}^{-} d {y'}_{i}^{-} d^2 y^i_\perp  d^2 {y'}^i_\perp 
\lc p | A^{+}  (\vec{y}^i_\perp )  A^{+}  ( \vec{y'}^i_\perp ) | p \rc
e^{-i x_D^i p^+ y_i^-}  e^{ i p^i_\perp \x y^i_\perp}   
e^{i {x'}_D^i p^+ {y'}_i^-}  e^{ - i {p'}^i_\perp \x {y'}^j_\perp}  \nn \\
&=& (2\pi)^2 \kd^2( {\vp}^{\,i}_\perp - {\vec{p'}} ^i_\perp )  \int dy_{i}^{-} d {y'}_{i}^{-} d^2 y_\perp  e^{-i x_D^i p^+ ( y_i^-  -  {y'}_i^- )} 
% \frac{-g^{\A \B}_{\perp}}{2} \nn \\ 
%
\times 
 e^{ i p_\perp \x y_\perp} \lc p | A^{+} (\vec{y}_\perp/2 )  A^{+}  ( - \vec{y}_\perp/2 ) | p \rc . \label{two_gluon_cor}
 \eea 
Given this pairing of interactions, the largest length enhancement that may be obtained is $L^{n}$ for 
$n$ scatterings in the amplitude and complex conjugate.

The other approximation is that the scatterings transfer small amounts of transverse momentum
compared to the transverse momentum of the original quark, though both are at the same scale 
$\lambda Q$.
The possibility of scattering over $n$ nucleons, greatly enhances the transverse momentum exchanged 
between the jet and the medium. 
Thus the transverse momenta from these scatterings cannot be ignored. However, since each is small 
compared to the transverse momentum of the jet, we may Taylor expand in them. 
In the case of spin independent cross sections, the leading term in the Taylor expansion is the 
second derivative of the $\kd$-function in terms of each of the small momenta $k_{\perp}^{i}$: 
\bea
\kd^2 ( \vec{l_q}_\perp - \sum_i\vec{k^i}_\perp ) 
\simeq \prod_{i=1}^{n} \frac{\lt( k^{i}_{\perp} \cdot \nabla_{ {l_{q}}_{\perp} } \rt)^{2} }{2} \kd^{2}(\vec{l_q}_\perp) 
\label{transverse_delta}
\eea
The two factors of $k_{\perp}^{i}$ can be combined with the two gluon vector potentials in 
Eq.~\eqref{two_gluon_cor} to convert them into the correlator of field strengths 
$\lc p |F^{+ \perp}(y_{i}^{-}) F_{\perp}^{+} (y_{i}'^{-}) | p \rc$. Higher transverse momentum 
derivatives will involve more factors of $k_{\perp}$ which may be converted through by-parts 
integration into derivatives of the field strength. Thus these are further suppressed without any 
extra length integration to enhance them. Thus we stop at just the double derivative in the Taylor 
expansion of the $\kd$-function in terms of the transverse momentum. 
 
Incorporating these approximations, one obtains the differential hadronic tensor as,
\bea 
\mbox{} \!\!\!\!\frac{d^2 W_n^{\mu \nu}}{d^2 l_\perp} \!\!\!\!\!\!&=& \!\!\!\!
W_0^{\mu \nu} \frac{1}{n!} \left[ \{\nabla^2_{l_\perp}\}^n \kd^2 (\vec{l}_\perp) \right] 
 \left[ \frac{\pi^2 \A_s}{N_c} L^- \int \frac{dy^-}{2\pi} \frac{d^{2} y_{\perp} }{(2 \pi)^{2}}  
e^{-i \lt( \frac{k_{\perp}^{2}}{2 q^{-}} + k_{\perp}\cdot y_{\perp} \rt)   } 
\lc p | {F^a}^{+ \A}(y^{-},y_{\perp}) {F^a}_{\A,}^{\,\,\, +}| p \rc \right]^n.  \label{W_n}
\eea
The latter quantity in square brackets is referred to as $D$. This series may be resummed by 
noting that $\phi({l_{q}}_{\perp}) = d W^{\mu \nu}/d^{2} {l_{q}}_{\perp}$ obeys the diffusion equation, 
\bea
\frac{\prt \phi }{\prt L^{-}} = D \nabla^{2}_{{l_{q}}_{\perp}} \phi,
\eea
which has a normalized solution given as 
\bea
\phi(L^-,\vec{l_{q}}_\perp) = \frac{1}{4 \pi D L^-} \exp \left\{-  \frac{{l_{q}}_\perp^2}{4 D L^-} \right\}. \label{solution}
\eea
The transverse momentum squared gained by a parton as it traverses a length $L^{-}$ is given as 
$|{l_{q}}_{\perp}|_{L^{-}}^{2} = 4 D L^{-} = \hat{q} L^{-}$. 
Thus the jet quenching transport coefficient $\hat{q} = 4 D$.

%%%%%%%%%%%%%%%%%%%%%%%%%%%%%%%%%%%%%%%%%%
%%%%%%%%%%%%%%%%%%%%%%%%%%%%%%%%%%%%%%%%%%
%%%%%%%%%%%%%%%%%%%%%%%%%%%%%%%%%%%%%%%%%%

\subsubsection{{Multiple scattering of quark and radiated gluon}}~\label{single_gluon:multiple_scattering:radiation}

%%%%%%%%%%%%%%%%%%%%%%%%%%%%%%%%%%%%%%%%%%
%%%%%%%%%%%%%%%%%%%%%%%%%%%%%%%%%%%%%%%%%%
%%%%%%%%%%%%%%%%%%%%%%%%%%%%%%%%%%%%%%%%%%

\begin{figure}[tb]

\epsfysize=9.0cm

\begin{center}

\begin{minipage}[t]{8 cm}

\epsfig{file=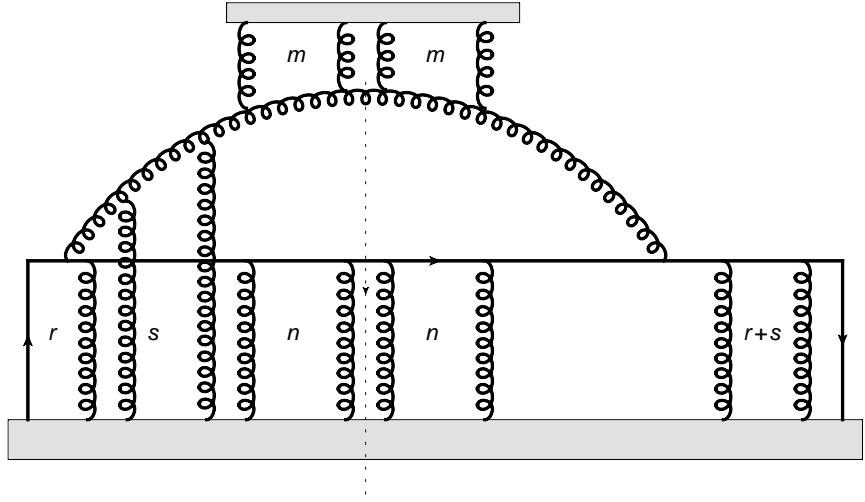,scale=0.8}

\end{minipage}

\begin{minipage}[t]{16.5 cm}

\caption{DIS on a nucleus leading to the formation of a quark jet with a radiated gluon. \label{fig5}}

\end{minipage}

\end{center}

\end{figure}

With the derivation of the single gluon emission cross section and the multiple scattering cross section 
for a single quark without radiation, we can surmise the form of the multiple scattering single 
gluon emission cross section. Consider the general process depicted in Fig.~\ref{fig5}. 
A quark is produced in the DIS on a large nucleus at location $y_{0}$. 
We consider the process where, in the amplitude, this quark will radiate a gluon without a re-scattering 
before the vertex at which the hard gluon is radiated. The produced 
quark and gluon will then scatter $r$ and $s$ times, respectively, on the incoming nucleons. 
The first location where the gluon scatters will be denoted as $\zeta_{C}$. The first 
location where the quark scatters will be denoted as $y_{C}$ 
There 
will be additional scatterings but the $q=r+s$ scatterings are special as they represent the 
interference contributions. 
On the complex conjugate side, the 
produced quark will itself scatter $r+s=q$ times on the same nucleons and then radiate the gluon. 
The location of the last such scattering which occurs on the parent quark in the complex conjugate and on either the 
produced quark or radiated gluon in the amplitude will be denoted as $y_{E}$. 
Thus the radiated gluon is produced at two separate points: it is just after $y_{0}^{-}$ in the amplitude and 
just after $y_{E}^{-}$ in the complex conjugate.
After this cross scattering, the gluon will scatter $m$ times and the quark $n$ times in both the amplitude 
and the complex conjugate. 
The location of the first independent scattering on the gluon is at 
location $\zeta_{I}^{-}$ and that on the quark at location $y_{I}^{-}$. 

The Feynman integral for this process may be expressed as (see Ref.~\cite{Majumder:2009ge} for a 
detailed derivation of this equation), 
\bea
\Op \!\!\!\!\!&=& 
\!\!\!\!\!\int\!\!\frac{dy d^2l_\perp d^2 {l_q}_\perp }{2\pi^2} \frac{\A_s C_F P(y)}{ y}
C_A^m C_F^n (C_F - C_A/2)^r (C_A/2)^s 
\kd^2 \!\!\lt( {l_q}_\perp - \sum_{i=1}^s k^i_\perp - \sum_{j=1}^r p^j_\perp  
- \sum_{l=1}^m k^l_\perp - \sum_{k=1}^n p^k_\perp \rt)  \nn \\
\ata \frac{ l_\perp -  \slm_{i=1}^s k^i_\perp  - \slm_{l=1}^m k^l_\perp  }
{ \lt( l_\perp -  \slm_{i=1}^s k^i_\perp -  \slm_{l=1}^m k^l_\perp\rt)^2 }
\x \frac{ l_\perp -  y\slm_{i=1}^{r+s} k^i_\perp -  \slm_{l=1}^m k^l_\perp  }
{ \lt( l_\perp - y \slm_{i=1}^s k^i_\perp -  \slm_{l=1}^m k^l_\perp\rt)^2 }  \nn \\
\ata \prod_{i=1}^{q} \int d y_i^- \frac{\int d^{3} \kd y_{i} \rho \lc p | A^{+}(y_i^-+\kd y_{i}^{-},0)  A^{+}(y_{i}^{-},-\kd y_{\perp}^{i}) | p \rc }
{2 p^{+}(N_c^2 - 1)}  e^{i k_{\perp}^{i} \kd y_{\perp}^{i}}
\nn \\
\ata 
\prod_{j=1}^{n}  \int d y_j^- \frac{ \int d^{3} \kd y_{j} \rho \lc p | A^{+}(y_j^- + \kd y_{j}^{-},0)   A^{+}( y_{j}^{-}, -\kd y_{\perp}^{j} ) | p \rc }
{ 2p^{+}(N_c^2 - 1)} e^{i k_{\perp}^{j} \kd y_{\perp}^{j} } \nn \\
\ata \prod_{l=1}^{m} 
\int d\zeta_l^- \frac{ \int d^{3} \kd \zeta_{l} \lc p | A^{+}(\zeta_l^- + \kd \zeta_{l}^{-} , 0 )   A^{+} ( \zeta_{l}^{-} ,- \kd \zeta_{\perp}^{l} ) | p \rc} 
{(N_c^2 - 1)}  e^{i k_{\perp}^{l} \kd \zeta_{\perp}^{l}} \nn \\
\ata \lt[ \h(\zeta_I^- - y_E^-)  \lt\{ e^{-ip^+ x_L y_E^-   }  
-  e^{ - ip^+ x_L \zeta_I^-  }  \rt\} 
-  \h( \zeta_I^- - y_I^- )  e^{ - ip^+ x_L y_I^-  }
-  \h( y_I^- - \zeta_I^- )  e^{ -ip^+ x_L \zeta_I^-   }  \rt] \nn \\
\ata \lt[ \h( \zeta_C^- - y_0^- )  \lt\{ e^{ i p^+ x_L y_0^-   }  
-  e^{  i p^+ x_L  \zeta_C^-  }  \rt\} 
- \h( \zeta_C^- - y_C^- )  e^{ i p^+ x_L  y_C^- } 
- \h( y_C^- - \zeta_C^- )  e^{ i p^+ x_L \zeta_C^- }  \rt].  \label{all_twist_non_amy}
\eea
In the equation above $q=r+s$.
While not explicitly mentioned, all scattering points on the same line (quark or gluon) are 
strongly ordered. The only unspecified orderings are scatterings on the gluon versus those on the 
quark after emission.
This expression may be easily generalized from Eqs.~(\ref{W_mu_nu_gluon},\ref{O_00_3},\ref{W_mu_nu_simple})
The $\kd$-function $\kd(l_{\perp} + { l_{q} }_{\perp})$ is shifted to $\kd({l_{q}}_{\perp})$ and is then 
broadened by multiple scattering as in Eq.~\eqref{W_mu_nu_simple}. There are now $r+s+m+n$ two 
gluon matrix elements corresponding to the number of scatterings.

While there is no simple means to derive the second line of Eq.~\eqref{all_twist_non_amy}, it {can be qualitatively understood as follows.} Consider the case of gluon 
radiation without scattering as described in Sect.~\ref{single_gluon:vacuum_radiation}. 
The factor of $1/l_{\perp}^{2}$ in 
Eqs.~(\ref{O_00_3},\ref{W_mu_nu_gluon}) may be more naturally expressed as 
\bea
\frac{1}{l_{\perp}^{2}} = \sum_{\lambda} \lt( \frac{  l_{\perp} \cdot \varepsilon_{\lambda}^{*} }{l_\perp^2} \rt) \lt( \frac{ \varepsilon_{\lambda} \cdot l_{\perp} }{l_{\perp}^{2}} \rt),
\eea
where $l_\perp$ is the transverse momentum of the gluon just after emission and $\varepsilon$ is 
the polarization vector of the produced gluon. This expression naturally arises in the case of 
one gluon bremsstrahlung from a hard quark by evaluating the square of the matrix 
element $\mathcal{M} \sim J^{a \mu} A^{a}_{\mu}$. 
This will be modified in the multiple scattering case. 
In $A^{-}=0$ gauge with the gluon moving in the $(-)$-direction, the dominant 
components of the projection vectors are the transverse components.
In the case of no scattering the transverse momentum at emission 
is identical in the amplitude and complex conjugate. In the case of multiple scattering, 
the transverse momentum of the radiated gluon in the amplitude and complex conjugate is identical 
at the cut line but not at the point of emission. Thus, at the point of emission, the factor is modified as,  
\bea
\sum_{\lambda} \frac{ \lt( l_\perp -  \slm_{i=1}^s k^i_\perp  - \slm_{l=1}^m k^l_\perp   \rt)  \cdot 
\varepsilon^{*}_{\lambda}  }
{ \lt( l_\perp -  \slm_{i=1}^s k^i_\perp -  \slm_{l=1}^m k^l_\perp\rt)^2 }
\frac{ \varepsilon_{\lambda} \x \lt( l_\perp -  y\slm_{i=1}^{r+s} k^i_\perp -  \slm_{l=1}^m k^l_\perp \rt) }
{ \lt( l_\perp - y \slm_{i=1}^s k^i_\perp -  \slm_{l=1}^m k^l_\perp\rt)^2 }.  \label{num-factor}
\eea
Summing over the projections of the radiated gluons, yields the second line of Eq.~\eqref{all_twist_non_amy}.
In the amplitude part (the right side of the expression), the gluon is radiated after the $r+s$ 
scattering on the original quark. Of the transverse momenta gained by the parent quark in the 
$r+s$ scattering, only a fraction $y$ is transferred to the radiated gluon. This is the reason for the 
factor of $y$ in the amplitude terms in Eq.~\eqref{num-factor}.

The third, fourth and fifth line of Eq.~\eqref{all_twist_non_amy} contain the products of the
two-gluon matrix elements at the locations of the various scatterings. The first set on the 
third line contains 
the collection of correlated scatterings where the parent quark scatters in the amplitude at the 
locations $i=1$ to $q$ and the produced quark and gluon scatter off these same locations
in the complex conjugate. The next set of scatterings on the fourth line 
at locations $j=1$ to $n$ represent the 
scatterings of the final produced quark in both amplitude and complex conjugate, i.e., there is
no interference in these scatterings. The last set on the fifth line represent the 
independent scatterings on the final outgoing gluon line.

\begin{figure}[tb]

\epsfysize=9.0cm

\begin{center}

\begin{minipage}[t]{8 cm}

\epsfig{file=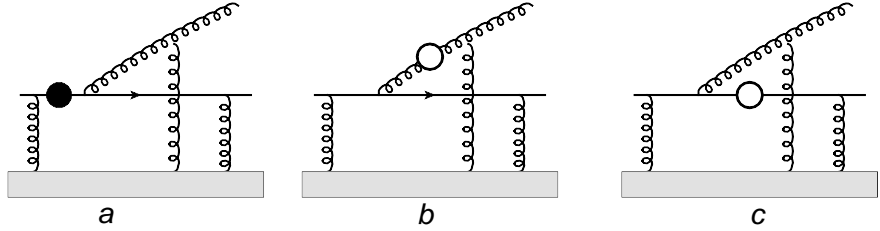,scale=0.8}

\end{minipage}

\begin{minipage}[t]{16.5 cm}

\caption{Diagrams which lead to LPM interference. \label{fig6}}

\end{minipage}

\end{center}

\end{figure}

The last two lines of Eq.~\eqref{all_twist_non_amy} contain the phase factors that arise from the 
interference between the gluon emission from different locations in the amplitude and complex 
conjugate. 
The phase factors may be somewhat more straightforwardly motivated. 
There are in principle three types of vertices that will 
lead to the radiation of a hard gluon. These are illustrated in Fig.~\ref{fig6}. The left-most 
diagram represents the case where the incoming quark is taken time-like off-shell, indicated 
by the filled dot. 
In the first term of the phase factors in Eq.~\eqref{all_twist_non_amy} [sixth line of Eq.~\eqref{all_twist_non_amy}] 
the quark in the complex conjugate 
is taken time-like off-shell by the scattering at $y_{E}^{-}$. This immediately introduces a phase factor of 
$\exp( -i p^{+} x_{L} y_{E}^{-} )$; the quark goes off-shell and immediately radiates a gluon. The 
$\h$-function expresses the fact the the other propagators are all on shell and thus the location $\zeta_{I}$ 
is far ahead of $y_{E}^{-}$. 

The second diagram in Fig.~\ref{fig6} expresses the possibility that the quark after the scattering remains 
on shell and 
radiates a space-like gluon which is then brought on-shell by the first independent scattering on the 
gluon. In the second term of the phase factor of Eq.~\eqref{all_twist_non_amy}, this occurs in the 
complex conjugate at location $\zeta_{I} $. 
The exponential factor is $\exp( -i p^{+} x_{L} y_{E}^{-} )$. The
 negative sign is due to the fact that the propagator goes space-like. This in indicated by the hollow 
blob in the figure. Such a contribution occurs twice in the calculation: it occurs in the case where
the initial quark propagator is taken on shell first or when the final quark propagator is taken on shell  
first. The fourth exponential factor is the instance of the latter pole in the complex conjugate. The third 
phase factor corresponds to the case where the on-shell quark radiates an on-shell gluon and goes 
off-shell. It is brought back on-shell by a hard scattering. In the case of the quark in the complex 
conjugate in Eq.~\eqref{all_twist_non_amy}, this occurs at location $y_I^-$ and has the 
{corresponding} exponential factor.
The remaining phase factors for the gluon emission in the amplitude have similar explanations and 
thus similar phase factors.

\begin{figure}[tb]
\epsfysize=9.0cm
\begin{center}
\begin{minipage}[t]{8 cm}
\epsfig{file=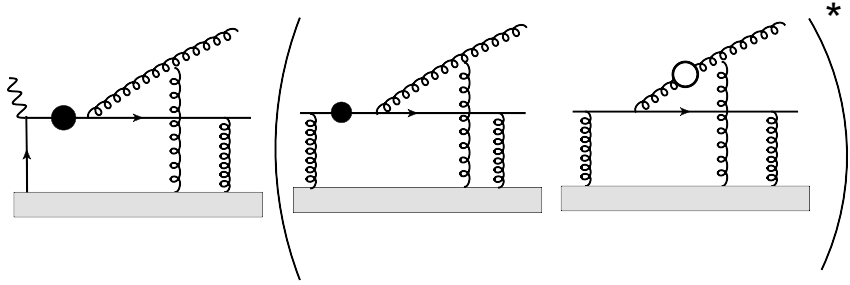,scale=0.8}
\end{minipage}
\begin{minipage}[t]{16.5 cm}
\caption{Diagrams which are leading in the higher twist approach. \label{fig7}}
\end{minipage}
\end{center}
\end{figure}

In the higher twist formalism, one takes the hard scattering hierarchy, in the sense that the hard scattering
scale $Q$ is assumed to be much larger than the medium scale $\sim T $ or the scale of Debye screening. 
Thus, one tries to isolate terms 
which are minimally suppressed by the hard scale $Q$. This is obtained, as in the case of no emission, by
Taylor expanding the small transverse momenta $k^i_\perp$. The expansion of the $\kd$-function 
can be shown to yield a similar diffusion equation as for the case of no emission. The emission yields 
a factor of the splitting function $P(y)$ and phase factors which appear in the last line of 
Eq.~\eqref{all_twist_non_amy}. Of these, the phase factor which is dependent on the origin ($e^{  i p^{+} x_{L} y_{0}^{-} }$) represents the terms where the jet exits the hard interaction off-shell and immediately 
radiates a hard gluon. This is represented by the leftmost diagram in Fig.~\ref{fig7}. Terms containing 
this factor in a product with the complex conjugates of similar diagrams as well as those where the 
hard parton is driven off shell later in its history are representative of vacuum like emission and 
vacuum medium interference. These are the leading contributions in the 
higher twist formalism. A subset of these are represented by the diagrams of Fig.~\ref{fig7}. One also 
has to include contributions where the quark propagator after emission becomes space-like off-shell. 

Including these contributions along with the contributions where all three types of scattering may not 
be present and then Taylor expanding in $k^i_\perp$, we obtain the next-to-leading power suppressed contributions, 
i.e., those suppressed by one extra power of $Q^2$ (or rather $l_\perp^2$) as a sum of three terms each with a 
different power of the gluon momentum fraction $y$. The leading contribution to the radiated gluon spectrum in 
$y$ arises from purely gluon scattering, and is given as 
\bea
\frac{dN_1}{dy dl_{\perp}^{2}} &=& \frac{\A_{s} C_{F}}{2 \pi l_{\perp}^{4}} P(y)   \int_{0}^{L^{-}} d \zeta^{-} 
\hat{q} (\zeta^{-} ) \lt[  2  - 2\cos \lt( p^{+}  x_{L}  \zeta^{-}  \rt) \rt]. \label{final_res} 
\eea
We have dropped the subscript $I$ from the position $\zeta_I^-$ in writing the above equation. 
The next contribution is proportional to $y$ and includes the cross-scattering terms between the 
final radiated gluon and the quark before the split. This contribution to the radiated gluon 
spectrum is given as
\bea
\frac{dN_2}{dy dl_{\perp}^{2}} &=& -\frac{\A_{s} C_{F}}{2 \pi l_{\perp}^{4}} y P(y)  \int_{0}^{L^{-}} d \zeta^{-} 
\frac{\hat{q} (\zeta^{-} )  }{2}\lt[  2  - 2\cos \lt( p^{+}  x_{L}  \zeta^{-}  \rt) \rt].\label{final_res_y}
\eea
The last contribution at this level of power suppression, suppressed by two powers of $y$, 
is the set of diagrams where the initial quark scatters in the amplitude and the 
complex conjugate. These diagrams were not included in the Eq.~\eqref{all_twist_non_amy} or in the corresponding 
diagrams in Fig.~\ref{fig5}. The contributions from these are
\bea
\frac{dN_3}{dy dl_{\perp}^{2}} &=& \frac{\A_{s} C_{F}}{2 \pi l_{\perp}^{4}} y^{2}P(y)  \int_{0}^{L^{-}} d \zeta^{-} 
 C_F \hat{q}(\zeta^{-} ) \lt[  2  - 2\cos \lt( p^{+}  x_{L}  \zeta^{-}  \rt) \rt].\label{final_res_3} 
\eea
Summing the above three equations yields the next-to-leading power correction due the single 
gluon emission cross section due to multiple scattering of the parent quark, radiated gluon as 
well as the produced quark. The argument for terminating at this order of power correction 
is the hard factorization limit of $Q^2 \gg \hat{q} L^-$, i.e., higher power corrections are 
further suppressed.

%%%%%%%%%%%%%%%%%%%%%%%%%%%%%%%%%%%%%%%%%%
%%%%%%%%%%%%%%%%%%%%%%%%%%%%%%%%%%%%%%%%%%
%%%%%%%%%%%%%%%%%%%%%%%%%%%%%%%%%%%%%%%%%%

\subsubsection{{Length dependence of energy loss and extensions to heavy-ion reactions}}~\label{single_gluon:multiple_scattering:length_dependence}
%
%%%%%%%%%%%%%%%%%%%%%%%%%%%%%%%%%%%%%%%%%%
%%%%%%%%%%%%%%%%%%%%%%%%%%%%%%%%%%%%%%%%%%
%%%%%%%%%%%%%%%%%%%%%%%%%%%%%%%%%%%%%%%%%%

{In general, the application of Eqs.~(\ref{final_res},\ref{final_res_y},\ref{final_res_3}) to the 
computation of a medium modified fragmentation function tends to yield a complicated 
dependence on the length $L^{-}$ of the medium. \amnew{However, in earlier 
publications on the higher twist method,  one presents a simple 
and approximate relation for the fractional energy lost in the small $y$ limit to 
illustrate the length dependence of the energy loss. This has recently been criticized in 
Refs.~\cite{Aurenche:2008hm,Aurenche:2008mq}. In the next few paragraphs we 
outline this estimate and review the critique against it.}

\amnew{In the small $y$ limit, one may focus solely on the first term of the differential gluon 
radiation spectrum, i.e., Eq.~\eqref{final_res}. The fractional energy loss may then 
be estimated as, 
\bea
\frac{ \lc \Delta E \rc }{ E } = \lc y \rc = \int_{y_{min}}^{y_{max}}  dy 
\int_{0}^{\infty} d l^{2}_{\perp} \int_{0}^{L^{-}}  d \zeta^{-} y \frac{dN_{g}}{dy dl_{\perp}^{2}} . 
\eea
We say ``estimated'' as the integral over $l_{\perp}$ is integrated from zero to infinity which is never carried out 
in realistic HT calculations. Note that both limits violate the basic assumptions of the higher twist 
method: Being an expansion in inverse powers of the hard scale $l_\perp^2$, the radiated gluon's transverse
momentum cannot be taken to zero without resumming all higher order contributions which have been dropped in 
arriving at Eq.~(\ref{final_res}). Also the maximum radiated transverse momentum is limited by the virtuality $Q$ of the 
jet, which in turn is limited by the energy $q^-$, i.e., $l_\perp < Q \ll q^-$. Thus $l_\perp$ cannot be taken to 
infinity with $q^-$ remaining finite. }

\amnew{Ignoring these constraints, a simple minded estimate may be obtained by carrying out 
the $l_{\perp}$ integration from $0$ to $\infty$ first and 
retaining only the leading term in $y$. We obtain, 
\bea
\lc y \rc = \int_{y_{\min}}^{y_{max}}d y \frac{\A_{s} C_{F}P(y)}{4 q^{-}} \int_{0}^{L^{-}} d\zeta^{-} \hat{q} \zeta^{-}
\eea
In a medium the range of $y$ is limited by the fact that the momentum fraction 
$x_{L} = \frac{ Q^{2} }{ 2 q^{-} p^{+}y (1-y) } <1$, i.e., a proton can at most impart its entire light-cone 
momentum to the jet in a single gluon exchange. Thus the $y$ integration is not divergent. Assuming 
that $\hat{q}$ is independent of the location $\zeta^{-}$ we obtain the simple and well known 
relation that, 
\bea
\lc \Delta E \rc \propto \hat{q} {L^{-}}^{2},
\eea
with the constant of proportionality depending on the range of $y$ integration. This, in turn, depends on 
the maximum of  $l_{\perp}^{2} = Q^{2}$  and on the energy $q^{-}$ of the hard jet. 
Note that one cannot replace $Q^2$ by $\infty$ in this part of the calculation.
We should 
point out that this relation is quite approximate and in any realistic calculation 
the modification of the fragmentation function 
has a much more involved length dependence that what is suggested by the above equation.}

\amnew{Hand waving estimates such as the above have come under some criticism recently for an entirely different reason. 
In obtaining the phase factors in Eq.~(\ref{all_twist_non_amy}) all factors of $k_\perp$ in the phases have been 
dropped. This is justified as $k_\perp$ dependent phase factors are generically of the form $\exp[i \frac{k_\perp^2}{2q^- y(1-y)} \zeta^-]$,
and differentiating these with respect to $k_\perp^2$ leads to multiplicative factors of the form $\zeta^- / q^-$,  which
in the large $q^-$ limit are small. This assumes that the coefficients of these terms are small. As the authors of  
 Refs.~\cite{Aurenche:2008hm,Aurenche:2008mq} have shown, this is no longer the case when one integrates over $l_\perp$  
and allows $l_\perp$ to become large as in the estimate above. When $l_\perp$ is integrated up to $\infty$ the transverse 
momentum derivatives of the phase factors cancel those from the hard part. However in the limit that $\frac{l_\perp^2 \zeta^-}{2 q^- y (1-y)} \lesssim 1  $ derivativatives of the phase factors are small and can be ignored.} 

As pointed out previously, the HT approach is cast in the framework of 
DIS on a large nucleus. The medium in this case is confined. However the basic formalism 
may be straightforwardly extended to the case of a jet propagating through a deconfined 
medium. This extension is illustrated with the case of transverse broadening as in 
Eq.~\eqref{W_n}. In this case the part of the hadronic tensor (or cross section) where the 
virtual photon strikes an incoming quark and sends it into the nucleus is entirely 
contained within $W_0^{\mu \nu}$. This part, which is specific to DIS, has been factorized from the remaining 
process which represents the multiple scattering of the hard quark in the medium.

Extending this formalism to the case of jet propagation in a hot deconfined medium consists of 
replacing the hard cross section  to produce a hard quark in DIS: 
\bea 
\frac{E_{L_2} d \sigma_0 } {d^3 L_2 d^2 l_{q_\perp}  d^2 l_\perp dy  } &=&
\frac{\A_{em} ^2}{2\pi s  Q^4}  
\frac{L_{\mu \nu}  d W_0^{\mu \nu}}{d^2 l_{q_\perp} d^2 l_\perp dy} \;  
= \; \frac{\A_{em} ^2}{2\pi s  Q^4}  
\frac{L_{\mu \nu}  (- g_\perp^{\mu \nu})\sum_q 2 \pi Q_q^2 f_q (x_B) \kd^2 (\vec{l_q}_\perp) }
{d^2 l_{q_\perp} d^2 l_\perp dy} ,
\label{LO_cross_2}
\eea
with the cross section to produce a hard quark or gluon in a heavy-ion collision with a momentum $\hat{p}_T$,
\bea
\frac{d^2 \sg^h}{dy d^2 \hat{p}_T} = \frac{1}{\pi}\int dx_a  \, G_a(x_a,Q) G_b(x_b,Q) 
\frac{d \sg_{ab \ra cX} (Q,x_{a}P,x_{b}P, \hat{p}_{T} )}{d \hat{t}} .
\eea
What remains is the final state multiple scattering of the produced 
quark. This is assumed to be formally identical to the case of jet 
propagation in cold matter. The only change is that the expectation 
of the two gluon operator which is an input in the expression for the 
transport coefficient $\hat{q}$ Eq.~\eqref{W_n} will now be evaluated in a hot deconfined 
medium, i.e., 
\bea
\hat{q}_{QGP} = \frac{ 4 \pi^2 \A_s C_R  }{N_c^2 - 1} \int \frac{dy^- d^2 y_\perp}{2\pi} 
e^{-i \lt( \frac{k_\perp^2}{2 q^- } y^-  - k_\perp \x y_\perp \rt)  }
\sum \lc n | e^{ - \B\mathcal{H}} {F^a}^{+ \mu} (y^-,y_\perp)    {F^a}^+_\mu (0,0) | n \rc \label{qhat_FF}
\eea
In the equation above, $| n \rc$ represents a state in a thermal ensemble and $\mathcal{H}$
is the hamiltonian operator and $\B$ is the inverse temperature. The exact 
value of $\hat{q}$ obtained depends on the methodology used in modeling the medium. For 
example, assuming that the medium can be represented as a weakly coupled plasma of quarks 
and gluons in the Hard Thermal Loop (HTL) approximation yields the expression in Eq.~\eqref{HTL_qhat}.

\newpage
\subsection{Other approaches and pictures of the medium}~\label{single_gluon:other_approaches}

\subsubsection{AMY approach: {Hard Thermal Loop field theory}}  \label{single_gluon:other_approaches:AMY}

Unlike the HT approach, the formalism of jet modification 
based on the work of Arnold, Moore and Yaffe (AMY) starts with a rather precise 
definition of the medium. The medium is assumed to be composed of quark gluon 
quasi-particles with dispersion relations and interactions given by the HTL effective
theory. Thus all quasi-particles in the medium have thermal masses $\sim gT$ and 
their scattering is dominated by soft scattering. The hard jet is assumed to have a 
virtuality scale comparable to the Debye mass or thermal mass. Thus there is 
no jet-like hard scale in the problem. One constructs a consistent perturbation 
theory at the $T \ra \infty$ limit and applies it to a realistic temperature.

In spite of these differences, at the diagram by diagram level, the AMY approach evaluates diagrams 
which are topologically equivalent to those evaluated by the higher twist (HT) approach. To illustrate 
this we consider a generic diagram in the HT approach and convert it to a diagram similar 
to those evaluated in the AMY scheme. In the left top panel of Fig.~\ref{fig8}, we consider the 
case that a hard quark produced in a hard scattering then radiates a hard gluon and both final 
partons scatter once off the medium. While in the case of HT, the dominant contribution contains 
the case where the initial quark could be quite off-shell and radiates immediately, this is not the case in 
AMY. The incoming parton is close to on-shell and will radiate a hard gluon only on being 
stimulated by scattering. As a result, the transverse momentum of the radiated gluon is always of the 
the order of $gT$. 

\begin{figure}[ht!]
\epsfysize=9.0cm
\begin{center}
\begin{minipage}[t]{8 cm}
\epsfig{file=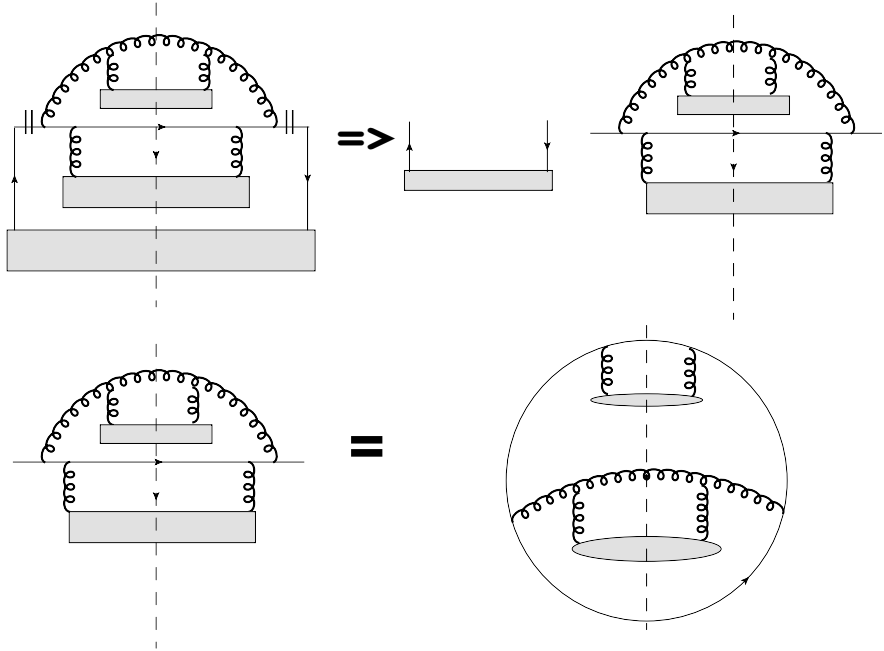,scale=0.8}
\end{minipage}
\begin{minipage}[t]{16.5 cm}
\caption{Starting with a higher twist diagram and obtaining the equivalent AMY diagram. \label{fig8}}
\end{minipage}
\end{center}
\end{figure}

As a result of the initial quark being on-shell, the momentum of the quark in the amplitude and the complex conjugate 
is identical, indicated by the double notches on these lines in the left top diagram of Fig.~\ref{fig8}. In this limit, 
the final state scattering part can be factorized from the initial structure function, as indicated by the right top 
diagram. The final state scattering part now contains an incoming quark which is on-shell and then is driven off shell by 
the scattering experienced by the remnant quark and the radiated gluon. The complex conjugate also contains the 
same process. The blobs which connect to the scattering gluons can be replaced by the imaginary parts of the 
HTL self-energies. Given that the initial quark in the amplitude and the complex conjugate are identical, we 
can consider these to also arise from a cut diagram as shown in the bottom right panel. This final 
diagram corresponds to the type of self energy diagrams evaluated by the AMY scheme.

While the diagrams considered by AMY and HT are topologically similar, the limit in which these are 
evaluated, as well as the pole structure, is quite different. 
As pointed out in the previous section, the leading contribution in the higher twist 
always contains, either in the amplitude or in the complex conjugate, a term which represents the 
quark being produced far off-shell in the initial hard scattering and immediately radiating a gluon. 
This is missing in the AMY approach. The dominant contributions in the AMY approach consist of 
interferences between different medium stimulated emissions along with the squares of the amplitudes 
themselves. As the single gluon emission cross section is evaluated in a different limit in the 
AMY approach, the formalism used to iterate this is also different and will be discussed in the next chapter.

\subsubsection{GLV approach: {Opacity expansion}}\label{single_gluon:other_approaches:GLV}

This scheme named after Gyulassy, Levai and Vitev is very closely 
related to the HT approach. The primary difference is that the GLV scheme makes a 
very specific assumption regarding the structure of the medium. The medium is modeled as 
separated heavy static scattering centers with color screened Yukawa potentials. 
This model of the medium was originally introduced in the seminal work of Gyulassy 
and Wang~\cite{Wang:1991xy,Gyulassy:1993hr,Wang:1994fx} and is often 
referred to as the Gyulassy-Wang model.
 The cross 
section for the interaction of the jet parton with one of these heavy medium partons is 
\bea
\frac{ d \sigma_{el}  }{ d^{2} k_{\perp}} = \frac{ C_{R} C_{2} }{ (2 \pi)^{2} (N_{c}^{2} - 1)  }  \lt| \frac{4 \pi \A_{s}}{ k_{\perp}^{2} + \mu^{2}} \rt|^{2}.  \label{GLV_sigma}
\eea
Where, $C_{R} $ is the Casimir for the jet parton and $C_{2}$ is for the medium parton. Alternatively, 
one may think of this as a specific model to evaluate the gluon field strength correlator of Eq.~\eqref{qhat_FF}.
The quantity $\mu$ represents the Debye mass which screens the potential of the scattering centers and 
represents one of the parameters of the model. The other parameter is the mean free path of the 
jet $\lambda = (\rho \sigma)^{-1}$, where $\sigma$ can be obtained from integrating Eq.~\eqref{GLV_sigma} 
and $\rho$, the density of scattering centers in the medium can be obtained from entropy considerations.

Beyond this, the assumptions regarding the hierarchy of scales where the forward light-cone momentum 
of the jet $q^{-} \gg l_{\perp}$  the transverse momentum of the radiated gluon [which in turn is much 
larger than the ($-$)-components] are the same as in HT. However, unlike the HT, there is no concept of 
scale evolution, thus the model of the medium as incoherent scattering centers is meant at the momentum 
scale of the medium. 
The first term to be considered is the vacuum radiation term as in the HT scheme. 
This is followed by terms that involve the interference of medium induced radiation and vacuum radiation. 
As in the HT scheme, the leading contribution in the GLV approach also involves one scattering with the 
medium and is proportional to the number of 
scattering centers as seen by the jet. This is expressed as $L/\lambda$, where $L$ is the path length of the hard jet and 
$\lambda$ is the mean free path for a scattering. The former quantity depends on the medium in question 
while the latter is a parameter of the calculation. The ratio of $L/\lambda$, called the opacity, is similar to the 
length enhancement factor in the HT approach and, combined with the cross section for a scattering, 
represents the expansion parameter for this calculation.

Thus calculations which include only the vacuum radiation and one scattering, which are proportional to one 
power of $L/\lambda$ and are called up to first order in opacity. The additive term in the cross section to 
produce a single gluon at first order in opacity is
\bea
\frac{d \sg}{dy dl_{\perp}^{2}}  = \frac{\A_{s} C_{F}}{2 \pi l_{\perp}^{2}} P(y) \frac{L}{\lambda} 
\int_{0}^{q_{MAX}^{2}} d^{2} k_{\perp}^{2} 
\frac{ \mu^{2}}{\pi ( k_{\perp}^{2}  +  \mu^{2})^{2} } \int_{0}^{\infty} d\zeta n(\zeta) 
\lt[ 2 - 2 \cos{ \lt( (l_{\perp} - k_{\perp})^{2}/2q^{-}y \rt)}  \rt].
\eea
The major difference with the next to leading twist term in the HT expansion is the integral over 
the exchanged gluon transverse momentum $k_{\perp}$ and the explicit appearance of the 
the Yukawa cross section to scatter off a heavy scattering center. In the equation above,
$n(\zeta) = \theta(L - \zeta)/L$ is the longitudinal profile.

Terms with two scatterings are proportional to 
$(L/\lambda)^{2}$ and so on. Note that leading order in opacity does not mean that the jet has 
scattered only once in the dense medium. Rather, it represents the case that at most one scattering 
was involved in a single gluon radiation. The basic methodology of the GLV approach is to compute the 
single gluon emission kernel, order by order in opacity. For dilute media or short path lengths one uses only 
the leading order in opacity, while denser or more extended media require higher orders. {Higher orders in 
opacity have been explored analytically \cite{Gyulassy:2001nm}, but} it has, so far, 
not been possible to analytically resum the full series in opacity.

\subsubsection{ASW approach: {Multiple-soft gluon exchange}}\label{single_gluon:other_approaches:ASW}

Similar to the GLV scheme, the ASW approach named after Armesto, Salgado and Wiedemann also 
assumes that the medium may be modeled as a collection of heavy static scattering centers with 
Debye screened potentials. In the calculation of the single gluon spectrum, the ASW scheme also 
invokes the small $y$ limit where $y$ is the longitudinal momentum fraction of the radiated 
gluon. As in the case of the HT scheme, it also finds that the leading contributions in this limit 
consist of diagrams where primarily the gluon scatters or diagrams with cross scatterings where 
the quark in the amplitude scatters and interferes with gluon scattering in the complex conjugate. 

In the ASW approach, the multiple scattering analysis is carried out in impact parameter 
space. The square of the amplitude where the gluon scatters $m$ times in both amplitude and 
complex conjugate, is at a transverse position $u_{\perp}$ in the amplitude and at a 
transverse position $\bar{y}_{\perp}$ in the complex conjugate is given as 
\bea
\propto \frac{1}{m!} \lt( -\hf  \int_{\zeta_{I}}^{\zeta_{F}} d \zeta  n (\zeta) \sg (u_{\perp} - \bar{y}_{\perp})  \rt)^{m} 
e^{ - i k_{\perp} \x (u_{\perp} - \bar{y}_{\perp} )}, \label{ASW-m-scattering-on-gluon}
\eea
where, $\zeta_{I}$ and $\zeta_{F}$ indicate the longitudinal positions over which the 
scatterings occur. The quantity $n(\zeta)$ is the opacity at the location $\zeta$ and 
$\sg$ is the dipole cross section of two gluons separated by a transverse distance 
$u_{\perp} - y_{\perp}$. The gluon propagation from $\zeta_{I}$ to $\zeta_{F}$ in the 
amplitude represents one part of the dipole, where as the propagation of the gluon in the
complex conjugate is the other part of the dipole. The case of multiple scattering is 
obtained by summing over $m$ to obtain the obvious exponential with the argument 
in parenthesis in Eq.~\eqref{ASW-m-scattering-on-gluon}. 

In the region before $\zeta_{I}$ there is cross scattering with scattering on the incoming 
quark in the amplitude matched with scattering on the gluon in the complex conjugate and 
vice versa. This case is more complicated; starting at the location $\bar{\zeta}_{I}$, the 
quark propagates in the amplitude and a quark gluon pair in complex conjugate. 
The result is given by a path integral where the dipole moves in the two 
dimensional space of relative transverse location. The diagrams involved in this 
part are near identical to those of the AMY scheme. However, in the AMY case, 
the entire calculation 
is carried out in momentum space, as a result all positions are integrated from $-\infty$ to $\infty$. 
As the calculation here is carried out in position space, the numerator factor 
of $l_{\perp} \x l'_{\perp}$ 
in Eq.~\eqref{num-factor} is replaced by the appropriate derivatives over 
transverse position. The final expression for the cross section to radiate a gluon 
with transverse momentum $l_{\perp}$ and energy $\omega$ and momentum 
fraction $y$ (or alternatively the 
distribution of radiated gluons in transverse momentum and energy), is given as
\bea
x\frac{d^{3} I }{dx  d l_{\perp}^{2}} = \frac{\A_{s} C_{F}}{2 \pi^{2} \omega^{2}} 
 2 Re \int\limits_{\zeta_{0}^{-}}^{\infty}\!\!\! d y_{l}^{-} \!\!
\int\limits_{y_{l}^{-}}^{\infty} \!\!\!d {y'}_{l}^{-}\!\! \int \!\!\!d^{2} u_{\perp} e^{ - i l_{\perp} \x u_{\perp}} 
 e^{ \lt[ \frac{- i}{2} \!\!\int\limits_{{y'}_{l}^{-}}^{\infty} \!\!d \zeta n(\zeta) \sg(u_{\perp})\rt] } 
  \frac{\prt^{2}}{\prt u_{\perp} \prt y_{\perp} }
\int \limits_{y_{\perp}=0}^{u_{\perp}} \!\!\mathcal{D} r_{\perp} 
e^{ \int\limits_{y_{l}^{-}}^{{y'}_{l}^{-}} \!\!\! d \zeta^{-} i 
\frac{\w \dot{r}_{\perp}^{2} }{2}  - \frac{n(\zeta^{-}) \sg(r_{\perp})}{2} } . \label{ASW-formula}
\eea
For a short range potential one may take the saddle point approximation around the  $r_{\perp} \ra 0$ limit
by replacing, 
\bea
n(\zeta) \sg(r_{\perp}) \simeq \frac{\hat{q}(\zeta)}{2} r_{\perp}^{2}. \label{ASW-saddle}
\eea
This approximation allows the above path integral to be carried out. While introduced in this 
way, the one transport coefficient $\hat{q}$ has the standard meaning as in any other 
formalism: the transverse momentum gained by a hard parton (in this case a gluon) per 
unit length.

Unlike the HT or the GLV approach, 
there is no explicit introduction of the vacuum contribution or vacuum medium interference. The fact 
that a hard jet in DIS or a heavy-ion collision is not created at $y_{l}^{-} = -\infty$ is incorporated 
by introducing a finite starting point to the longitudinal integrals. In this case starting $y_{l}^{-}$ integral 
at $\zeta_{0}^{-}$ usually set to be the origin. There is no systematic insistence that the off-shellness 
from the vacuum contribution be much larger than that from the medium scatterings. Given all these 
approximations, the single gluon cross section from multiple scattering is found to be dependent 
on only {one} parameter: $w_{c} = \int d \zeta \zeta \hat{q} (\zeta) = \dfrac{1}{2}\hat{q}L^2$, 
as (in the $\w_{c}L \ra \infty$ limit),
\bea 
\frac{d \sg}{d\w} = \frac{\A_{s}  C_{F} P(y)}{\pi} \ln \lt[ \cos \lt[   (1 + i) \sqrt{ \frac{\w_{c}}{2 \w} } \rt]  \rt]. 
\eea

\subsection{Parametrizing the effect of the medium: $\hat{q}$, $dN/dy$, $T$ and $\A_s$}\label{single_gluon:medium_parameters}

All the different schemes above depend on one or two parameters. In the case of the HT, the 
unknown quantity is the two gluon field strength correlation in the medium in question:
\bea
\int d y^{-} d^{2} y_{\perp}   e^{ - i \frac{ k_{\perp}^{2}   }{ 2 q^{-} } y^{-} + k_{\perp} \x y_{\perp}     }
 \lc   F^{+ \mu} (y^{-},y_{\perp} )  F_{\mu}^{+} (0)  \rc.
\label{FF_corr}
\eea
Expressing $F^{+ \mu} = F^{\nu \mu} v_{\nu}$ where $v_{\nu}$ is the
light-like velocity of the hard parton we note that the above quantity
is simply the Lorentz force correlator in the system. Since this
quantifies the transverse force experienced by a jet as it passes
through a medium, it may be straightforwardly related to the transport
coefficient $\hat{q}$ as in Eq.~\eqref{qhat_FF}.  Since $k_{\perp} \gg
\Lambda_{QCD}$, the integrals over transverse location $y_{\perp}$ are
rather limited. On the other hand $k^{+} = k_{\perp}^{2}/2q^{-}$ may
well be non-perturbative, so that the $y^{-}$ integral is limited by
non-perturbative effects, such as the the confinement length in cold
nuclear matter and the screening length in hot deconfined matter.

To obtain some form for the two-gluon field strength correlator in
Eq.~\eqref{FF_corr}, one requires a microscopic model of the medium.
The situation in the ASW approach is similar. Although a specific
medium model of static scattering centers is used in the derivation,
the final result depends only on the transport coefficient $\hat{q}$,
suggesting that the energy loss within the multiple soft scattering
approximation (i.e. when the saddle point approximation of
Eq.~\eqref{ASW-saddle} is valid) does not depend on the specific details of the medium,
but measures directly the mean transverse momentum exchange per unit
path length.

In the AMY scheme, which is cast completely within the HTL formalism of finite temperature 
field theory, the scattering is not factorized from the gluon emission process. Since HTL 
perturbation theory is cast in the limit that there exist separate scales $T, gT$, the two 
obvious parameters of the theory are $T$ and $g$ (or rather $\A_s$). Note that this 
formalism may only be applied to hot deconfined matter. 
The coupling constant 
$\A_s$ is the same for the jet and for the medium. It is allowed to vary as a fit parameter and set by fitting one data point.

Since the AMY 
formalism makes a precise assumption regarding the microscopic structure of the produced 
matter, it provides a well-defined way to calculate the the transport coefficient $\hat{q}$, given a $T$ and an 
$\A_s$.  Following Ref.~\cite{Arnold:2008vd} we note that $\hat{q}$ may be defined as,
\bea
\hat{q} = \int d^2 k_\perp \frac{d \Gamma }{d^2 k_\perp}, \label{qhat-from-rate}
\eea
where, $d \Gamma /d^2 k_\perp$ is the differential rate for elastic collisions of 
a hard parton with quasi-particles from the medium. While the expression for 
the differential rate has slightly different forms depending on whether the 
exchanged gluon is soft or hard, we will approximate the rate with the expression 
for soft momentum transfer:
\bea
\frac{d \Gamma}{d^2 k_\perp} \simeq 
\frac{ C_R }{ (2 \pi)^2} \frac{ g^2 T m_D^2 }{ k_{\perp}^{2} ( k_\perp^2 + m_D^2 )},
\label{HTL_rate}
\eea
where $m_D$ is the Debye mass, $m_{D}^{2} = \frac{4 \pi \A_{s} T^{2}}{3} \lt(  N_{c} + \frac{N_{f}}{2} \rt)$. 
Substituting the expression for the differential rate in Eq.~\eqref{qhat-from-rate}, 
one may easily calculate $\hat{q}$ given the range of integration of the exchanged 
momentum. 

One notes immediately that the expression for $\hat{q}$ is bounded 
from below and thus the lower limit of integration may be taken to zero. 
This is often done in many different approaches. The upper limit is bounded by  
kinematics of the process, i.e, the outgoing two partons may not carry more energy 
than the incoming partons. This constrains the upper limit to $q_\perp^{MAX} \sim \sqrt{4 E T}$ 
where $E$ is the energy of the hard parton.
The energy loss calculation in the AMY is based on rate equations
similar to Eq. \eqref{HTL_rate} and thus uses the same transport
properties, but without explicitly calculating the transport coefficient~$\hat{q}$.

In the case of the GLV approach, one needs to set two parameters: the Debye screening mass and 
the mean free path of the radiated gluon $\lambda$. The mean free path 
may be obtained using $ \lambda = 1/(\rho \sg)$, 
where $\rho$ is the density of scattering centers and $\sg$ is the integrated cross section for 
a gluon to scatter off these. Because of the dependence on $\rho$, the
total number of gluons at mid-rapidity $dN/dy$ has often been quoted as a
measure of the medium density in GLV calculations \cite{Wicks:2005gt}. However, it
should be noted that the Debye screening mass and thus the temperature
also enters in the GLV formalism via the scattering potential 
Eq.~\eqref{GLV_sigma} and the cross section $\sg$.

In conclusion, this means that in all four formalisms, a microscopic model of the medium
is needed to set the input parameters for the energy loss
calculation, be it the two gluon field strength correlator, the transport
coefficient $\hat{q}$, or $m_D$ and $\lambda$. The AMY approach makes a
specific choice for a HTL plasma. Given such a model, the local medium
parameters can be related to single quantity in the
medium, such as the temperature $T$ in HTL. It is, however, worth to
keep in mind that the radiative energy loss in all models is driven by
transverse momentum exchanges, so that the energy loss is governed by
the transport coefficient $\hat{q}$ or a closely related quantity. 
In addition to the microscopic model of the medium, a realistic
calculation of energy loss in heavy collisions also requires a
macroscopic model of the medium, specifying the space-time dependence
of the local properties of the plasma. This aspect will be further
discussed in Chapter \ref{modeling}.

\section{Incorporating Multiple radiations}\label{multiple_radiation}

In the preceding chapter we described the calculation of medium induced 
single gluon radiation from a hard parton traversing a dense medium. 
Imagine a gluon radiated with a transverse momentum $l_{\perp}$ and 
a forward momentum $l^{-}$, its formation time is given as, $\tau = l^{-}/l_{\perp}^{2}$. For 
realistic values of $l_{\perp}^{2} = 3$ GeV$^{2}$ and $l^{-} = 6$ GeV say from a 
jet with forward momentum $q^{-} \sim 20$ GeV, we obtain a formation time of $\tau \sim 0.4$ fm. 
Thus in almost any medium with a length L $> 1/2$ fm there will be more than one 
such radiation. In this chapter we describe the different ways in which the single 
gluon emission is iterated in different jet modification schemes.

The incorporation of multiple emissions completes the theoretical calculation of radiative energy loss 
for light partons. The final form of the result depends on the observable in question. For the 
computation of the single hadron inclusive cross section one computes the medium modified 
fragmentation function $\tilde{D}^{h} (z)$, where $z$ is the momentum fraction of the hadron $h$ with 
respect to the momentum of the originating hard parton. 
We should point out that, unlike the 
case in vacuum where the hard parton produced in a hard collision continues to 
monotonously lose virtuality till the onset of the non-perturbative process of hadronization, the 
process of multiple radiation in a medium may be far more complicated, including a 
series of both virtuality increasing collisions and virtuality decreasing emissions. To date, the 
space-time structure of this process has not been completely elucidated. The different 
approaches of the various schemes represent approximations to the actual mechanism.

\subsection{Higher twist: In medium DGLAP}\label{multiple_radiation:dglap}

A hard parton from a large $Q^{2}$  collision is considerably off-shell when it is produced. 
This parton then proceeds to lose this virtuality by a series of gluon emissions ordered in 
virtuality or transverse momentum. The radiated gluons are themselves virtual and 
continue to radiate lower virtuality gluons. The progress of this shower may be 
described using pQCD as long as the virtuality $\mu$ is large compared to $\Lambda_{QCD}$. 
Beyond this, one describes the process with a phenomenologically fitted fragmentation 
function $D^{h}(z,\mu^{2})$ which contains all processes up to the scale $\mu^{2}$ which 
produce at least one hadron with momentum fraction $z$.  The effect of multiple 
gluon emissions from the scale $\mu^{2}$ up to some predefined virtuality or scale $Q^{2}$ 
is incorporated by evolving the fragmentation function using the Dokshitzer-Gribov-Lipatov-Altarelli-Parisi 
(DGLAP) evolution equation from $\mu^{2}$ up to $Q^{2}$.  

The DGLAP evolution equation may be understood as the resummation of 
an arbitrary number of ordered emissions which connect a hard parton with a virtuality $Q^{2}$ 
with a parton with lower virtuality $\mu^{2}$. For the non-singlet fragmentation function 
$D_{NS} = D^{h} - D^{\bar{h}}$, which is always vanishing when the fragmenting parton is a gluon, 
this process may be straightforwardly expressed as, 
\bea
D (z, Q^{2}) &=& D (z, \mu^{2}) + \frac{\A_{s}}{2\pi} \int_{\mu^{2}}^{Q^{2}} d l_{\perp}^{2} 
\frac{dy}{y} \frac{C_{F} P(y)}{l_{\perp}^{2}} D\left( \frac{z}{y}, \mu^{2} \rt)  \label{dglap-derivation} \\  
&+& \frac{\A_{s}}{2\pi} \int_{\mu^{2}}^{Q^{2}} d l_{\perp}^{2} \frac{d y}{y} \frac{C_{F} P(y)}{l_{\perp}^{2}}
\frac{\A_{s}}{ 2 \pi} \int_{\mu^{2}}^{l_{\perp}^{2}} d {l^{1}}_{\perp}^{2} \frac{d y^{1}}{y^{1}} \frac{C_{F} P(y^{1})}{{l^{1}}_{\perp}^{2}} 
D\left( \frac{z}{yy^{1}}, \mu^{2} \rt) + \ldots \nn.
\eea
We have ignored the ordering in the momentum fraction $y$ and have only written the first three 
terms in a series where each term contains a product of integrals over transverse momenta where 
the transverse momenta at each emission is limited by the preceding radiation. 
The factor $P(y)$ represents the $q \ra qg$ splitting function, i.e., the probability for a quark to 
radiate a gluon and retain a fraction $y$ of its momentum. 
This series can 
be resummed by noting that differentiating with respect to $\log(Q^{2})$ yields an identical series 
as above for the 
shifted fragmentation function $D(z/y)$, convoluted with the splitting function $P(y)$: 
\bea
\frac{\prt D(z, Q^{2})}{\prt \log Q^{2}}  = 
\frac{\A_{s}}{2 \pi} \int_{z}^{1} dy C_{F} P(y) D \lt( \frac{z}{y} , Q^{2}  \rt). \label{vac-frag-dglap}
\eea
For an arbitrary fragmentation function, the evolution equation for the quark-to-hadron fragmentation 
function is coupled with the evolution equation for the gluon-to-hadron fragmentation function. 

The jet modification formalism which is closest to this line of argument is the HT formalism. It assumes 
that while scattering processes may raise the intermediate virtuality in a given amplitude 
prior to or just after a gluon emission,  there is an overall drop in the average virtuality  between 
the incoming parton and the outgoing pair of radiated gluon and remnant parton in the single 
gluon emission cross section. Note that multiple scattering tends to progressively raise the
virtuality of the propagating partons. Thus the above assumption is valid in the regime of 
densities and incoming virtualities where the formation time of the gluon is so short 
that the drop in virtuality from the emission dominates over the rise due to multiple scattering 
over the same time, i.e.,
\bea
\hat{q} \tau = \frac{  \hat{q} 2 l^{-}}{ l_{\perp}^{2 } } \lesssim l_{\perp}^{2} \Longrightarrow
 l_{\perp}^{2} \gtrsim \sqrt{ 2\hat{q} l^{-}} .
\eea
We also note that gluons with a formation time $\tau$ longer than the length of the medium $L$ are not 
influenced by energy loss, {which sets a minimum virtuality: $\tau \leq L \Longrightarrow l_\perp^2 \gtrsim 2 l^-/L$}.
Thus for the formalism to remain consistent till the exit of the jet from the medium we require the 
minimal condition, 
\bea
\sqrt{2 \hat{q} l^-} = \frac{2 l^-}{L} \Longrightarrow \Delta E = \hf \hat{q} L^2 .
\eea
The last equality is obtained by relating the light-cone momentum of the radiated 
gluon with the light-cone loss and thus the energy loss. 

In a calculation one simply replaces the vacuum splitting in Eq.~\eqref{vac-frag-dglap}
with the in-medium splitting function derived in the section above. It is assumed that 
the hard parton exits the medium with a virtuality (related to the transverse momentum of 
its radiations) $\mu^2 \simeq 2Ey/L \sim E/L$, where $E$ is the energy of the initial hard 
parton, $L$ is the length of the medium and $y$ is a representative momentum fraction 
carried by the radiated gluon; this is approximated as $y \sim 1/2$ in the second part of 
the equality. As pointed out earlier, a dynamical calculation of the loss of virtuality with 
emission as a function of the distance travelled by a hard parton is still lacking. 

 Given the measured fragmentation function at the scale $\mu^2$, this is evolved up to the 
hard scale $Q^2$ of the process which represents the largest possible initial 
virtuality. Results of such an evolution in a hot QGP-brick will be presented at the 
end of this chapter. The result of the evolution is not very sensitive to the choice of the 
upper limit of the evolution; for instance, using $Q^2/4$ will yield very similar results as long as 
$Q^2 \gg \mu^2$.

\subsection{ASW and GLV: Poisson convolution}\label{multiple_radiation:poisson}

The ASW and the GLV schemes share a common microscopic picture of the medium: 
that of heavy static scattering centers. The also use an almost identical methodology 
for iterating the single gluon emission. Unlike the HT which computes the change 
in the distribution of hadrons (or the AMY scheme which computes the change in the 
distribution of partons), the central quantity {that is calculated} in these formalisms is the 
probability distribution  $P(\Delta E)$ for a hard parton to lose an energy $\Delta E$, via 
an arbitrary number of gluon emissions. 
This probability is then used to shift the 
fragmentation momentum fraction $z$ and define a medium modified fragmentation 
function.

Contrary to the HT approach, in these formalisms, the virtuality evolution of the parton 
is not taken into account. A hard 
jet propagating through the medium will progressively gain virtuality by collision and 
lose it by emission. Each of these emission processes makes a minimal shift in the 
energy of the jet and to leading order one may ignore this shift, this is often 
referred to as the eikonal limit. Thus all these 
emissions may be considered as independent of each other.  The fate of these emitted 
gluons is also not considered, the focus is solely on the emitting hard parton.

In either formalism, one calculates the differential 
distributions of radiated gluons, e.g., in Eq.~\eqref{ASW-formula}. To obtain the differential 
energy distribution of radiated gluons, these expressions have to be integrated over the 
transverse momentum $l_\perp$. 
{The formalism does not have a natural cut-off for transverse momentum integration, but instead} 
uses kinematic considerations to limit these integrations, e.g., $l_\perp^{Max} \leq $min$\{y,1-y\}E$ 
as used by the GLV scheme or in the small $y$ limit $l_\perp^{Max} \leq y E$ as used by the 
ASW scheme. The different choices of kinematic bounds, along with slightly different definitions 
of $y$ (light-cone momentum fraction in GLV and energy fraction in ASW) lead to somewhat 
different results~\cite{Horowitz:2009eb}.

Having limited the transverse momentum integrations, the differential energy or momentum 
fraction distribution $dI/d\w, dI/dx$ is well defined. The integral over this quantity yields the 
mean  number of gluons radiated from the jet:
\bea
N_g = \int dx \frac{dI}{dx} = \int d \w \frac{d I}{d \w}. 
\eea
Given that each of these emissions are independent, {the number of radiated gluons $n$ is 
assumed to follow a Poisson distribution, giving the following probability distribution for energy loss}:
\bea
P(\Delta E)  = \sum_{n=0}^{\infty} \frac{ 1  }{n !} \lt[  \prod_{i=1}^n  \int d \w_i \frac{ d I (\w_i)}{ d  \w_i} \rt]
\kd \lt(  \Delta E - \slm_{i=1}^n \w_i \rt) e^{- N_g}. 
\eea
Using the expression for $d I / d \w$ the above probability distribution may be numerically calculated. Having 
lost energy $\Delta E$, the jet now fragments a hadrons with momentum $p_h$ off a hard parton 
with energy $E- \Delta E$. The shifted or medium modified fragmentation function is defined as, 
\bea
\tilde{D} (z,Q^2) = \int_0^E d \Delta E P( \Delta E) \frac{ D \lt( \frac{z}{1-\Delta E/E}, Q^2 \rt) }{1-\Delta E/E}.
\label{frag-shift}
\eea
Under the assumption that the virtuality gain from the medium is comparable to that brought in by the 
hard parton, the scale used in the final fragmentation function is the same as the hard scale of 
the process. 

\subsection{AMY: rate equations.}\label{multiple_radiation:rate_equation}

In the AMY formalism, the hard jet is treated similar to a hard parton in the medium. The virtuality
is assumed to be the same as that for any hard thermal parton in the medium $\sim gT$. In this 
sense it is similar to the assumption made in the ASW and GLV calculations. In this formalism, the 
radiation of a gluon by a quark or a gluon is an order $\alpha_{S}$ suppressed process and thus it 
occurs rarely enough that one may consider sequential emissions to be independent of each other. 
The rates for a quark 
to decay into a quark and gluon, a gluon to decay into a quark anti-quark pair etc., are now used to 
set up a Fokker-Planck equation which describes the change in the distribution of hard partons with 
time. There are two sets of equations which describe the change of the distribution of the sum of 
quarks with anti-quarks and gluons. For example the $q+\bar{q}$ distribution may be 
described as
\bea
\frac{d P_{q \bar{q}} (p) }{ d t} = \int_{\infty}^{\infty} d l P_{q \bar{q}}( p+l  )  \frac{d \Gamma_{ q g }^q (p+l, l )}{dl dt}  
- P_{q \bar{q} } (p) \frac{d \Gamma_{ q g }^q (p, l )}{dl dt} + 
2 P_{g} (p+l) \frac{d \Gamma_{ q \bar{q} }^g (p+l, l )}{dl dt}  .
\eea
In the equation above, the first term on the $r.h.s.$ describes the process where a hard quark (antiquark) with 
a momentum $p+l$ decays in to a quark (anti-quark) with momentum $p$ and a gluon with momentum $l$. 
Hence, its contribution to the Fokker-Planck equation is proportional to the distribution of $q +\bar{q}$ at 
a momentum $p+l$. The second term on the $r.h.s.$ represents the decay of the quark (antiquark)
with momentum $p$ into lower momentum quarks and gluons and thus represents a depletion of the 
distribution at that momentum. The last term proportional to the population of hard gluons with momentum 
$p+l$ represents the process where the gluon decays into a quark anti-quark pair where one has a momentum 
$p$ and the other has a momentum $l$. The factor of 2 accounts for the case where the momenta of the 
two fermions are reversed. 

The integrals over $l$ run from $-\infty$ to $\infty$. When the momentum of the outgoing gluon or quark 
($l$ or $p-l$) becomes soft, of the order of $T$,
they will encounter Bose enhancement or Pauli blocking from the medium. These factors of distribution are already 
included in the expression for the rate $d\Gamma/dl dt$. 
Negative values of $l$ indicate absorption of thermal gluons from the medium, the distributions for these 
arise simply from the sign change in the functions which lead to Pauli blocking or Bose enhancement. 
The AMY formalism is different from the other three approaches as it naturally incorporates feedback 
from the medium which is missing from all other approaches. A similar, but slightly more complicated 
equation may be written down for the gluon distribution~\cite{Jeon:2003gi}. These two 
equations are solved in tandem.

In this set-up one represents an incoming quark jet by setting the distribution at $t=0$ to be a $\kd$-function, 
$P_{q+\bar{q}}(p,t=0) = \kd(p-E_{jet})$. The gluon distribution at this time is set to zero. These 
distributions are then evolved up to the time of exit by means of the coupled Fokker-Planck equations. 
In spite of the completeness of this approach, it does suffer from the rates being computed at a fixed 
temperature. Which implies that the medium remains static during the formation time of the radiated gluons. 
As a result, the application of this formalism to short or rapidly varying media is somewhat tenuous.

The final medium modified fragmentation function is obtained by convoluting the final distribution of 
hard quarks, anti-quarks and gluons with the respective shifted fragmentation functions, 
using Eq.~\eqref{frag-shift} with the probability of 
energy loss replaced by the distribution of hard partons. In most applications so far the scale at 
which the fragmentation functions are evaluated is chosen to be the hard scale of the process.

\subsection{Comparison of the different energy loss formalisms}\label{multiple_radiation:comparison}

In the previous sections, we have outlined the calculation of the medium modified fragmentation 
functions within the four different schemes that are currently in use. We now attempt a standard 
comparison between these different formalisms in an identical medium. 
In a recent effort by the 
TECHQM collaboration~\cite{techqm} this problem has been formulated as a computation 
of the modification of a single quark jet propagating through a homogeneous 
static medium of fixed length held at 
a constant temperature. This medium is often referred to as the ``QGP brick'', as 
temperatures are usually chosen to be high enough that deconfinement has set in.

In order to carry out computations in all four formalisms, their medium parameters 
have be defined in this standardized medium. We present our own particular 
prescription of a standardized medium in this review. 
One starts out by setting the temperature $T$ 
and a value of the in-medium coupling $\A_{s}=0.33$. Given these two parameters, 
calculations of jet modification may be carried out in the AMY formalism. Given these 
two parameters, $\hat{q}$ can be computed using Eq.~\eqref{qhat-from-rate}. 
Carrying out calculations for a quark-gluon plasma with 3
flavors of quarks ($N_f = 3$) using the HTL
approximation for small $k_{\perp}$, Eq. \eqref{HTL_rate}, we obtain, 
\bea
\hat{q} = \frac{C_{A} g^2 m^2_D T}{2 \pi} \ln\left(\frac{q_{\perp}^{\mathrm MAX}}{m_{D}}\right), \label{HTL_qhat}
\eea
with $q_{\perp}^{\mathrm {MAX}}=\sqrt{ET}$. 

The value of $\hat{q}$ obtained by the prescription above for the AMY scheme 
may now be directly substituted in the 
HT and the ASW approach. This will in some sense place these three calculations on the 
same footing, {by using} the $\hat{q}$ for a weakly coupled plasma of quark 
gluon quasiparticles in all three formalisms.

{The GLV formalism requires the specification of the Debye mass $m_D$ and the mean free path $\lambda$. The Debye mass is taken directly from HTL field theory $m_{D}^{2} = \frac{4 \pi \A_{s} T^{2}}{3} \lt(  N_{c} + \frac{N_{f}}{2} \rt)$. The mean free path $\lambda$ is calculated from the scattering rate}
\bea
\frac{1}{\lambda} = \int d^{2} k_{\perp} \frac{d^{2}  \Gamma }{ d^{2}  k_{\perp} },
\eea
{using the same HTL
approximation for small $k_{\perp}$, Eq. \eqref{HTL_rate}.}
Unlike the calculation for $\hat{q}$, the equation above has an infra-red divergence. 
This is controlled by restricting the lower limit to $m_{D}$, with the result:
\[
\frac{1}{\lambda} = \int_{m_D}^{q_{\perp}^{\mathrm{MAX}}} d^2k_{\perp}
  \frac{d\Gamma^2}{d^2k_\perp} \approx 3  \alpha_s T \ln(2),
\] 
with the further approximation $q_{\perp}^{\mathrm{MAX}} >> m_{D}$.

\begin{figure}
\centering
  \epsfig{file=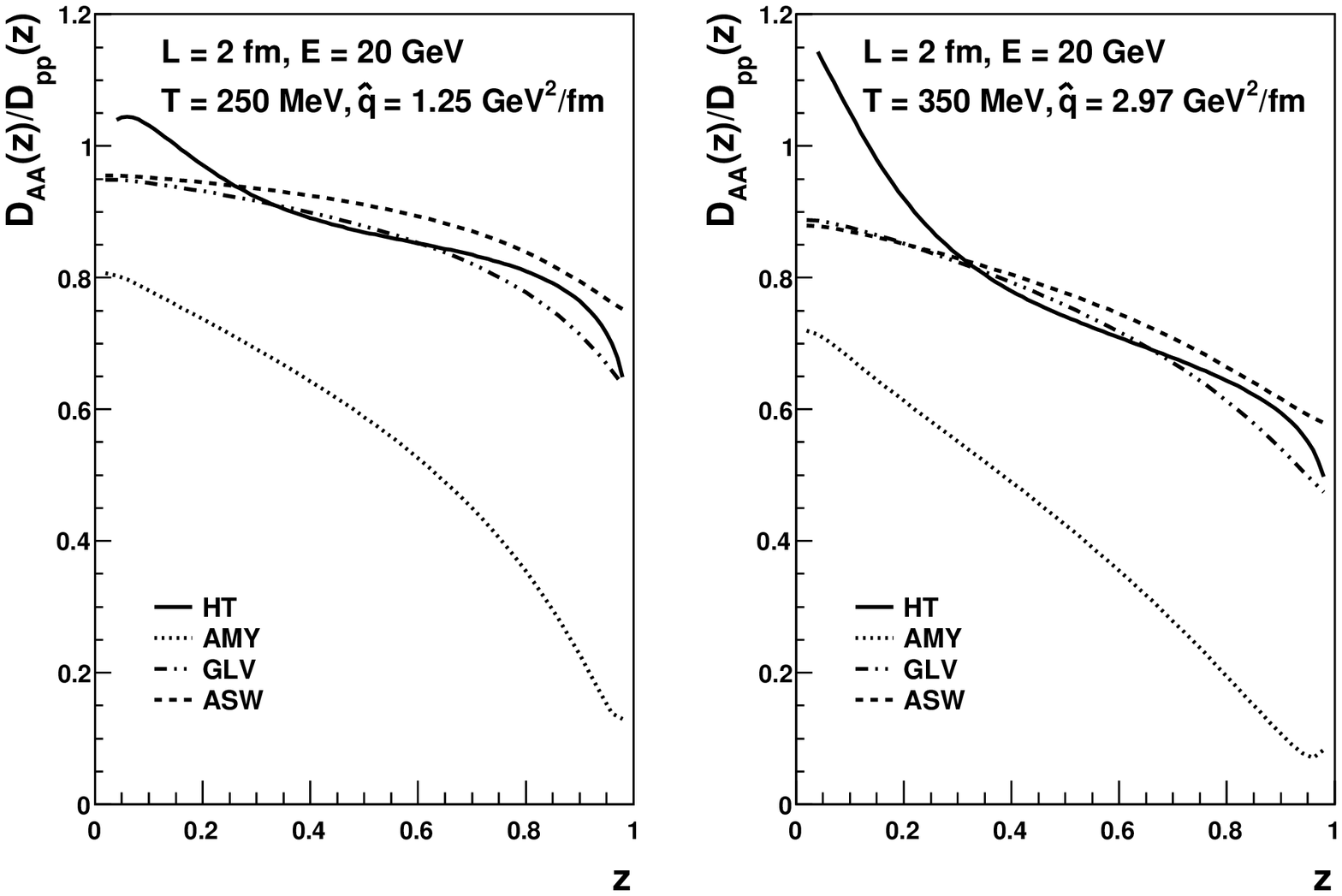,width=0.8\textwidth}
  \epsfig{file=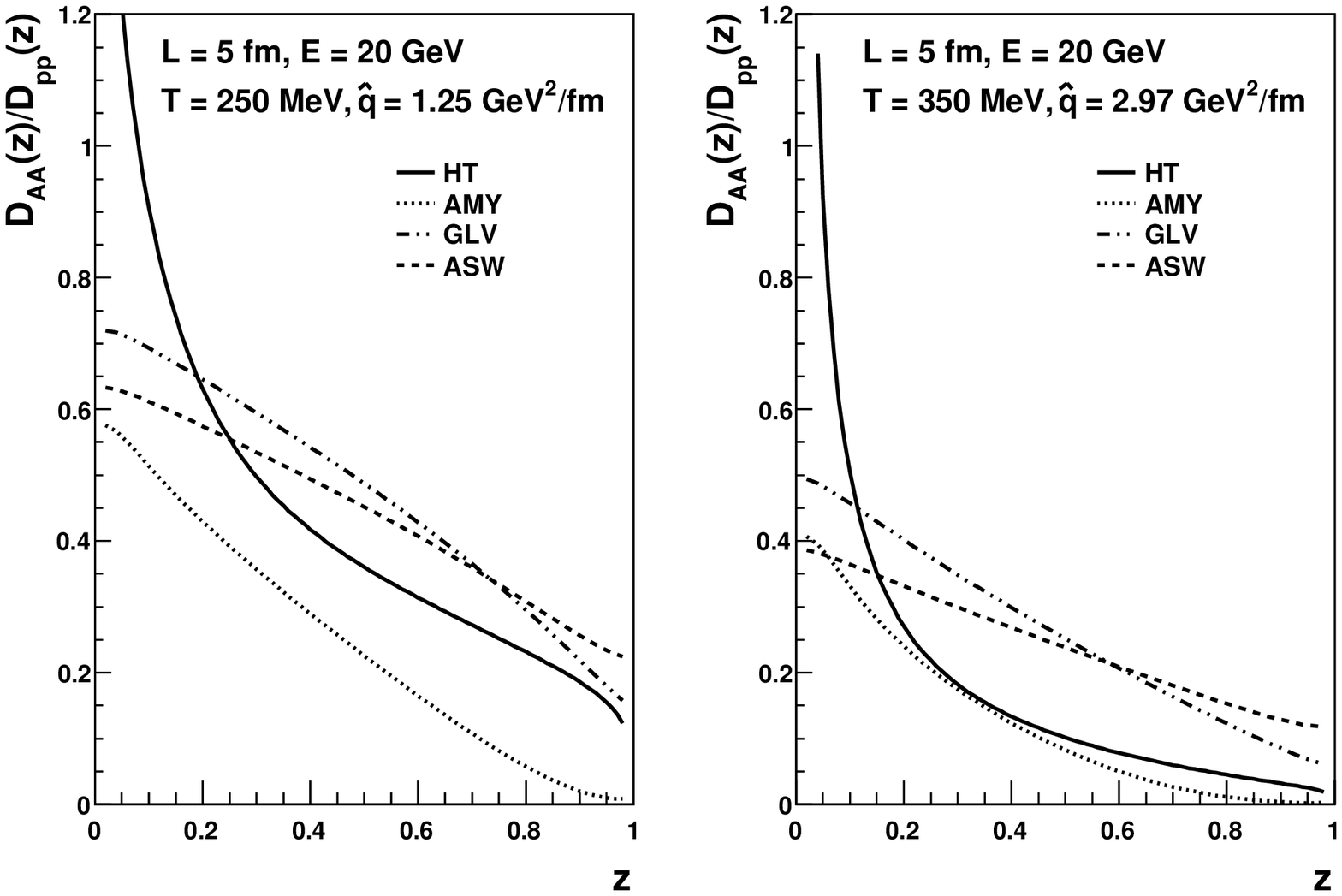,width=0.8\textwidth}
\caption{\label{fig:ff_comp} Comparison of quark fragmentation function ratios
  using four different formalisms for a uniform medium with $L = 2$ fm
  (upper panels) and $L = 5$ fm (lower panels). For both upper 
  and lower panels the left plot is at $T = 250$ MeV and the right plot is at $T = 350$ MeV. 
  For details, see text.}
\end{figure}

Figure \ref{fig:ff_comp} shows a comparison of quark fragmentation
function ratios using four different formalisms to calculate energy loss
(HT, AMY, WHDG radiative and ASW) for a uniform medium with 2
different path lengths $L$ and temperatures $T$ in the regime relevant
to RHIC.
The Higher Twist (HT)
uses $\hat{q}$ and $L$ as input to calculate a fragmentation function
in the medium. The Figure shows the ratio between the medium-modified
fragmentation function and the fragmentation function in the vacuum.

In the AMY formalism, the Fokker-Planck equations are solved
to calculate the distribution of outgoing quarks and gluons starting
from a mono-energetic distribution of quarks, the incoming quark with
$E=20$ GeV. The outgoing quarks are then convoluted with the KKP
fragmentation function to give a medium-modified fragmentation
function. Gluon fragments were not included in this calculation for the 
purpose of comparison to the opacity expansion and multiple soft
scattering approximations.

For the multiple soft scattering approach, ASW, the publicly available
quenching weights code was used to calculate the energy loss
probability distribution $P(\Delta E)$ given the transport coefficient
$\hat{q}$ and medium length $L$. The medium modified fragmentation
function is calculated by convoluting the energy loss probability
distribution with KKP fragmentation functions using Eq.~\eqref{frag-shift}.
Note that only quark fragments are taken into account,
fragmentation of the radiated gluons is not calculated.

For the opacity expansion, we used the DGLV expressions from Appendix
B of Ref.~\cite{Wicks:2005gt}, with an in-medium gluon mass of $m_g =
m_{D}/\sqrt{2}$ and a quark mass $m_q=m_{D}/2$. The Debye
screening mass $m_D$ and $1/\lambda$ are defined above. Multi-gluon
emission is treated using a Poisson convolution for multi-gluon
emission. The medium modified fragmentation function was calculated
using Eq.~\eqref{frag-shift}.

Comparing the fragmentation functions in Fig. \ref{fig:ff_comp}, we
see a characteristic rise in the ratio at low $z$ for the HT
calculation that can be identified with gluon fragments
\mvlnew{(including secondary medium-induced radiations)}, which are
not included in the other calculations.
Because of the steeply falling
parton spectrum, the large $z$ behaviour is most important for
high-\pt{} hadron production. For short path length $L = 2$ fm, HT,
GLV and ASW give very similar results at high $z > 0.5$, for both
$T=250$ MeV and $T=350$ MeV. The AMY calculation shows a much larger
suppression. This may be due to the absence of vacuum-medium
interference in AMY.

At larger $L = 5$ fm (bottom panels of Fig. \ref{fig:ff_comp}), larger
differences between the various formalism are seen. The magnitude of
the suppression in the GLV opacity expansion and the ASW multiple-soft
gluon emission formalisms are similar, but the $z$ dependence is different
between the two. The HT result approaches the AMY result for $L=5$ fm
and $T=350$ MeV. It should be noted, however, that in this case, the
suppression becomes very large, more than a factor 10 at $z >
0.5$, thus violating the $\Delta E << E$ limit in which all the
formalisms have been derived.

\section{Heavy flavors and elastic energy loss}~\label{heavy_flavor}

In the preceding chapters we explicitly discussed the case of light
quark and gluon energy loss. It is now almost established that for
light partons, transverse scattering induced radiative energy loss,
plays a dominant role {(see Section
\ref{experiment_path_length})}. There is, however, an entirely orthogonal
mechanism: elastic energy loss. While the nomenclature is somewhat
confusing, this refers to a process where the jet transferred energy
and momentum to a constituent of the medium and the interaction with
the medium did not stimulate the jet to radiate a hard gluon, i.e., it
was not inelastic for the jet. On the medium side, the effect may be
both elastic or inelastic. {We should point out that
existing calculations of elastic energy loss
\cite{Thoma:1990fm,Braaten:1991we,Mustafa:2003vh,Qin:2007rn} have
focused specifically on 2-to-2 scattering, where the medium parton
also stays more or less on shell.}

{For heavy quarks, radiative energy loss is much
  reduced, due to the dead cone effect, which suppresses collinear 
radiation. As a result, elastic energy loss may be more important for
  heavy quarks than for light quarks.} In the remaining part of this chapter we will review 
the dead cone effect and how it reduces the energy lost by a hard heavy quark. 
Following this we will consider the inclusion of elastic energy loss within the jet modification 
formalism.

\subsection{Dead cone effect in radiative energy loss}~\label{heavy_flavor:dead_cone}

Imagine a hard heavy quark produced in a DIS event about to radiate a hard gluon. 
Physically this will occur in a process where a hard heavy quark in the nucleon (produced in 
large $Q^{2}$ fluctuation) is struck by the 
virtual photon. An alternative to this process is when the incoming electron encounters a 
charge current interaction via a $W^{-}$ exchange which converts a $u$ quark in the nucleon to a
$b$ quark. In the interest of simplicity and given the factorization of final state 
effects from the hard cross section and the initial PDF, we ignore the specifics of the 
charge current interaction and only focus on the final state radiation of a hard gluon from the 
produced heavy quark. 

The simplest means to see the effect of the mass of the heavy quark is to introduce a mass $M$ 
in the fermion propagators in Eq.~\eqref{O_00}. This converts the equation to 
\bea 
\Op^{00} &=& \tr \left[ \frac{\g^-}{2} \widehat{\Op^{00}} \right]  
 = \int \frac{d^4 l}{(2\pi)^4}   d^4 z d^4 z'
\frac{d^4 l_q}{(2\pi)^4}\frac{d^4 p_0}{(2\pi)^4} \frac{d^4 p'_0}{(2\pi)^4}  \label{O_00_heavy}\\ 
\ata e^{i q \x y_0 }e^{ -i (p_0 + q) \x ( y_0 - z) } 
e^{-i l \x (z - z')} e^{-i l_q \x (z - z')} 
 e^{ -i (p_0'+ q) \x z' }   
g^2 \tr \left[  \frac{\g^-}{2} \frac{-i (\f p_0 + \f q + M )}{ (p_0 + q )^2 -M^{2} - i \e }  \right. \nn \\
\ata \left. i\g^\A  \lt( \f l_q + M \rt)  2\pi \kd (l_q^2 - M^2)  
 G_{\A \B} (l)   2\pi \kd (l^2)  (-i\g^{\B})
\frac{ i (\f p_0'+ \f q + M )}{ (p_0'+ q )^2 - M^{2} +  i \e }  \right] . \nn
\eea
Carrying out the integrations over $z$ and $z'$ yields the usual constraints, 
$l_q = p_0 + q - l$ and $p_0 = p_0'$. These are used to perform the $p'_0$
and $l_q$ integrals. The difference with the case of the light quarks are essentially 
the factors of $M$ that appear in the denominators, in the Dirac matrix structure 
and in the on-shell constraint $\kd(l_q^2 - M^2) $. The argument of the 
$\kd$-function may be simplified as (using $l_{q} = p_{0} + q  - l$ and $l$ is on-shell),
\bea
\mbx & & 2p_{0}^{+} q^{-} + 2q^{+} q^{-} + 2p^{+} l^{-} + 2q^{+}l^{-} + 2q^{-} l^{+}  - M^{2} = 0  \nn \\
&\Rightarrow& 2P^{+} q^{-} \lt( x_{0}(1-y) - x_{B}(1-y) - \frac{ l_{\perp}^{2} }{y} - M^{2} \rt)
= 2P^{+}q^{-} (1 - y) \lt[ x_{0} - x_{B} - x_{L} - x_{M} \rt].
\eea
Where, $x_{L} = l_{\perp}^{2}/2P^{+}q^{-}y(1-y)$  has the usual definition, while, $x_{M} = M^{2}/2q^{-}P^{+} (1-y)$.

To simplify the trace, we take the 
small $y$ limit i.e., the soft gluon limit, which reduces this expression to 
a form similar to that of Eq.~\eqref{LO_cross}. 
In this limit, the 
only projection of the gluon propagator that has to be retained is $G^{++} = l_{\perp}^{2}/(q_{-}^{2})$.
As a result, one only needs to retain the terms proportional to $\g^{-}$ in the gluon emission vertices. 
This simplifies the trace over Dirac matrices: 
\bea
\tr \lt[   \frac{\g^{-}}{2}  \g^{+} q^{-}  \g^{-} \g^{+} q^{-} \g^{-} \g^{+} q^{-}  \rt]  = 8 {q^{-}}^{3}.
\eea
 With these simplifications, the operator $\mathcal{O}^{00}$ [defined for the case of light quarks in Eq.~\eqref{O_00}] which is directly proportional to the 
 cross section may be expressed as, 
 \bea
 \mathcal{O}^{00} = \frac{\A_{s} C_{F}}{2 \pi} \int dy \frac {2}{y} \frac{l_{\perp}^{2}}{ \lt( l_{\perp}^{2}  + y^{2} M^{2} \rt)^{2} }.
 \eea
Defining the angle of radiation as $\theta = l_{\perp}/ \w^{-}$ where $\w^{-} = q^{-} y$, and the mass dependent 
angle $M/\w^{-}$ as $\theta_{0}$ we obtain the gluon radiation distribution to be 
\bea
\frac{d N_{Q\ra Qg}}{d \theta^{2}} =  \frac{\A_{s}  C_{F}}{ \pi}  \frac{d\w^{-}}{\w^{-}}  \frac{ \theta^{2} }{ \lt( \theta^{2} + \h_{0}^{2} \rt)^{2} }
= \frac{d N_{q \ra q g}}{d \h^{2}} \lt( 1+ \frac{\h_{0}^{2}}{\h^{2}}\rt)^{-2},
\eea
where $dN_{q \ra q g}/d \h^{2}$ is the angular distribution from a light quark which has a logarithmic 
divergence as $\h \ra 0$. The angular distribution of gluon radiation from the heavy quark is cut off 
at small angles. This is called the dead cone effect~\cite{Dokshitzer:2001zm}.
The origin of this effect is essentially the shielding of the collinear singularity by the large mass of the 
quark. {Since gluon radiation from quarks occurs dominantly at 
smaller values of $l_{\perp}$, the suppression of radiation in this region by the mass of the 
quark, leads to a smaller amount of radiative energy loss from a heavy 
quark compared with a light quark.}
 
In the case of gluon radiation from multiple scattering there will be multiple instances of the 
heavy quark mass as there will be multiple quark propagators. So far, the calculation of medium 
induced gluon radiation from a heavy quark has only been performed up to one scattering per radiation 
in the HT formalism~\cite{Zhang:2003wk,Qin:2009gw}, up to 1$^{\rm st}$ order in opacity in a variant of the 
GLV approach (called the DGLV)~\cite{Djordjevic:2003zk} and in the multiple soft scattering approximation of the 
ASW approach~\cite{Armesto:2003jh}. 
As in the case of light flavors we will follow the derivation in the higher twist approach.
The expression for the 
radiated gluon distribution from a massive quark is given, following Ref.~\cite{Zhang:2003wk} as a shift 
in the formation time $\tau_{f}$ of the radiated gluon and an overall multiplicative factor $f$ as, 
\bea
\tau_{f} = \frac{2 q^{-}  y (1-y)}{ l_{\perp}^{2 }  + y^{2} M^{2}}, \,\,\, \mbox{and} \,\,\,  
f = \frac{l_{\perp}^{8}}{ \lt[ l_{\perp}^{2}  + y^{2} M^{2} \rt]^{4}}.
\eea
In both the GLV and the HT calculations, the above factors lead to a considerable 
reduction in the amount of energy loss by radiation of gluons. 
This has led to the realization that other sources 
of energy loss, such as elastic energy loss,  which is a smaller effect for light 
partons, may be more important in the case of heavy quarks.  

While the radiative energy loss calculations in the GLV, HT and ASW schemes 
have all been modified to incorporate the effect of the heavy quark's mass, 
it was found that in all three cases, purely radiative processes 
cannot describe the observed suppression in the yield of non-photonic
electrons from $D$ and $B$ meson decay (see also Section \ref{sect:heavy_flavour_exp}).
Attempts to incorporate elastic energy loss within the energy loss formalism and 
describe the heavy flavor suppression have been carried out by the DGLV and the 
HT schemes. In the last section of this chapter we will describe 
the inclusion of elastic energy loss in jet modification calculations.

\subsection{Longitudinal drag and diffusion: other transport coefficients}~\label{heavy_flavor:other_coeffs}

In the derivation of the multiple scattering of a hard parton and the ensuing 
gluon radiation in Chapter 2, we specialized to the case where the negative 
light-cone momentum of the exchanged gluon was assumed to be vanishingly 
small, i.e., $k^- \ra 0$. All factors of  the exchanged negative light-cone momenta $k_i^-$
were ignored from the expressions and this led to the restriction of the entire 
process to the $y^+ = 0$ plane. In reality, depending on the structure of the 
dense medium with which the jet interacts, one can have a non-vanishing amount of 
$k^-$ exchanged. This is true not only for a massive quark but also for light 
partons. In this section we derive the leading correction to the differential cross section for 
a hard light quark traveling through a large nucleus encountering multiple scattering 
which may change both its transverse as well as its longitudinal momentum. 
These may then be easily extended to the case of a heavy quark. The final 
expression will be written in a factorized form where the hard dynamics of the 
quark propagation will be separated from the in-medium expectation of gluonic 
operators which will codify the respective in-medium transport coefficients. 
Once defined in such a factorized form, the results may be immediately 
extended to the case of jets in a heavy ion collision by replacing the 
hard production cross section in DIS with that in $A$-$A$ and replacing the 
transport coefficients with those calculated in a hot deconfined environment. 
In real comparisons to date, these are often fitted to {light hadron
production data points}. 

The reader may wonder about the focus on light cone loss as opposed
to actual energy loss. This emanated from our insistence on defining as boost invariant a
quantity as possible. For a large magnitude of energy loss in an extended time $\Delta t$, 
the measure $\Delta E/ \Delta t$
changes with boost, but the light cone ratio is boost invariant:
\bea
\frac{\Delta q^- }{ \Delta L^-} = \frac{\Delta E - \Delta p_z}{ \Delta t - \Delta z} 
\longrightarrow^{\!\!\!\!\!\!\!\!\!\!\!\!\!\!\!{\rm boost}} 
\frac{ \g (1 - \B)  ( \Delta E - \Delta p_z  ) }{ \g (1 -\B) ( \Delta t - \Delta z ) }.
 \eea

Starting again from the case of DIS on a large nucleus, we now define the 
triple differential distribution for the hadronic tensor for the case of 
no rescattering and no final state radiation where a quark with $x_B$ momentum fraction is 
struct by a hard virtual photon and sent back through the nucleus as 
\bea
\frac{ d^3 W_0^{\mu \nu}  }{d^2 {l_q}_\perp d l_q^- } =  W^{\mu \nu}_0 
\kd^2(\vec{l}_{q \perp}) \kd(l_q^- - q^- ),
\eea
where $W_0^{\mu \nu} = -\gmn 2 \pi \sum_q Q_q^2 f^A_q (x_B)$. The two $\kd$-functions 
indicate that the three momentum components of the outgoing quark have a narrow distribution 
around $l_q^- \sim q^-$, the large light cone momentum of the incoming photon, and ${l_q}_\perp \sim 0$.
Being close to onshell, $l^+ \simeq l_\perp^2/2 l^-$

Considering the case of one scattering, the modified differential distribution may 
be expressed as
\bea
W^{\mu \nu} \!\!\!\!&=& \!\!\! W_0^{\mu \nu} 
% (-g_\perp^{\mu \nu }) C_{p,p_2}^A  \!\!\int d y_0^- e^{-ix_B p^+ y_0^-}  
%
% \lc p | \psibar(y_0^-) \frac{\g^+}{2} \psi(0) | p \rc \nn \\
%
\!\!g^2 \int \frac{d l^- d^2 l_\perp}{ (2\pi)^3 } 
d Y^- d y^- dy^+ d^2 y_\perp \frac{dk^- d^2 k_\perp}{(2\pi)^3} 
\!\!\frac{(2\pi)^3 \kd(l^- - q^ - - k^- ) \kd^2 (\vl_\perp - \vk_\perp)}{2(q^- + k^-)}
\frac{\tr [t^a t^b]}{N_c} \nn \\ 
\ata \!\!\frac{\tr}{4} \left[ \g^+ \g^\A \left\{ (\f q^- \!+\f k^- ) + 
 \frac{ \g^- k_\perp^2}{2 (q^- \!+k^-)} - \f k_\perp  \right\} 
\g^\B  \right] 
\exp{ \left[ - i \frac{k_\perp^2}{2q^-} (y^-) + i y_\perp \x k_\perp -i y^+ k^- \right] } \nn \\
\ata \!\!\lc P | A^a_\A (Y^- + y)    A^b_\B (Y^-)   | P \rc. \label{W_mu_nu_heavy}
\eea
One immediately notices the shifts in the three $\kd$-functions which describe the 
final outgoing quark distribution. These are, as expected, proportional to the expectation of 
the two gluon operator in the nucleus. The Dirac structure is obvious from the fact that the 
quark immediately after the hard scattering on the virtual photon is almost on shell in 
both the amplitude and the complex conjugate and has a magnitude only in the negative light 
cone direction $l = (0,q^-,0,0)$. 
The cut line emanates after the scattering and has a momentum [$k^+ , q^- + k^- , k_\perp$] with 
$k^+ = k_\perp^2/2(q^- + k^-)$.  We make the same set of approximations on the gauge fields as 
for the case of transverse broadening, i.e., $A^{+} \gg A_{\perp}$ (recall that we are calculating 
in $A^{-} = 0$ gauge) as in Section \ref{single_gluon:multiple_scattering}. 

Given the scaling of the ${A^{a}}^{\mu}$ field, we obtain that the dominant contribution in the 
Dirac trace as given by
\bea
\tr \lt[  \g^{+}   \g^{-} \g^{+} (q^{-}  - k^{-})  \g^{-}   \rt].
\eea
In the case of transverse broadening, one Taylor expands in $k_{\perp}$; to obtain the distribution 
in light cone momentum one Taylor expands in $k^{-}$. In the interest of simplicity, we will ignore 
the factors of transverse momentum and simply focus on the light cone components. Taylor expanding 
the hard part, we obtain that the leading terms emanate from the expansion of the $\delta$-function, 
\bea
\kd (l^- - q^- - k^-) &=& \kd(l^- - q^-) - \frac{ \prt \kd(l^- -q^-)  }{ \prt l^- } (k^-) 
+ \frac{1}{2} \frac{\prt^2}{ \prt {l^-}^2} \kd (l^- - q^-) [k^-]^2  + \ldots  .\label{delta_func}
\eea
\amnew{In both the current case of longitudinal drag and diffusion, as well as in the 
case of transverse broadening in Eq.~(\ref{transverse_delta}), the $\kd$-function is 
meant in the limit of a narrow Gaussian distribution and should not be thought of as 
a singular distribution. The small shifts $k^{-}$ are infinitesimal compared to the 
jet energy $q^{-}$. In the limit of multiple such interactions we, once again, appeal 
to the central limit theorem and compute only the mean and the variance of the 
distribution.}

Unlike the case of transverse broadening there is no cylindrical symmetry to 
insist that the term proportional to $k^{-}$ is vanishing. One thus obtains two sets of 
terms involving derivatives of the $\kd$-function, one which involves single derivatives, which 
lead to a drag effect on the light-cone momentum distribution and 
the other which depend on double derivatives which lead to a diffusion in 
the light cone momentum. The coefficients of these 
terms will yield the elastic light-cone loss per unit light cone length $\hat{e}$ 
and the diffusion in light-cone loss $\hat{e}_2$. 

The coefficient of the the single derivative of the $\kd$-function may be 
simplified as, 
\bea
C_1 &=& \int dY^- d^4 y \frac{d k^- d^2 k_\perp}{(2 \pi)^3} 
e^{-i \frac{k_\perp^2}{2 q^-} y^- - i k^- y^+ + i k_\perp \x y_\perp} 
   \frac{4 \pi \A}{ 2 N_c} \lc P |  \left[ i \prt^- {A^a}^+(y) {A^a}^+ (0) \right] | P \rc.
\eea
In the equation above $| P \rc$  represents the nuclear state. We have dropped the 
$Y^{-}$ in the arguments of the gluon field based on translation invariance in a large 
nucleus.
Color confinement insists that the quark struck by the hard virtual photon emanate 
 from the same nucleon in both the amplitude and complex conjugate. Similarly the 
 gluon off which the outgoing quark scatters must also be restricted to the 
 same nucleon in both the amplitude and complex conjugate. In the model of a 
 nucleus made up of almost free nucleons (in the high energy limit), the correlation 
 between these two nucleons is $\rho (y^{-})/2p^{+}$ where 
 $\rho $ is the nucleon density evaluated along the path of the propagating out-going 
 quark and $p^{+}$ is the mean light cone momentum of a nucleon in the nucleus $p^{+} = P^{+}/A$.

Incorporating the above simplifications one obtains the light-cone loss per unit light cone 
path travelled by a struck quark without radiation in cold nuclear matter as $\hat{e} \simeq C_{1}$. In the 
case of hot deconfined matter the light-cone loss per unit path is obtained by generalizing the 
expression in cold matter as 
\bea
\hat{e}_{hot} &=& \frac{ 4 \pi \A_{s} \int d y^- \lc n | e^{-\B \hat{H}} [ i \prt^- {A^a}^+(y^-) {A^a}^+(0) ] | n  \rc
}{ 2 N_c } .  \label{e_loss_thermal}
\eea
Where  $|n\rc$ is a state in the thermal ensemble. In writing the above equation, we have 
made use of the boost invariance of the derived expressions along the direction of the jet. 
Unlike the expression for $C_{1}$ which is derived in the Breit frame, Eq.~\eqref{e_loss_thermal}
is intended in the rest frame of the medium. In this frame, one relates the light cone loss to the 
energy lost by a hard jet per unit length, for a jet traveling in the $-z$ direction 
\bea
\frac{d q^{- }}{ d L^{-} } \sim \frac{dE + |d p_{z} |}{ dt + |dz|} = \frac{dE (1 + | v |^{-1})}{ |dz|(|v|^{-1} + 1 ) } = \frac{d E}{d|z|}. 
 \eea
 
 The other transport coefficient of importance to the light-cone or longitudinal propagation of a hard 
 jet without radiation is obtained from the double derivative of the $\kd$-function in Eq.~\eqref{delta_func}. 
 This yields the diffusion in the light cone momentum per unit light cone path, denoted as the coefficient $\hat{e}_{2}$:
\bea
\hat{e}_2 &=&  \frac{ 4 \pi \A \int d y^- \lc n | e^{-\B \hat{H}}  [ {F^a}^{-+} {F^a}^{- +} ] | n  \rc }{ 2 N_c } ,
\eea
Calculating the change in the light-cone fraction of the leading parton 
due to only these two coefficients amounts to incorporating a drag and a 
diffusion term in the energy distribution of the hard parton. This is often 
referred to as the Gaussian approximation with the location of the peak 
of the Gaussian given by $\hat{e}$ and the width given by $\hat{e}_{2}$.

While operator products, in general, are evaluated at the hard scale at which
they interact with the jet, they may be calculated at the scale of the medium 
if there exists a well defined microscopic model of the medium. These 
may then be evolved up to the hard scale of the jet. 
For example, for an HTL plasma, the elastic loss 
coefficient may be evaluated as 
\bea
\hat{e} \!\!\!\!\!&=&\!\!\!\! 4 \pi \A_s\!\!\!\! \int\limits_0^{Q^2_{MAX}} \frac{d|\vk|^2 }{ 2 \pi  } \int\limits_0^1 \frac{dx}{2\pi} 
\frac{|\vk|^2 (-x) (1-x^2)(N_c^2 - 1) }{4 N_c}   
\left[ \left\{ \rho_T (|\vk|,k^0)  - \rho_L (|\vk|,k^0) \right\}
\left\{ 1 + n_B(k^0) \right\} \frac{\mbx}{\mbx} \right]_{k^0 = -|\vk|x} \!\!\!\! . \label{e_final}
\eea 
The equation above is obtained by evaluating Eq.~\eqref{e_loss_thermal} in the HTL limit. 
The factors $\rho_{L}$ and $\rho_{T}$ are the longitudinal and transverse HTL spectral densities 
(see Ref.\cite{Majumder:2008zg} for details). The limits of the integral $Q_{MAX} = \sqrt{ET}$ is 
set identically as for the calculation of $\hat{q}$ in an HTL medium.  The elastic energy loss experienced 
by a hard jet with energy $E$ propagating through an HTL plasma at a 
variety of temperatures and couplings are presented in 
Fig.~\ref{fig:elastic_loss}.

\begin{figure}
\centering
  \epsfig{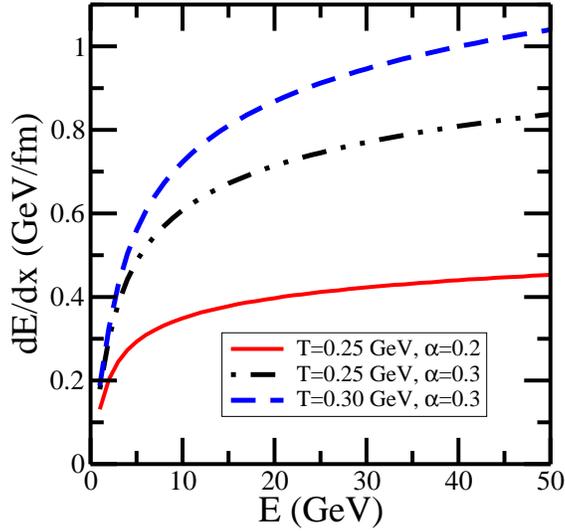}
 \caption{\label{fig:elastic_loss} The drag experienced by a hard quark jet in an HTL plasma. Figure adopted 
 from Ref.~\cite{Majumder:2008zg}. }
\end{figure}

With the discussion of elastic energy loss we complete the theoretical section of this review. 
In the remaining, we will outline the phenomenological modeling of the media 
encountered in jet modification experiments and review
{the main} experimental 
measurements.

%%%%%%%%%%%%%%%%%%%%%%%%%%%%%%%
%%%%%%%%%%%%%%%%%%%%%%%%%%%%%%%
%%%%%%%%%%%%%%%%%%%%%%%%%%%%%%%

\section{Parton energy loss in collisions: modeling of the medium}~\label{modeling}

%%%%%%%%%%%%%%%%%%%%%%%%%%%%%%%
%%%%%%%%%%%%%%%%%%%%%%%%%%%%%%%
%%%%%%%%%%%%%%%%%%%%%%%%%%%%%%%

Experimental measurements of parton energy loss are performed in cold
nuclear matter ($eA$ and $pA$ collisions) and in hot nuclear matter
produced in heavy ion collisions. In the experiment, the matter
density is not uniform. In heavy ion collisions, there is a strong
longitudinal expansion which causes a rapid decrease of the medium
density. In addition, the point of origin of the hard parton within
the matter is not known and therefore has to be integrated over.
Due to these effects, it is necessary to include some form of
averaging over the collision geometry in parton energy loss
calculations when comparing to experimental results.

We will first discuss the relevant aspects of the
collision geometry, before turning to a more detailed discussion of
the geometry averages used in the various calculations.

\subsection{Nuclear overlap geometry}
\label{sect:glauber_geom}
In central high-energy nuclear collisions, all of the incoming nucleons
traverse the entire nucleus that moves in the opposite direction. The
collision can be conveniently modeled in the approximation where the
incoming nucleons travel along straight lines. 

The density distribution of nucleons in the nucleus follows the
Woods-Saxon density profile~\cite{Woods:1954zz}:
\[\label{woods_sax}
\rho(\vec{r}) = \frac{\rho_0}{1+\exp{\frac{|\vec{r}|-R}{d}}},
\]
where $R$ is the radius of the nucleus and $d$ is the 'skin depth';
$\rho_0$ is a normalization constant, such that the integrated density
is equal to the atomic number $A$.
The density in the transverse plane, {the  thickness function $T$}, is then
\[
{T}(\mathbf s) = \int dz \rho(z,\mathbf s),
\]
where $z$ is along the longitudinal direction and $\mathbf s$ is a vector in
the transverse plane.

In a collision of two nuclei, the transverse distance between the
centers of the nuclei is called the impact parameter {$b$. Using
the transverse impact vector $\mathbf b$, we define the thickness function for the overlapping nuclei $T_{AB}$ and the collision density $\rho_{\mathrm coll}$:
\begin{equation}
T_{AB}(\mathbf s) =
T_A({\mathbf s}-\tfrac{1}{2} \mathbf b) \times T_B(\mathbf
s+\tfrac{1}{2} \mathbf b) \;\;\;\; \mathrm{and} \;\;\;\; \rho_{\mathrm coll} = T_{AB} \, \sigma.  
\end{equation}}

{Another measure of the overlap is the participant density 
\begin{equation}
\rho_{part}(\mathbf s) =
T_A({\mathbf s}-\tfrac{1}{2}{\mathbf b})\cdot\left(1-e^{-T_B({\mathbf
s}+\tfrac{1}{2} {\mathbf b})\; \sigma}\right) +
T_B({\mathbf s}+\tfrac{1}{2}\mathbf b)\cdot \left(1-e^{-T_A(\mathbf
s-\tfrac{1}{2} \mathbf b) \;\sigma}\right).
\end{equation}
$T_A$ and $T_B$ are thickness functions of the two colliding nuclei and $\sigma$ is the cross section}
for the process one is interested in. By integrating these two
densities {over the transverse coordinate $\mathbf s$ and taking $\sigma$ as the total inelastic cross section we
obtain the total number of binary (inelastic) collisions $N_\mathrm{coll} (b)$
and the total number of participants $N_\mathrm{part}(b)$, the number of
nucleons that had at least one (inelastic) collision at a given impact parameter $b$.}

The number of participants and the number of binary collisions are the limiting
cases of scaling of the particle production in heavy ion collisions, in a way that is related to the production cross section.
\begin{itemize}
\item{{\it Participant scaling} occurs for large-cross section
  processes, where every nucleon can produce particles at most
  once. This is also called wounded nucleon scaling
  \cite{Bialas:1976ed}. The multiplicity distribution in Au+Au
  collisions at RHIC follows this limit rather closely.}
\item{{\it Collision scaling} occurs for small cross section
  processes, where every individual encounter between two
  nucleons contributes the same probability to the total production
  cross section. High-\pt{} production of non-interacting probes at RHIC
  (photons) follows this limit (see Section \ref{sect:photons}).}
\end{itemize}

In parton energy loss models, $\rho_{part}$
or $\rho_{coll}$ can be used as a first approximation for the medium
density profile.  Since the soft
particle production scales more closely with the number of
participants, $\rho_{part}$ is the most
intuitive choice for a medium density profile. The distribution of
initial hard scatterings, which are the points of origin for the
partons that traverse the medium, is expected to follow $\rho_{coll}$.

In the above, we have used a continuum approach, the `optical limit',
where the nuclei are modeled as smooth density
profiles. Alternatively, one can model the collisions with localized
nucleons leading to a non-uniform density. This leads to important
fluctuations of the density profile in peripheral collisions. For more
details we refer the reader to \cite{Miller:2007ri} and references
therein.

The density profiles discussed above are suitable descriptions of the
initial state geometry. The initial state density profile has pressure
gradients that will lead to expansion, both in the transverse and
longitudinal direction. In an expanding medium, the density decreases
with time. The dominant effect is longitudinal expansion, which
results in a density decreasing approximately as $1/\tau$, where
$\tau$ is the proper time. Transverse expansion also leads to a
reduction of the local density, but at same time increases the system
size, so that the overall effect on energy loss observables is
smaller. Transverse and longitudinal expansion can be modeled using
hydrodynamic evolution.

%%%%%%%%%%%%%%%%%%%%%%%%%%%%%%%
%%%%%%%%%%%%%%%%%%%%%%%%%%%%%%%
%%%%%%%%%%%%%%%%%%%%%%%%%%%%%%%

\subsection{Modeling the produced matter: hydrodynamic evolution}~\label{modeling:hydro}

%%%%%%%%%%%%%%%%%%%%%%%%%%%%%%%
%%%%%%%%%%%%%%%%%%%%%%%%%%%%%%%
%%%%%%%%%%%%%%%%%%%%%%%%%%%%%%%

Relativistic fluid dynamics at low viscosity has emerged as the
leading theoretical setup with which to describe the space-time
evolution of the soft produced matter at RHIC. Global fits of a number
of simulations to the experimental data have placed an upper bound on
ratio of the viscosity $\eta$ to the entropy density $s$ to be at most
five times the absolute minimum allowed from quantum mechanics and
predicted by the AdS/CFT conjecture~\cite{Policastro:2001yc}:
\bea
\frac{1}{4\pi} {\leq} \frac{\eta}{s} \leq \frac{5}{4\pi}. 
\eea
Even ideal hydrodynamics where $\eta = 0$ yields a more or less consistent 
description of the soft spectrum~\cite{Nonaka:2006yn}. In these simulations, 
one solves the local differential equation, 
\bea
\prt_{\mu} T^{\mu \nu} (x,y,z,t) = 0,
\eea
where, $T^{\mu \nu}$ is the energy momentum tensor, for the 
ideal case:
\bea
T^{\mu \nu} = \lt(  \e + p \rt) u^{\mu} u^{\nu} - p \gmn.
\eea
In the equation above, $\e$ is the local energy density, $p$ is the local pressure and $u$ is the fluid four velocity.
The energy density is related to the pressure by the equation of state which is usually 
obtained by parametrizing lattice results. The set of equations are
solved with the additional constraint 
of local baryon number conservation,
\bea
\prt_{\mu} ( n_{B}  u^{\mu} ) = 0.
\eea

A solution to the equations also requires a starting proper time $\tau_{0}$. 
Fits to data have placed this around $\tau_{0} = 0.6$ fm/$c$. 
One also requires an initial configuration for the energy density and the baryon number density as a 
function of both the transverse location as well as the longitudinal location or rapidity. 
In the calculations presented in this review, $\e$ and $n_{B}$ were assumed to scale with each other with an 
overall normalization fit from experimental data. For the energy 
density, one assumes the impact parameter dependent profile.
\bea
\e (\tau_{0}, \mathbf{s},\eta) = \e_{MAX} \rho(\mathbf{s};b) H(\eta).
\eea
Based on the equation of state and lattice data, the initial energy density can be related to the entropy density 
which may then be inferred from the produced multiplicity. The normalization of the the Baryon number 
density is also obtained by fitting to the final experimental data on $p/\pi$ ratios. The transverse profile 
$W(x,y;b)$ is obtained as a combination of the wounded nucleon density and the binary collision 
density. In 2+1 D hydrodynamics, the rapidity profile is assumed to be a constant while in 3+1 D 
simulations it is assumed to be a broad Gaussian, given by the form 
$H(\eta) = \exp \lt(- (\eta )^{2}/(2\sg_{\eta}^{2}) \rt)$
with the width $\sg_{\eta}$ set from experimental data.

The solution of the fluid dynamical equations results in a space-time profile of the energy momentum tensor 
and, as a result, of the energy density and pressure. As the temperature drops below the transition 
temperature, one enters the hadronic phase, at this point either the fluid dynamical simulation is continued 
with the hadron gas equation of state or one may switch to a hadronic cascade such as 
URQMD~ \cite{Hartnack:1997ez}. Eventually as the temperature drops below the kinetic freezeout 
temperature the hadrons have to be decoupled from the calculation. In the case where hydrodynamics 
is continued into the hadronic phase, one uses a Cooper-Frye prescription~\cite{Cooper:1974mv} 
to convert the energy and momentum density in a unit cell into a distribution of free hadrons. 

The various input parameters along with the initial thermalization time $\tau_{0}$ and final freezeout time 
are dialed to obtain the best fit with experimental data on soft hadrons of various flavors and from various 
centralities. Once all these parameters have been set, the space-time profile of the energy density can 
be used to calculate the quenching of hard jets.

%%%%%%%%%%%%%%%%%%%%%%%%%%%%%%%
%%%%%%%%%%%%%%%%%%%%%%%%%%%%%%%
%%%%%%%%%%%%%%%%%%%%%%%%%%%%%%%

\subsection{Energy loss in a non-uniform medium}~\label{modeling:non-uniform}

%%%%%%%%%%%%%%%%%%%%%%%%%%%%%%%
%%%%%%%%%%%%%%%%%%%%%%%%%%%%%%%
%%%%%%%%%%%%%%%%%%%%%%%%%%%%%%%

Hard partons are produced at all locations in the transverse plane, with a
probability distribution following $\rho_\mathrm{coll}$ and no
preference in azimuthal angle, and then propagate through the dense
matter undergoing energy loss. In order to take into account the
non-uniform, dynamic medium, one has to integrate over all production
points and azimuthal angles, taking into account the medium density
profile that the partons experience along their path out of the
medium.

All energy loss formalisms contain an integral over the scattering
centers that the partons encounter along their paths. To obtain
analytical results, the medium density profile is assumed
to be either uniform or of a specific analytical shape ({\it e.g.}
exponential decay in GLV). For a non-uniform medium, one can evaluate
the full integral over the non-uniform medium density for every
parton. This is what was done in the case of the HT formalism in
\cite{Bass:2008rv}.

For the ASW and GLV approaches, analytical results have been derived
for a homogeneous expanding medium, where the density falls as
$1/\tau$ along the path length
\cite{Baier:1998yf,Salgado:2003gb,Zakharov:1997uu}. It has been shown
in that in the infinite-energy regime those analytical results are
well approximated by
\begin{equation}
\Delta E \propto \int dx\, x\, \hat{q}(x), 
\end{equation}
where $\hat{q}(x)$ is a local transport coefficient and $x$ is a
coordinate along the path of the parton. Note that the
integral is equal to $\tfrac{1}{2}\hat{q}L^2$ for a uniform medium. In addition to the scaling of the average energy loss, it
has also been shown that in the homogeneous expanding medium, the
differential gluon emission spectrum can be approximated using
the calculation for a homogeneous static medium with effective
transport coefficient \cite{Salgado:2003gb}
\[
\hat{q}_\mathrm{eff} = \frac{2}{L} \int_{x_0}^{x_0+L} dx\, x\, \hat{q}(x).
\]
This is often referred to as the 'dynamical scaling law'. A more
general version of this scaling law was recently derived in
\cite{Arnold:2008iy}.

To calculate energy loss in a non-uniform, dynamical medium using the
ASW formalism, it is assumed that the dynamical scaling law can be
extended to this scenario. The first example of such a calculation is
the Parton Quenching Model (PQM) by Dainese, Loizides and Paic
\cite{Dainese:2004te}, where $\rho_{coll}$ was used based on
saturation model arguments \cite{Eskola:2001bf}. More recent
calculations use a hydrodynamical medium and where the local
$\hat{q}(x)$ is proportional to the temperature $T^3$, energy density
$\epsilon^{3/4}$ or the entropy density $s$
\cite{Bass:2008rv,Renk:2006pk,Armesto:2009zi}.

The AMY energy loss calculation is carried out in the finite
temperature field theory approach and therefore directly relates the
energy loss to the temperature $T$. In a hydrodynamical evolution, the
local temperature $T$ is known. The rates for a hard parton to decay
into a quark or a gluon are then calculated using the local
temperatures and the change in the 
distribution is computed using the Fokker-Planck equation over all 
paths \cite{Turbide:2005fk,Bass:2008rv}.

%%%%%%%%%%%%%%%%%%%%%%%%%%%%%%%
%%%%%%%%%%%%%%%%%%%%%%%%%%%%%%%
%%%%%%%%%%%%%%%%%%%%%%%%%%%%%%%

\subsection{Calculating observables}

%%%%%%%%%%%%%%%%%%%%%%%%%%%%%%%
%%%%%%%%%%%%%%%%%%%%%%%%%%%%%%%
%%%%%%%%%%%%%%%%%%%%%%%%%%%%%%%

Now we have all {the} ingredients for a calculation of parton energy loss
that can be directly compared to {experiment}. For example, to
calculate the nuclear modification factor $R_{AA}$, one starts from
the factorized calculation in the vacuum, Eq.~\eqref{fact_1}. Parton
energy loss is then included by substituting the vacuum fragmentation
function with the medium-modified fragmentation function, which is
given in Section \ref{multiple_radiation}, averaged over the medium
density profile, as explained in the previous section. In general, the
average over the medium density profile involves integrating out 
the parton production point and direction.

In the next {Chapter}, we will compare energy loss calculations to
experimental data on the nuclear modification factor $R_{AA}$ for
light and heavy hadrons, and di-hadron and $\gamma$-hadron
measurements. The {di-hadron} measurements involved pairs of partons,
which are produced at a single point in the medium and then propagate
outward in opposite directions in the transverse plane
(back-to-back). As a result, the pathlength and density profile that
both partons see are not independent. For example, if the pair
originates from the periphery of the collisions zone {and travels outward}, one parton is
likely to see a long pathlength in the dense medium, while the other
will see a short pathlength. These correlations have to be included in
the model calculations as well.

Early comparisons of energy loss calculations to experimental data
were based on a simplified medium geometry, using hard spheres or
cylinders \cite{Wang:2002ri,Vitev:2002pf,Eskola:2004cr}, instead of the Woods-Saxon geometry in
Eq. \eqref{woods_sax} and no
hydrodynamical evolution. There are also calculations using a
Woods-Saxon overlap geometry, without hydrodynamical evolution \cite{Dainese:2004te,Wicks:2005gt}. These
are more realistic than the hard-sphere geometries in the sense that
they include the low-density regions at the edge of the
nuclei. In this review, we will focus on calculations which sample the
space-time evolution using a hydrodynamical calculation for the medium
density evolution \cite{Renk:2006pk,Bass:2008rv,Armesto:2009zi}.

\section{Comparison to experiment}

\subsection{Choosing observables}
\label{sect:observables}
In the following, we review a selected set of experimental data
and compare to energy loss calculations to show how
the measurements constrain the energy loss formalisms and the
properties of the medium.

A wealth of different measurements have been performed to provide
insight in the energy loss mechanism. In general, care should be taken
to select observables that are sensitive to the process that one wants
to investigate. This selection process requires a careful exchange of
ideas between theory and experiment. For example, the first
measurements at RHIC of the nuclear modification factor
\cite{Adler:2002xw,Adcox:2001jp} and di-hadron correlations
(disappearance of the away-side jet) \cite{Adler:2002tq} have clearly
shown that energy loss is large. It was subsequently realized,
however, that a variety of models can describe the data, because these
measurements are not very sensitive to the details of the energy loss
distribution. It has been shown that it is in practice impossible to
extract the energy loss probability distribution from such
measurements \cite{Renk:2007mv} even if the experimental uncertainties
would be exceedingly small.
It is therefore key to identify observables that are sensitive to
specific aspects of the energy loss process. For example, it has been
argued that comparing the away-side suppression to the single hadron
suppression allows to determine the path-length dependence of energy
loss ($L$ vs $L^2$) \cite{Renk:2007id}.

In Figure \ref{fig:ff_comp} it was shown that the different formalisms
for parton energy loss calculations predict different medium modified
fragmentation functions. Such differences can only be determined from
experiments if the initial parton energy can be controlled or
measured. This technique has been used successfully in Deeply
Inelastic Scattering experiments off nuclei to measure parton energy
loss in cold nuclear matter (see Section \ref{sect:comp_hermes}). In
heavy ion collisions, there are two basic ways in which one can
control the initial parton energy. One is to use $\gamma$-jet events
(or $Z^0$-jet at LHC) where the transverse momentum of the photon and
the initial parton (jet) are equal (as long as initial state
broadening and higher order large angle radiations are negligible). The
alternative is to reconstruct jets in the final state. To what extent
the reconstructed jet energy is equal to the initial parton energy
depends both on the jet reconstruction algorithm and on the effect of
medium modifications on the jet. Several measurements are likely
needed to disentangle various effects (jet spectra, jet fragmentation
functions and away-side jet rates have been discussed).

The primary consideration when selecting observables should be the
intrinsic sensitivity: the larger the effect of a given model aspect
on the observed quantity, the more promising the observable is. This
type of optimization is performed based on theoretical insight and
model calculations. 

A secondary consideration is the experimental precision. Experimental
uncertainties vary from observable to observable, depending mainly on
the cross section and the amount of background. Experience at RHIC
shows that for the least complicated measurement, such as pion
spectra, an experimental uncertainty in the 5-10\% range can be
reached. More complicated measurements, with larger backgrounds that
cannot always be accurately modeled in heavy ion events, have
correspondingly larger uncertainties. Further reduction of the
uncertainties often requires a large effort, often requiring dedicated
detector construction and calibration for a specific
measurement, and is therefore only pursued by
experiments if there are compelling theoretical reasons to do so ({\it
e.g.} precision tests of the electroweak sector of the standard
model). \am{A few specific examples of how the experimental precision may evolve
at RHIC are given in Section \ref{sect:outlook_precision}}.

\subsection{Perturbative QCD in p+p collisions}
\begin{figure}
\centering
\epsfig{file=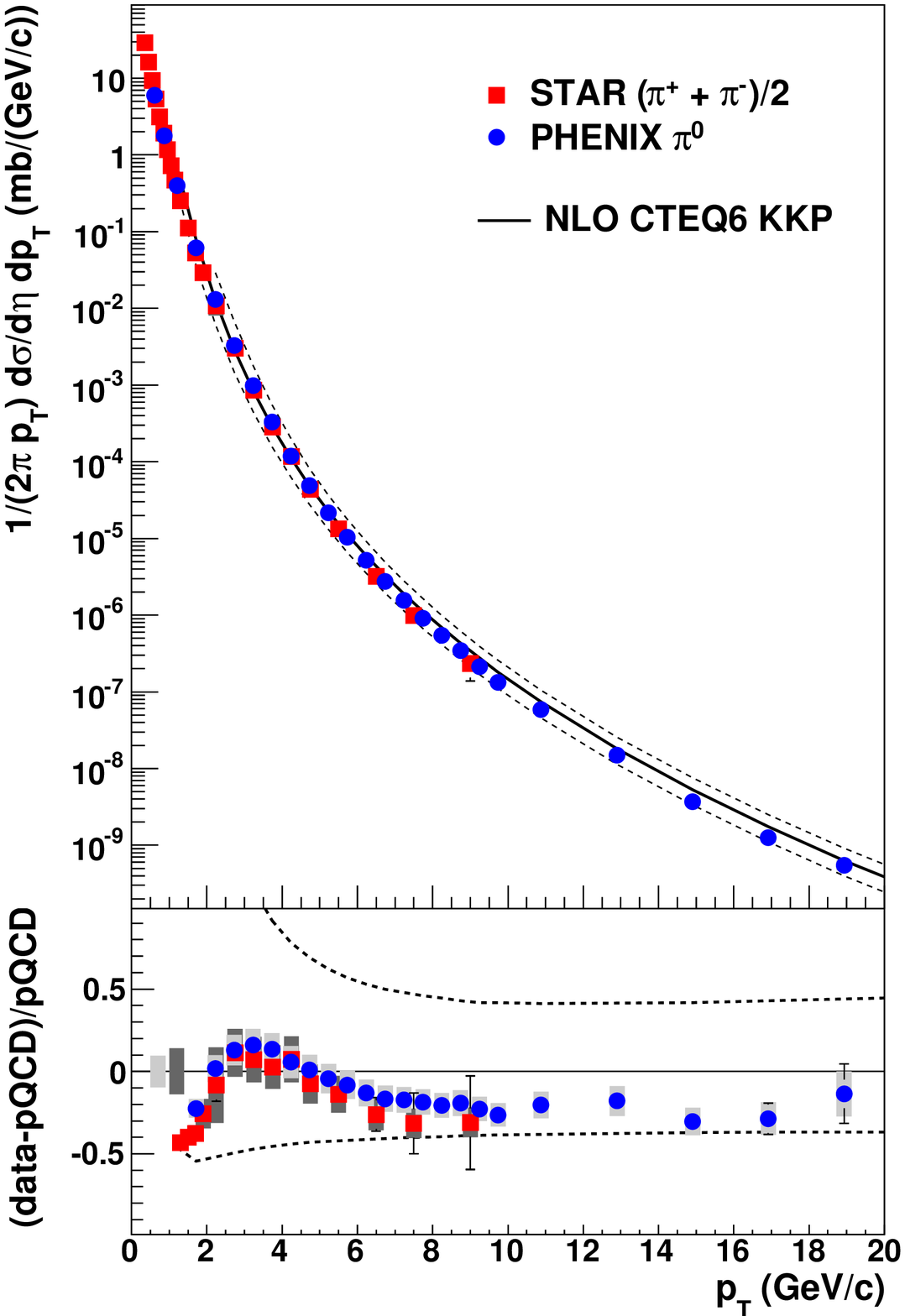,width=0.45\textwidth}\hfill%
\epsfig{file=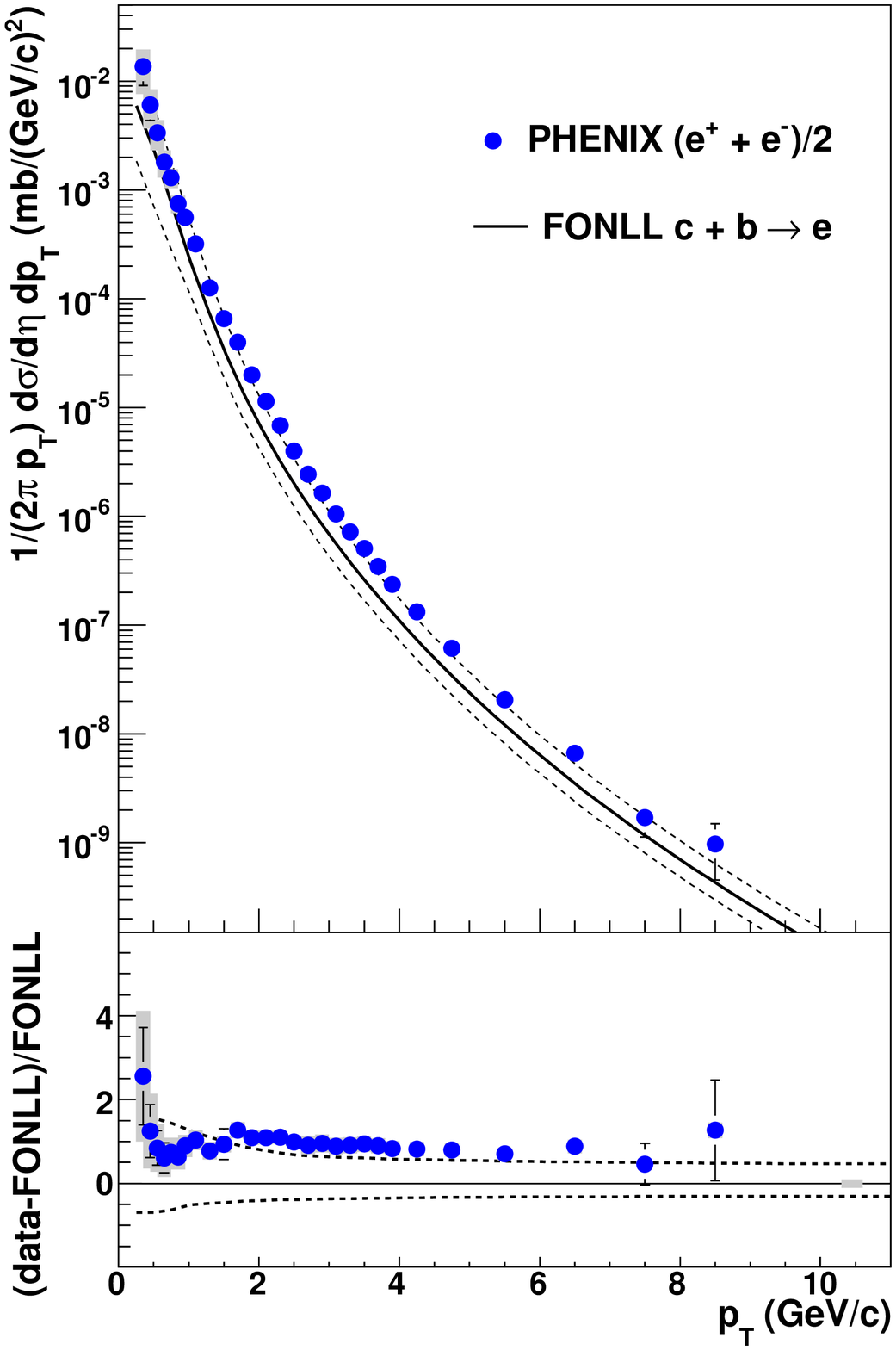,width=0.45\textwidth}
\caption{\label{fig:pp_spectra}Left panel: Comparison of the $\pi^0$
  \pt-spectrum in p+p collisions at $\sqrt{s} = 200$ GeV as measured
  by PHENIX \cite{Adare:2007dg} and charged pions measured by STAR
  \cite{Adams:2006nd} to NLO pQCD calculations \cite{Jager:2002xm} at
  three scales $\mu=\pt$ (solid line) and $\mu=0.5\pt$ and $\mu=2\pt$
  (dashed lines). The uncertainty bars on the left side of the lower
  panel show the overall normalization uncertainty in the STAR (dark
  grey band) and PHENIX (light grey band) measurements.  Right panel:
  Comparison of the \pt-spectrum of non-photonic electrons as measured
  PHENIX \cite{Adare:2006hc} to a NLO pQCD calculation with resummed
  Next-to-leading logarithms (FONLL) \cite{Cacciari:2003uh}. The
  dashed lines indicate the uncertainty on the FONLL calculation,
  which is estimated by varying the factorization scale and the quark
  masses. The lower panels show the relative difference with the
  central curves. The normalization uncertainty is indicated in the
  lower panel on the right. }
\end{figure}

Factorized perturbative QCD calculations have been compared
extensively to experimental data \am{on $p$-$p$ collisions} at
energies ranging from ISR $\sqrt{s} \approx 60$ GeV to Tevatron
energies $\sqrt{s}=1.9$ TeV
\cite{Aurenche:1999nz,:2008hua,Aaltonen:2009ty,Pumplin:2009nk}. RHIC
measurements at mid-rapidity generally agree well with perturbative
QCD calculations. Two examples are shown in this section.

Figure \ref{fig:pp_spectra} (left panel) shows a comparison of pion
spectra measured by STAR and PHENIX, with STAR measuring charged pions
in a Time Projection Chamber (TPC) \cite{Adams:2006nd} and PHENIX
measuring neutral pions with an Electromagnetic Calorimeter (EMCal)
\cite{Adare:2007dg}. The measurements are compared to Next-to-Leading
Order (NLO) pQCD calculations by J\"ager et al. \cite{Jager:2002xm},
using the CTEQ6M parton density functions \cite{Pumplin:2002vw} and
fragmentation functions by Kniehl, Kramer and P\"otter (KKP)
\cite{Kniehl:2000fe}. The measurements by STAR and PHENIX agree well
with each other and with the NLO pQCD calculation. The uncertainties
on the calculation are estimated by varying the factorization and
renormalization scales between $\mu = 0.5 \pt$ and $\mu = 2 \pt$ and
are indicated by dashed lines on the figure.

Heavy quark production is measured at RHIC via the semi-leptonic
decays $D \rightarrow e$ and $B \rightarrow e$. The inclusive electron
distributions that are measured in the experiment contain substantial
contributions from photon conversions and Dalitz decays of $\pi^0$ and
$\eta$ mesons which are subtracted to obtain the `non-photonic'
electron yield. The \pt-dependent inclusive cross section of
non-photonic electron production as measured by PHENIX
\cite{Adare:2006hc} in p+p collisions at $\sqrt{s}=200$ GeV is shown
in the right panel of Fig. \ref{fig:pp_spectra} and compared to NLO
pQCD calculations including resummed Next-to-Leading Logarithms
(FONLL) \cite{Cacciari:2003uh}. \mvlnew{A similar measurement was
performed by the STAR experiment \cite{Abelev:2006db} and showed a
larger yield of non-photonic electrons. This measurement, however, has
recently been superseded by a new, still preliminary, result
\cite{xie_dis2010}.} The uncertainties on the theory expectation are
estimated by varying the renormalization and factorization scale and
the quark masses. The yields measured by PHENIX are larger than the
nominal theory expectation, but close to the upper bound on the
expectation.

\subsection{Cold nuclear matter effects}
Before turning to nucleus-nucleus collisions to study parton energy
loss in hot nuclear matter, we will briefly discuss the relevant
effects in cold nuclear matter. We distinguish initial state and final
state effects in cold nuclear matter. The dominant initial state
effect is the modification of the parton density distribution in the
nucleons, or the 'EMC-effect', named after the first experiment that
measured this effect \cite{Ashman:1992kv}. The most important final
state effect is parton energy loss in cold nuclear matter, which has
been found to be significant in the HERMES experiment
\cite{Airapetian:2007vu}. Both effects are discussed below.

For completeness, we also mention here the 'Cronin effect', a possibly
mass-dependent enhancement of hadron production at intermediate \pt{}
in proton-nucleus collisions \cite{Cronin:1974zm} which is generally
attributed to transverse momentum broadening due to multiple
scattering of partons on the nuclear target. For a recent review, see
\cite{Accardi:2002ik}.

\subsubsection{Energy loss in cold nuclear matter}
\label{sect:comp_hermes}
\begin{figure}
\centering
\epsfig{file=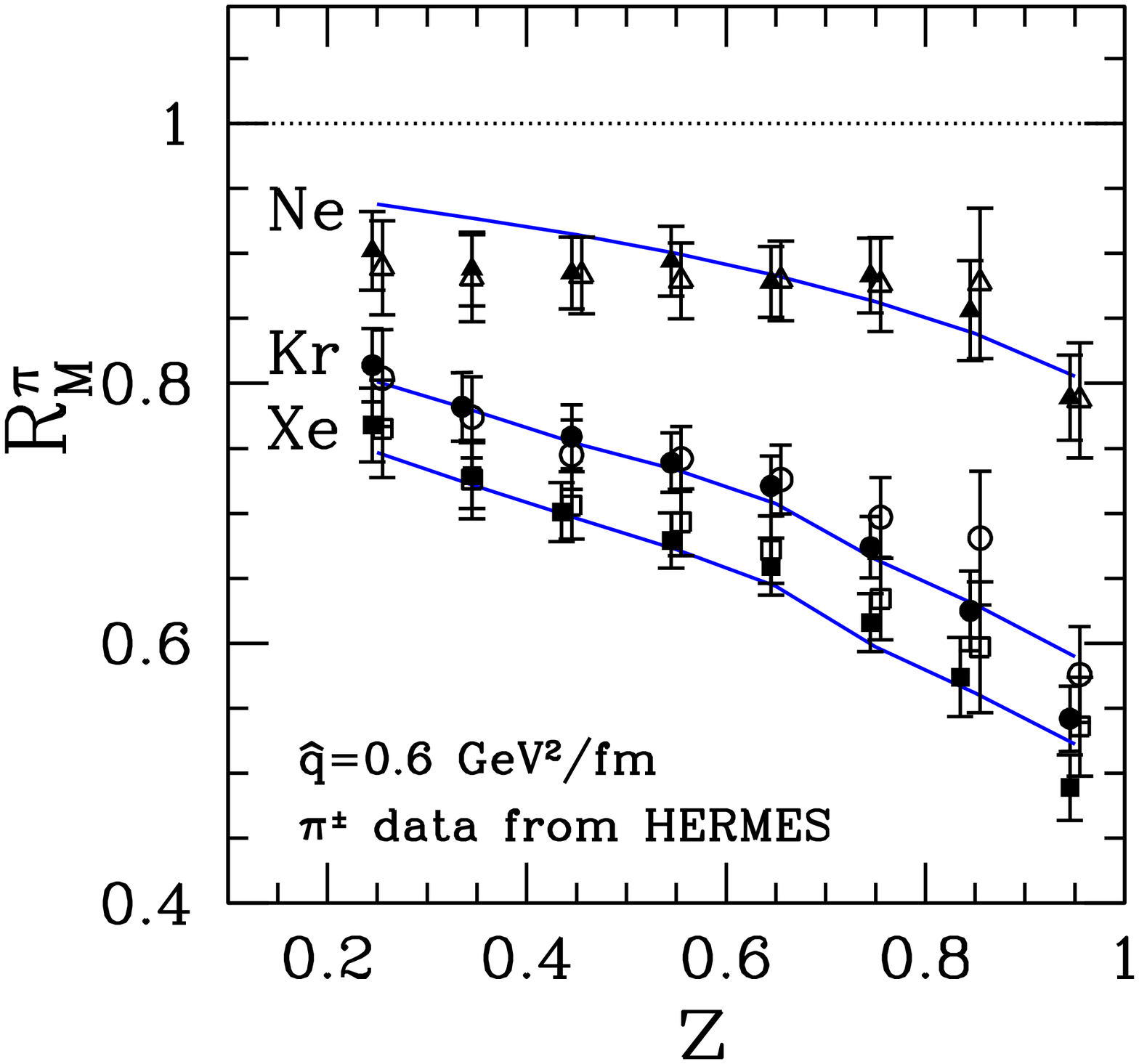,width=0.48\textwidth, bb=18 182 592
720}\hfill%
\epsfig{file=HERMES_review_plot.eps,width=0.45\textwidth}
\caption{\label{fig:hermes_comp}Comparison of the measured
  suppression of pion (full symbols $\pi^+$, open symbols $\pi^-$)
  production in $eA$ collisions by HERMES \cite{Airapetian:2007vu} with
  calculations using ASW-BDMPS quenching weights (left panel) and the
  higher twist formalism (right panel). Figures from
  \cite{Accardi:2009qv} (\copyright Societa Italiana di Fisica)  and~\cite{Majumder:2009zu}.}
\end{figure}

Figure \ref{fig:hermes_comp} shows a comparison the measured
suppression of pions from fragmentation in cold nuclear matter in $eA$
collisions by the HERMES experiment \cite{Airapetian:2007vu} with
calculations using the BDMPS-ASW quenching weights (left panel) and
the higher twist formalism (right panel). In deeply inelastic
scattering experiments, the jet energy can be determined from the
momentum vector of the scattered electron; the measurement is
presented in terms of the momentum fraction $z=E_{h}/\nu$ where $E_h$
is the hadron energy and $\nu$ the energy of the virtual photon. The
extracted transport coefficient for the BDMPS quenching weights is
$\hat{q} = 0.6$ GeV$^2$/fm in the center of the nucleus. Using the
higher twist formalism, \am{including the effect of multiple emissions, 
a quark $\hat{q} \simeq 0.08$ GeV$^{2}$/fm is obtained. 
In an earlier calculation, involving only one emission in medium~\cite{Wang:2002ri}, a $\hat{q} = 0.12$
GeV$^2$/fm was obtained \cite{Accardi:2004gp}.}

\subsubsection{Experimental study of initial state: d+Au}
\begin{figure}
  \centering
\epsfig{file=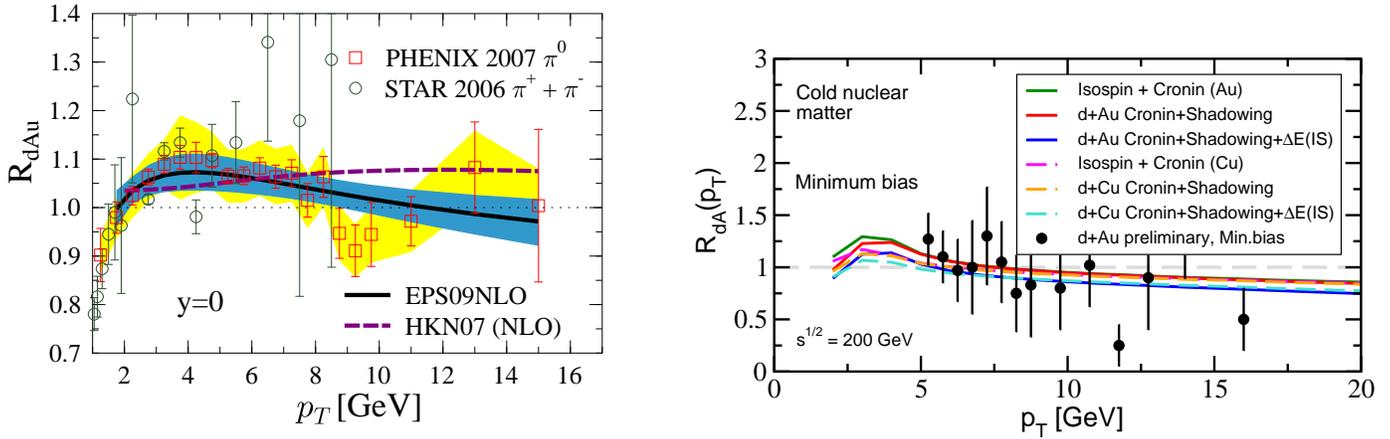,width=0.49\textwidth}\hfill%
\epsfig{file=vitev_gamma_dau.eps,width=0.49\textwidth,bb=38 446 728
  843,clip=}
\caption{\label{fig:dau}Nuclear modification factor $R_{dAu}$
  for pions (left panel) and photons (right panel) in d+Au collision
  at $\sqrt{s_{NN}} = 200$ GeV, compared to calculations with nuclear
  parton density functions (left panel) and various other cold nuclear
  matter effects (right panel). Figures
  reproduced with permission from \cite{Eskola:2009uj} and
  \cite{Vitev:2008vk} (right panel \copyright American Physical Society).}
\end{figure}
At RHIC, d+Au collisions are used to measure the effect of possible
initial state effects on the pion and photon spectra. The left panel
of Fig.  \ref{fig:dau} shows recent measurements of the nuclear
modification factor $R_{dAu}$ for $\pi^0$ from PHENIX
\cite{Adler:2006wg} and charged pions from STAR \cite{Adams:2003im}
and compares to calculations using a recent analysis of nuclear parton
density functions (nPDF) (EPS09 \cite{Eskola:2009uj} and HKN07
\cite{Hirai:2007sx}). It should be noted that the DIS results do not
fully constrain the nPDFs and the $R_{dAu}$ measurement has been
included in the determination of the nPDFs. The good agreement with
the data therefore indicates that the measured values of $R_{dAu}$ are
consistent with the nuclear modification of PDFs which is
measured in DIS off nuclei \cite{Ashman:1992kv}. No significant final
state interaction is apparent in the data.

The apparent absence of cold nuclear matter energy loss in the
$R_{dAu}$ in Fig. \ref{fig:dau} at first sight is surprising, given
the large suppression of hadron production at high momentum fraction
$z$ in the DIS data in Fig. \ref{fig:hermes_comp}. However, the $d+Au$
results are at $y=0$ in the laboratory\am{; in order to compare with
the HERMES results, we boost to the rest frame of the nucleus where
the jet with only a transverse momentum $p_{T}$ at midrapidity has an
extremely large energy due to the boost. The HERMES measurements show 
that the effect of energy loss reduces with jet energy \cite{Airapetian:2007vu} and is
expected to vanish at the very high energy of the boosted jet.}
A detailed modeling of this effect is given in
\cite{Accardi:2002ik}.

The right panel of Fig. \ref{fig:dau} shows a comparison of the
measured nuclear modification factor $R_{dAu}$ for photons in d+Au
collisions at RHIC ($\sqrt{s_{NN}} = 200$ GeV) with calculations
including various cold nuclear matter effects \cite{Vitev:2008vk}. The calculations show a
modest enhancement at $\pt < 5$ GeV/$c$ due to the Cronin effect and
modest suppression at higher \pt{} due to a combination of isospin
effects and shadowing. The curves with energy loss (labeled $\Delta
E(IS)$) refer to energy loss of the incoming parton as it propagates
through the nuclear matter before the hard scattering
\cite{Vitev:2007ve}. The total effect on the photon spectrum is around
20\%, which is slightly smaller than the current experimental
uncertainties in the measurement.

\subsection{Heavy ion collisions at RHIC}
At RHIC, several types of measurements that are sensitive to parton
energy loss in hot and dense QCD matter have been performed. In the
following, we will discuss the results on inclusive hadron
suppression, di-hadron suppression as well as results for heavy
quarks. Comparisons to model calculations of energy loss are made
where possible, with emphasis on calculations based on a full
hydrodynamic evolution of the density and radiative energy loss
according to the four formalisms discussed in \am{Chapter~\ref{single_gluon}}.

We also focus on the region of high \pt, where perturbative QCD is
most likely to be applicable. Experimental results at RHIC show
anomalously large baryon to meson ratios at $\pt \lesssim 6$ GeV/$c$
\cite{Adler:2003cb,Abelev:2006jr,Abelev:2007ra}, indicating that there
are important contributions to hadron production beyond factorized
perturbative QCD in that regime. We therefore limit the discussion to
$\pt > 6$ GeV/$c$ where possible.

\subsubsection{Verifying initial production rates: Photon production}
\label{sect:photons}

\begin{figure}
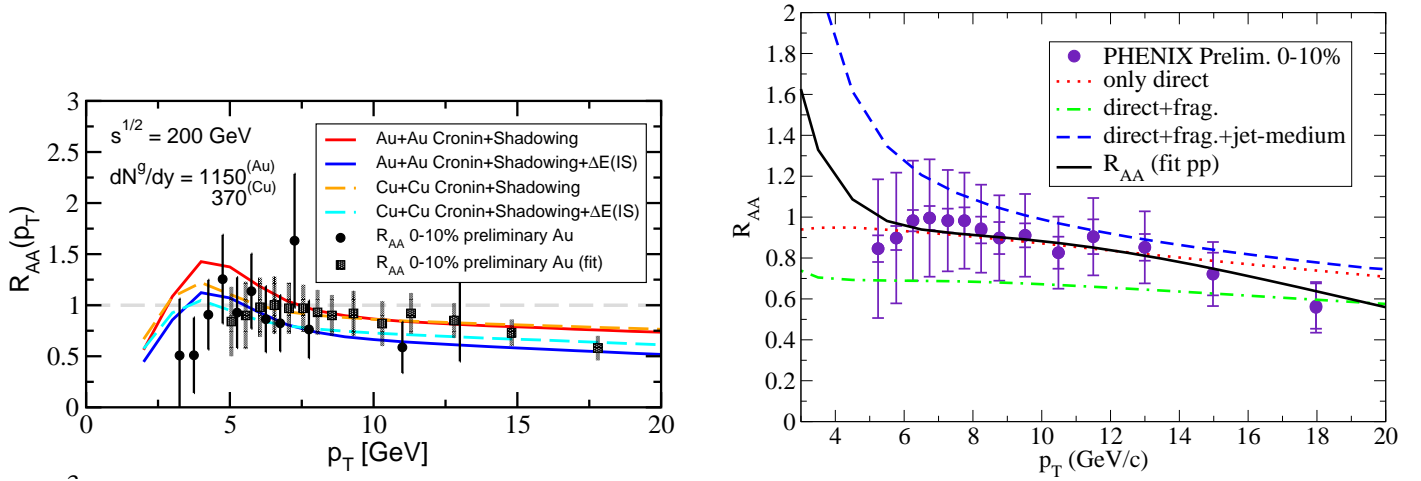

  \epsfig{file=vitev_photons_AA.eps,width=0.48\textwidth,bb=38 441
    728 843,clip=}\hfill%
\epsfig{file=qin_etal_photons_aa.eps,width=0.48\textwidth}
\caption{\label{fig:phot_raa}Nuclear modification factor $R_{AA}$ for
  prompt photons in
  central Au+Au collisions at $\rtsnn=200$ GeV as a function of \pt,
  \cite{Adler:2005ig} compared to various model calculations. Left
  panel: Model curves by Vitev, including initial state energy loss
  \cite{Vitev:2008vk}. Right panel: Model curves by Qin et al showing
  separate contributions of direct, fragmentation and medium induced
  photons \cite{Qin:2009bk}. The black solid line shows the effect of
  adjusting the $p+p$ reference in the calculation to the measured
  spectrum. Both panels \copyright American Physical Society.}
\end{figure}
Figure \ref{fig:phot_raa} shows the centrality dependence of the
nuclear modification factor $R_{AA}$ (Eq. \ref{raa})
for prompt photons in central Au+Au collisions as measured by PHENIX
at $\rtsnn=200$ GeV, compared to various model curves. The measured
$R_{AA}$ for photons is close to unity, as expected for hard probes
($N_{coll}$ scaling) without energy loss.

The model curves in Fig. \ref{fig:phot_raa} show the effect of initial
state energy loss (left panel) and final state energy loss for
fragmentation photons (right panel) and medium-induced photons,
including medium induced QED bremsstrahlung and parton-photon
conversion in scattering (right
panel). These mechanisms, together with isospin effects, are expected
to lead to a suppression of the photon yields by up to 25\% at high
$\pt \gtrsim 10$ GeV, with the exact value depending on details of the
modeling. The model predictions are in agreement with the experimental
data, but the current large uncertainties on the measurement preclude
quantitative conclusions about the contributions of the various effects.\\

\subsubsection{Light hadron suppression}

\begin{figure}
\centering
\epsfig{file=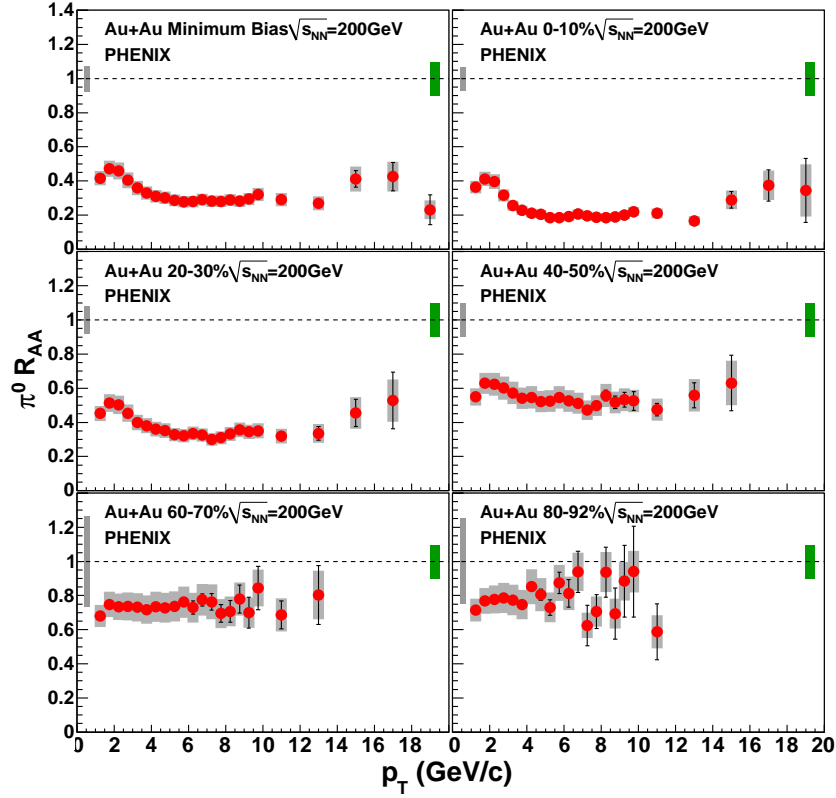,width=0.6\textwidth}
\caption{\label{fig:pi0_raa}Nuclear modification factor $R_{AA}$ as a
  function of \pt{} for $\pi^0$ as measured by PHENIX in Au+Au
  collisions at $\rtsnn=200$ GeV. The different panels show different
  centrality selections. The error bands around the dashed line at
  $R_{AA}=1$ indicate the uncertainty from the number of binary
  collisions (left band) and the normalization uncertainty in the
  $p+p$ reference data (right band). Figure reproduced with permission from \cite{Adare:2008qa} (\copyright American Physical Society).}
\end{figure}
Parton energy loss leads to a suppression of the pion yield at high
\pt . A recent measurement of $R_{AA}$ of $\pi^0$ by the PHENIX
experiment in shown in Fig. \ref{fig:pi0_raa}. A clear suppression is
visible at high \pt . The suppression increases with centrality and is
approximately \pt-independent for $\pt > 4$ \GeVc . The value of
$R_{AA}$ at high \pt{} for central collisions is around 0.2.

\begin{figure}
\centering
\epsfig{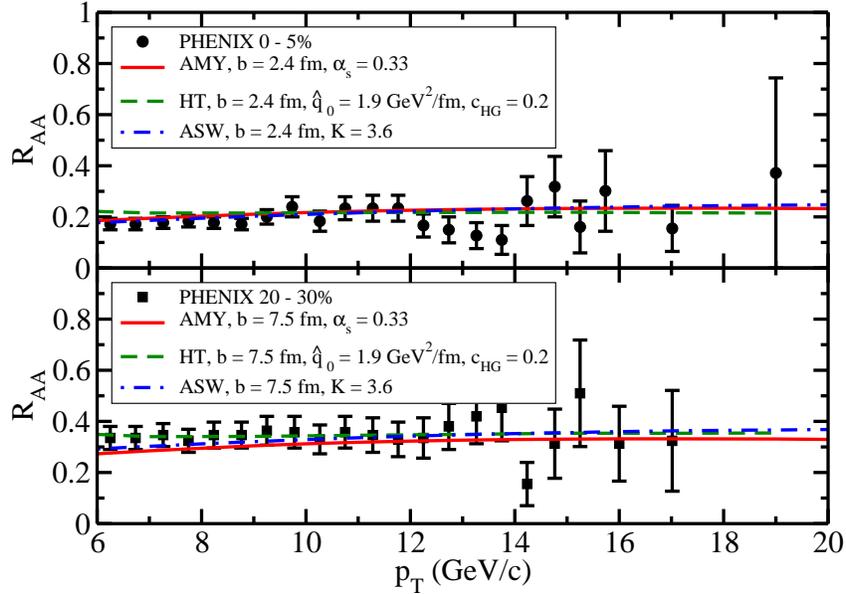}
\caption{\label{fig:raa_bassetal} Nuclear modification factor $R_{AA}$
  as a function of \pt{} as measured by PHENIX for $\pi^0$ compared to
  calculations using a 3D hydrodynamical medium evolution and 3 different
  energy loss formalisms: Higher Twist (HT), Hard Thermal Loop (AMY)
  and multiple soft gluon scattering (ASW). The upper panel shows
  results for 0-5\% central Au+Au collisions at $\rtsnn = 200$ GeV and
  the lower panel shows 20-30\% central collisions. Figure reprinted with permission from
  \cite{Bass:2008rv} (\copyright American Physical Society).}
\end{figure}

\begin{table}
\centering
\am{
\begin{tabular}{|c|c|c|c|}
\hline
$\hat{q} \propto$ & ASW & HT & AMY\\
 & $\hat{q}_0$ [GeV$^2$/fm] & $\hat{q}_0$ [GeV$^2$/fm] & $\hat{q}_0$
 [GeV$^2$/fm] \\
\hline
\hline
$T^3$ & 10 & 2.3 & 4.2 \\
$\epsilon^{3/4}$ & 18.5 & 4.5 & \\
$s$ & & 4.3 & \\
\hline
\end{tabular}}
\caption{\label{tab:qhat_bass_etal}Values of the initial transport
  coefficient $\hat{q}_0$ ($\tau = \tau_0 = 0.6$ fm/$c$) at the center
  of the \am{most central collision (0-5\%)} as determined by comparing calculations of $R_{AA}$
  with \am{ a medium density space-time profile taken from a 3D hydrodynamical evolution and three
  different parton energy loss models}~\cite{Bass:2008rv}.}
\end{table}

Figure \ref{fig:raa_bassetal} compares calculations of the nuclear
modification factor using a \am{3D} hydrodynamical evolution for the
medium density and three different parton energy loss formalisms to
experimental data on $\pi^0$ suppression from PHENIX in Au+Au
collisions at $\rtsnn = 200$ GeV and two different centralities.  \am{
The hydrodynamical simulation is tuned to fit RHIC data with a $p_{T}
< 2$ GeV and provides space-time profiles for the medium
properties.} The only free parameter in the calculation
is an overall proportionality constant between the local transport
coefficient and the medium density. This parameter was set using the
value of $R_{AA}$ at $p_{T} = 9.5$ GeV in central collisions. The
behaviour with $p_{T}$ and centrality are parameter-free
predictions. It is interesting to note that all three model
calculations show only a very weak dependence of $R_{AA}$ on \pt, in
agreement with the data. In the model calculations, the independence
of $R_{AA}$ on \pt{} does not have a simple interpretation; it comes
about as a result of the interplay between the shape of the parton
spectrum and the energy-dependence of energy loss, which is partially
due to the kinematic cut-off $\Delta E < E$.

The resulting values of $\hat{q}_{0}$, the transport coefficient at 
thermalization ($\tau=0.6$~fm/c) at the center of the most central
collisions (0-5\% centrality) are given in Table
\ref{tab:qhat_bass_etal}. The extracted values for $\hat{q}$ depend on
whether the local transport coefficient is assumed to be proportional
to the energy density $\epsilon$ (to the appropriate power 3/4), the
third power of the temperature $T$ or the entropy density $s$. In
addition, there are significant differences between the energy loss
formalisms: the Hard Thermal Loop formalism (AMY) gives $\hat{q}_0 =
4.1$ GeV$^2$/fm, while the higher twist formalism gives a lower value
of 2.3 GeV$^2$/fm and the multiple soft scattering approach (ASW)
gives a much larger value of 10 GeV$^2$/fm. 

It is interesting to note that in this comparison, the HT and AMY
formalisms give similar medium density parameters, while it can be
seen in Fig. \ref{fig:ff_comp} that AMY generally has a larger
suppression at given $T$ and $L$ than HT. Also, in
Fig. \ref{fig:ff_comp}, the suppression for ASW is closer to HT for
$L=2$ than to AMY, while in this comparison ASW needs a much larger
value of $\hat{q}_0$. The qualitative difference between the
comparison to $R_{AA}$ in Fig. \ref{fig:raa_bassetal} and the
fragmentation functions in the Fig. \ref{fig:ff_comp} is probably due
to differences in the treatment of the medium in the hydrodynamical
calculations of Fig. \ref{fig:raa_bassetal} and
Ref. \cite{Bass:2008rv}: for the AMY calculation, no energy loss was
taken into account for $\tau < \tau_0$ and \amnew{in} the hadronic phase at late
times, while for the HT calculation in the same paper, the density was
taken to be constant for $\tau < \tau_0$ and energy loss was also
calculated in the hadronic phase. The ASW calculation in that paper
calculated no energy loss for $\tau < \tau_0$ but did continue the
energy loss calculation in the hadronic phase.

The treatment of the
density for early times $\tau < \tau_0$, when the medium is very
dense, has a large impact on the extracted medium density. A factor 2
difference was found between the case where $\hat{q}=0$ is taken for
$\tau < \tau_0$ compared to constant $\hat{q}=\hat{q}_0$ for the
initial stage in Ref. \cite{Armesto:2009zi}. 
\amnew{The
treatment of late times may also have a significant quantitative
impact; if energy loss continues after the medium has hadronized. }
\amnew{Due to the rarefaction wave in hydrodynamic simulations, the path length 
does not increase much beyond the point of hadronization, however 
contributions from the hadronic phase may still have a noticeable effect on the energy
loss. A somewhat simple-minded study of this effect was carried out in Ref.~\cite{Majumder:2007ae}, 
where a rescaling parameter called $c_{HG}$ was used to dial the effective 
$\hat{q}$  in the hadronic phase compared to the $\hat{q}$  in the partonic 
phase at the same temperature. It was found that choosing $c_{HG}$ to be 
vanishing required almost a doubling of the effective $\hat{q}$ in the partonic phase 
for the results to remain consistent with the data. An alternative approach 
has been adopted by the authors of Ref.~\cite{Chen:2010te}, who used the 
$\hat{q}$ extracted from jet modification in DIS on a large nucleus to estimate the 
$\hat{q}$ in the hadronic phase of a heavy-ion collision. }

Thus far, the importance of the early \mvlnew{and late}
time dynamics in the modeling of parton energy loss has not been
\amnew{extensively studied} in the literature. Future studies will need to
explore this aspect further, for example by using the exact same
treatment for side-by-side comparisons of other model
aspects. Uncertainties in the initial density of the system should be
taken into account when determining the medium density from energy
loss measurements and will likely give a sizable contribution to the
overall uncertainty.  The present authors feel that taking $\hat{q}=0$
for $\tau < \tau_0$ is unrealistic, since it also implies a very sharp
rise of the density at thermalization $\tau \approx \tau_0$. A more
realistic lower limit of the medium density at early times is needed,
but there currently exists no commonly accepted estimate.

At the moment of writing, a comparable calculation of $R_{AA}$ with a
hydrodynamic medium evolution using the GLV energy loss formalism was
not available. Early work by Hirano and Nara \cite{Hirano:2003pw} uses
full hydrodynamic evolution, but only a schematic form of the GLV
energy loss formalism, and can therefore not be compared directly.

\subsubsection{Di-hadron correlations and $I_{AA}$}

\begin{figure}[!htb]
\epsfysize=5.0cm
  \epsfig{file=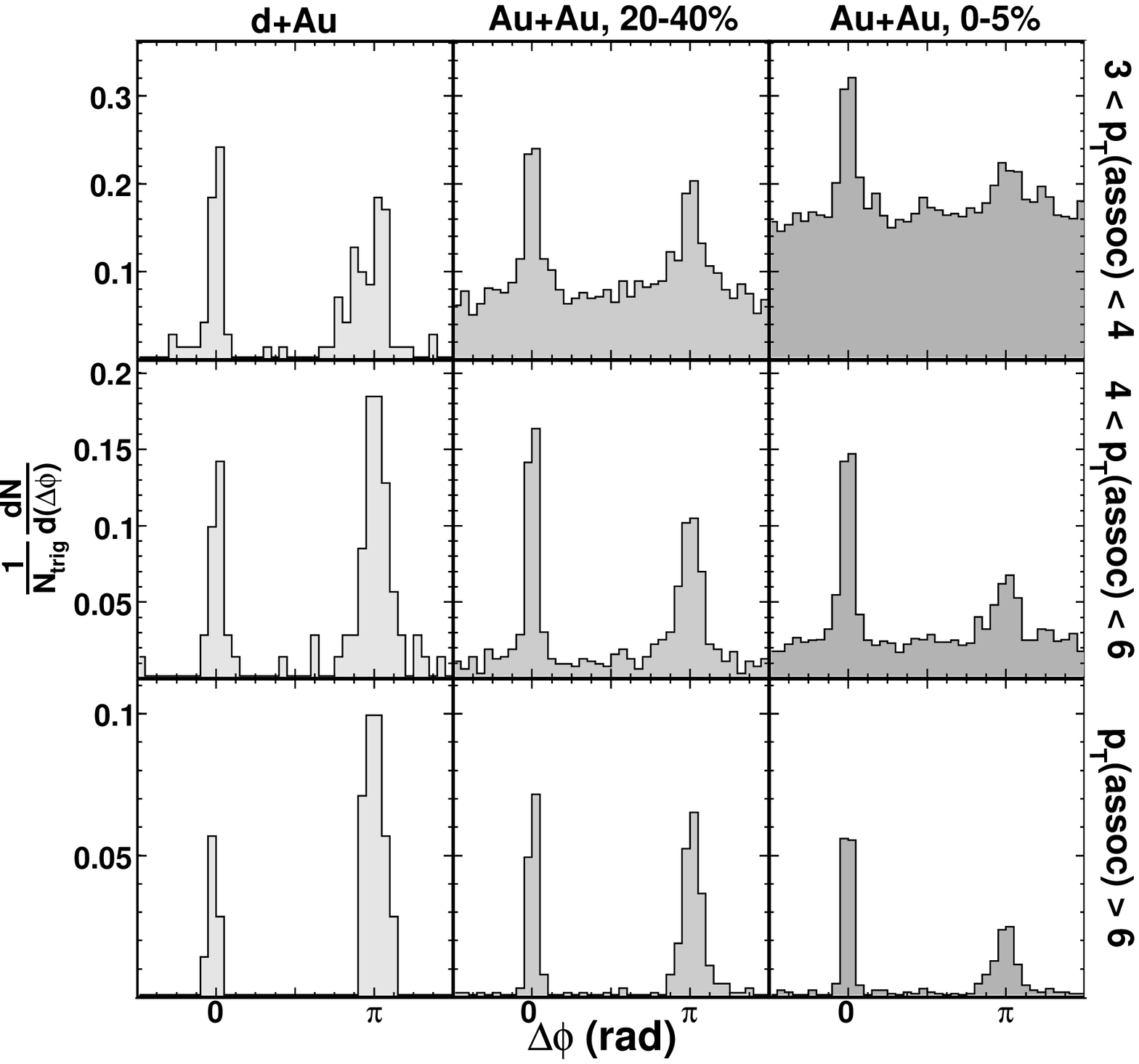,width=0.5\textwidth}
  \hspace{0.3cm}
  \epsfig{file=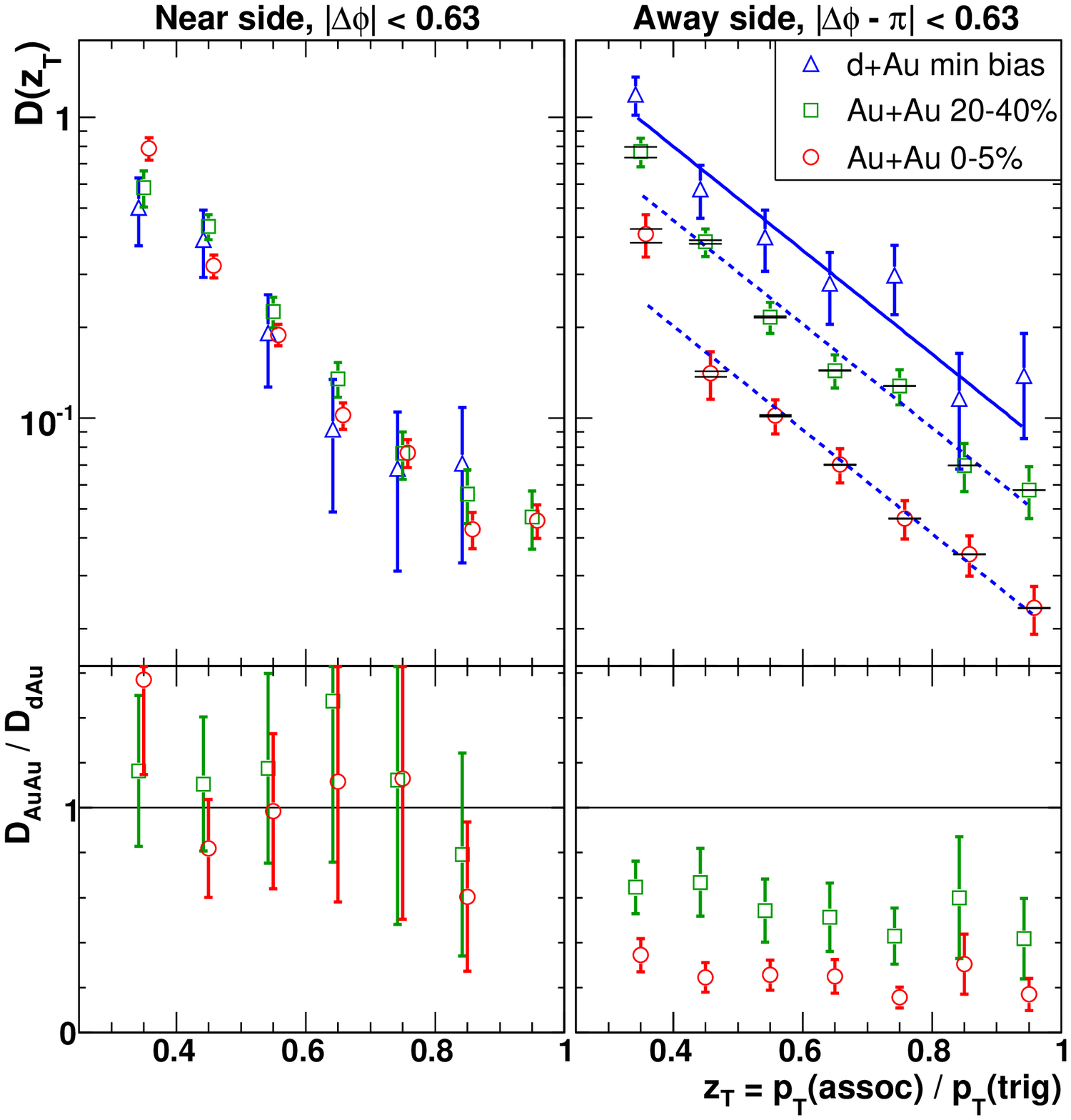,width=0.5\textwidth}
\caption{\label{fig:dihadr_dphi} (Left) Distributions of azimuthal angle
  difference $\dphi$ between charged associated particles and a
  high-\pt{} charged trigger particle ($8<\pt<15$ GeV/$c$) for
  different system sizes (d+Au, mid-peripheral and central Au+Au) at
  $\rtsnn=200$ GeV/$c$ and different \pt-ranges for the associated
  particles, including the uncorrelated background. The distributions are
  normalized per trigger particle. 
  \label{fig:IAA} (Right) Associated yield on the near side yield
  $|\Delta \phi| < 0.63$ (left panels) and recoil $|\Delta \phi - \pi|
  < 0.63$ (right panels), as a function of
  $z_{T}=p_T^{\mathrm{assoc}}/p_T^{\mathrm{trig}}$ for trigger
  particles with ($8<\pt<15$ GeV/$c$) and three different system sizes
  at $\sqrt{s_{NN}}=200$ GeV/$c$. The lower panels show the ratio of
  the Au+Au result to the d+Au reference
  measurement. Both panels are reprinted with permission from \cite{Adams:2006yt} (\copyright American Physical Society).
}
\end{figure}

The measurement of inclusive spectra and the nuclear modification
factor integrates over a large range of parton energies and over the
entire collision geometry. Di-hadron correlation measurements provide more
differential information on the partonic kinematics and a different
geometrical bias. Comparing results may therefore provide more insight
in for example the path length dependence of energy loss.

\am{The left panel of} Fig.~\ref{fig:dihadr_dphi} shows example
distributions of the azimuthal angle difference $\dphi$ between
charged {\it associated} particles and high \pt{} ($8<\pt<15$ GeV/$c$)
{\it trigger} particles. In all panels, two peaks are visible, which
is a clear di-jet signature. For the lower \pt{} associated hadrons
(upper panel), a clear increase is visible in the combinatorial
background when comparing d+Au (left column) and central Au+Au
collisions (right column). The near-side peak around $\dphi=0$ above
background has similar magnitude in d+Au and central Au+Au, while a
clear suppression of the yield is visible in the recoil at
$\dphi=\pi$. It is also interesting to note that the width of the
correlation peaks is similar in d+Au and Au+Au; no sign is seen of the
transverse momentum kicks from the medium, neither at the near side
(jet broadening) nor in the recoil (acoplanarity). Since both jet
broadening and acoplanarity are at some level unavoidable consequences
of the energy loss process, this leads to the conclusion that either
the transverse kicks are small compared to the trigger and associated
hadron momenta, so that the angular broadening is small, or that the
selection of high-momentum particles biases towards partons that had
little or no interaction with the medium. Note that only 20--30\% of
the recoil yield is visible in Au+Au, so a significant fraction of
back-to-back di-hadrons that is present in the d+Au reference
measurement does not pass the \pt-selection in Au+Au collisions due to
energy loss.

A more quantitative representation of the di-hadron results is shown
in \am{the right panel of} Fig. \ref{fig:IAA} which shows the
per-trigger associated yield, after subtraction of the combinatorial
background, on the near side ($\dphi \sim 0$, left) and recoil ($\dphi
\sim \pi$, right) with a trigger particle of $8 < \pt < 15$ \GeVc{}
for d+Au collisions and Au+Au collisions at two different
centralities. The lower panels show the ratio of the Au+Au results to
the d+Au reference.  One the near side, no modification of the
associated yield is seen, while the recoil yield is suppressed by a
factor $\sim 4$.

\am{ It is important to realize that the near-side and recoil yield in this
measurement carry different information due to the presence of a
trigger particle. The near side measures the yield associated with a
trigger particle in the same jet. The absence of suppression on the
near side does not necessarily imply that the jets emerge
without energy loss; it is also possible that partons lose energy and
subsequently fragment in the vacuum, while the lost energy is carried
largely by hadrons at lower \pt{} or outside the angular region of the
measurement. The trigger particle selection would then select partons
that have a momentum distribution {\it after energy loss} that is
similar to the d+Au measurement.}

Theoretical calculations for the near-side associated yield are quite
involved and require the introduction of a new non-perturbative
object, the dihadron fragmentation
function~\cite{Majumder:2004wh,Majumder:2004br} which contains the
correlation between the trigger and the associated hadron.  The scale
evolution of the dihadron fragmentation function mixes with the single
hadron fragmentation functions.  The calculation of medium-modified
di-hadron fragmentation has only been carried out in the HT formalism,
we refer the reader to Ref.~\cite{Majumder:2004pt} for
details. Figure~\ref{fig:dihadron}, compares theoretical calculations
for the case of one scattering and one emission in the medium with
experimental data on the centrality and associated $p_{T}$ dependence
of the associated yield. Note that since the transport coefficients
have been fit to the $R_{AA}$, there are no free parameters in this
calculation.  The agreement is excellent for the case of the highest
associated $p_{T}$ and gradually moves away from the data as the range
of $p_{T}$ for the associated hadron is lowered.

\begin{figure}[!h]
\centering
\epsfysize=5.0cm
  \epsfig{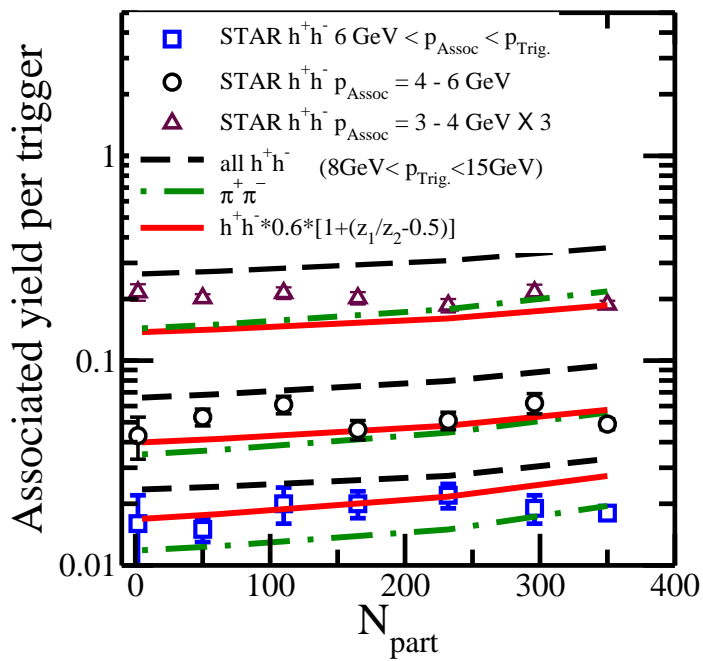}
  \caption{\label{fig:dihadron} The associated yield per trigger for a trigger with $8$ GeV$<p_{T}<15$ GeV for 
  three different ranges of associated $p_{T}$ and for all centralities. The dashed line represents the 
  results for the associated yield of charged hadrons where the initial condition for the dihadron fragmentation function 
  is taken from JETSET. The dot-dashed line is the result for charged pions where the initial condition is taken from 
  JETSET. The red lines are the results for the associated yield of charged hadrons where the initial condition 
  has been fitted to the associated yield of charged hadrons in $p$-$p$ collisions in STAR.   
  Reproduced from \cite{Majumder:2004pt} (\copyright American Physical Society). }
\end{figure}

The recoil yield, on the other hand,
consists of fragments from the recoil parton (back-to-back with the
triggered parton) which propagate independently. Energy loss of the
recoil parton must lead to reduced high-\pt{} recoil particle yield
per trigger. At lower associated \pt, enhanced yield has been
observed, suggesting that the radiated energy remains correlated with
the trigger particle \cite{Adams:2005ph,:2008cqb}.

It is also interesting to note that the away-side suppression in the
di-hadron measurement is independent of $z_{T}$ in the range shown in
Fig \ref{fig:IAA}. This suggests that the longitudinal momentum
distribution of high-\pt{} fragments is unmodified in Au+Au
collisions, which could mean that the measured yield is mostly from
partons that have little or no interaction with the medium, while, on
the other hand, partons that did interact with the medium only produce
associated yield at lower \pt. However, it should be kept in mind that
the steeply falling parton \pt{} spectrum, combined with approximately
exponential fragmentation functions imply that the trigger \pt{}
selection provides only limited control over the initial state parton
kinematics. More complete modeling is needed to provide quantitative
limits on the qualitative trends discussed above. First measurements
with experimentally controlled parton energies via $\gamma$-hadron
correlations and direct jet reconstruction are now becoming available,
which can be more straightforwardly interpreted in terms of parton
energies. These will be discussed in more detail in Sections
\ref{sect:gamma_jet}.

\subsubsection{Model comparisons of $R_{AA}$ and $I_{AA}$}
Early comparisons of theory to data
\cite{Wang:2002ri,Vitev:2002pf,Eskola:2004cr} have focused on the
nuclear modification factor $R_{AA}$. As shown in
Fig. \ref{fig:raa_bassetal}, most energy loss formalisms can be tuned
to describe $R_{AA}$, including the observed independence of $R_{AA}$
on \pt. To further validate energy loss calculations, a simultaneous
comparison to observables with different geometrical and fragmentation
biases is needed.

\begin{figure}
\centering
  \epsfig{file=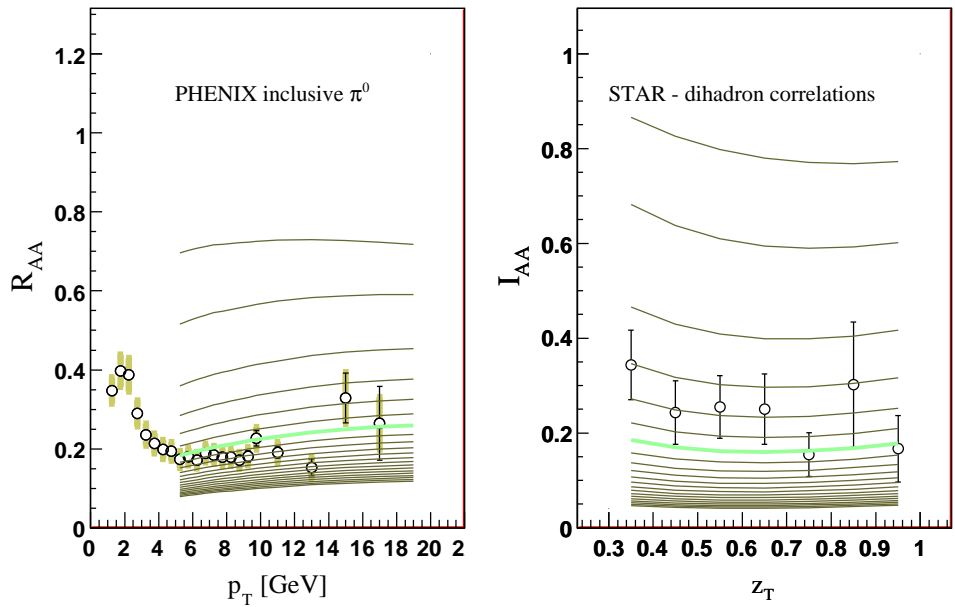,width=0.7\textwidth}
\caption{\label{fig:raaiaa_armesto} Model curves from a model with a
  hydrodynamical evolution of the medium and energy loss using the
  multiple soft scattering approximation (ASW quenching weights)
  compared to measurements of $\pi^0$ $R_{AA}$ (left panel) and
  charged particle $I_{AA}$. The broad green line indicates the best
  simultaneous fit to both experimental results Figure reproduced with
  permission from \cite{Armesto:2009zi} (\copyright IOP Publishing).}
\end{figure}
Figure \ref{fig:raaiaa_armesto} shows a comparison of a model using a
hydrodynamical evolution for the medium density profile and the ASW
multiple soft scattering formalism for gluon radiation
\cite{Armesto:2009zi}. The model can describe the measurements of the
nuclear modification factor $R_{AA}$ and the recoil suppression
$I_{AA}$ with a common scale factor $K$ between the energy density in
the hydrodynamical model and the transport coefficient $\hat{q}$.  The
obtained values for $K$ are in the range 3--5, indicating that the
transport coefficient is larger than the perturbative estimate of
Baier ($\hat{q} \approx 2 \epsilon^{3/4}$) \cite{Baier:2006fr}. These
values are in approximate agreement with the values of $\hat{q}_0$ in
Table \ref{tab:qhat_bass_etal} and also with an earlier fit by Renk
and Eskola \cite{Renk:2006pk}.

\subsubsection{Path length dependence of energy loss}~\label{experiment_path_length}
Conceptually, a measurement of the path-length dependence of energy
loss is very interesting, since it would directly show to what extent
the LPM interference is present. As discussed in Section \ref{single_gluon:multiple_scattering}, LPM
interference effects lead to energy loss proportional to $L^2$ in the
infinite energy limit. Elastic energy loss, on the other hand, would
be proportional to $L$.

\begin{figure}
\centering
\epsfig{file=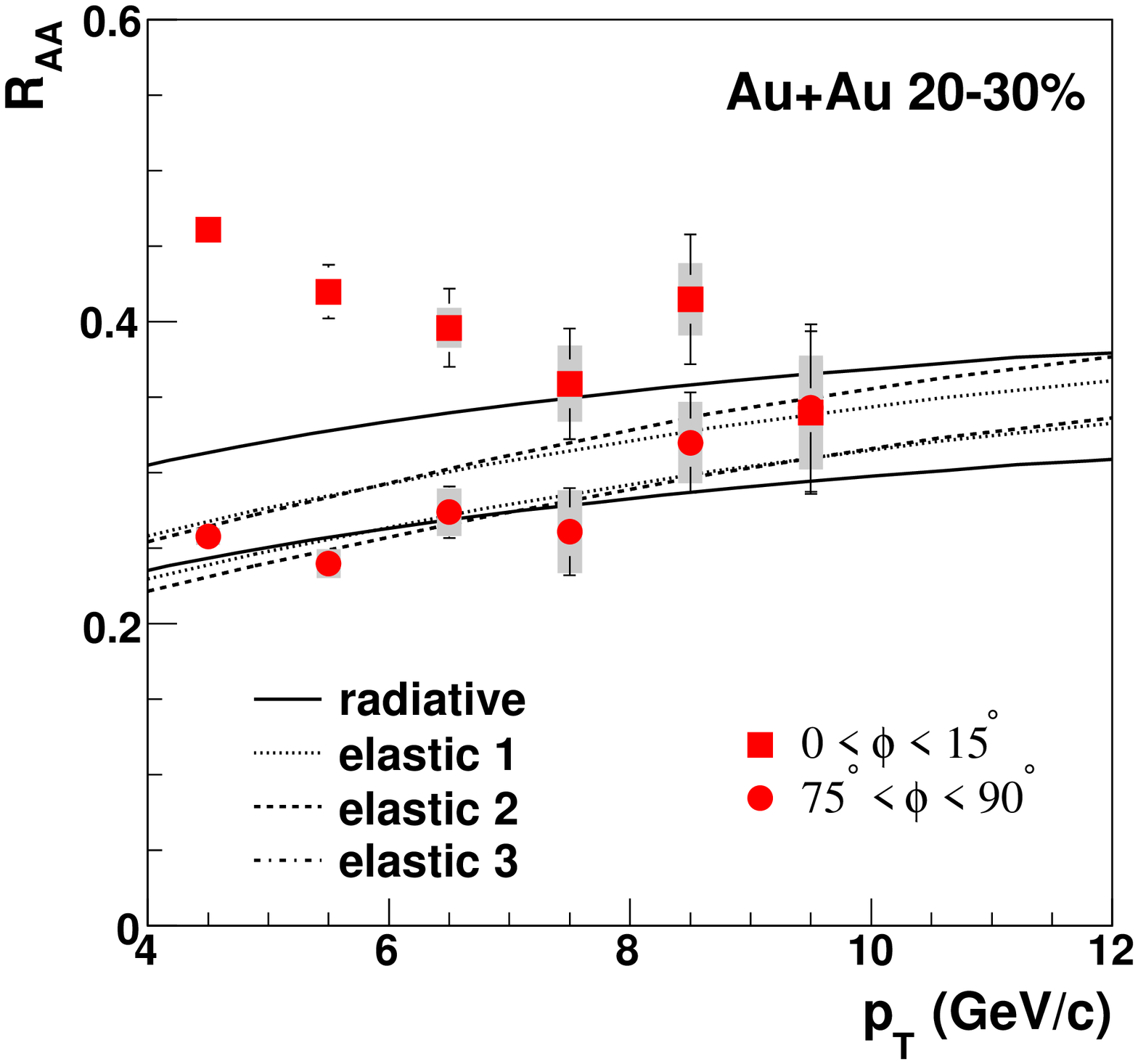,width=0.49\textwidth}\hfill%
\epsfig{file=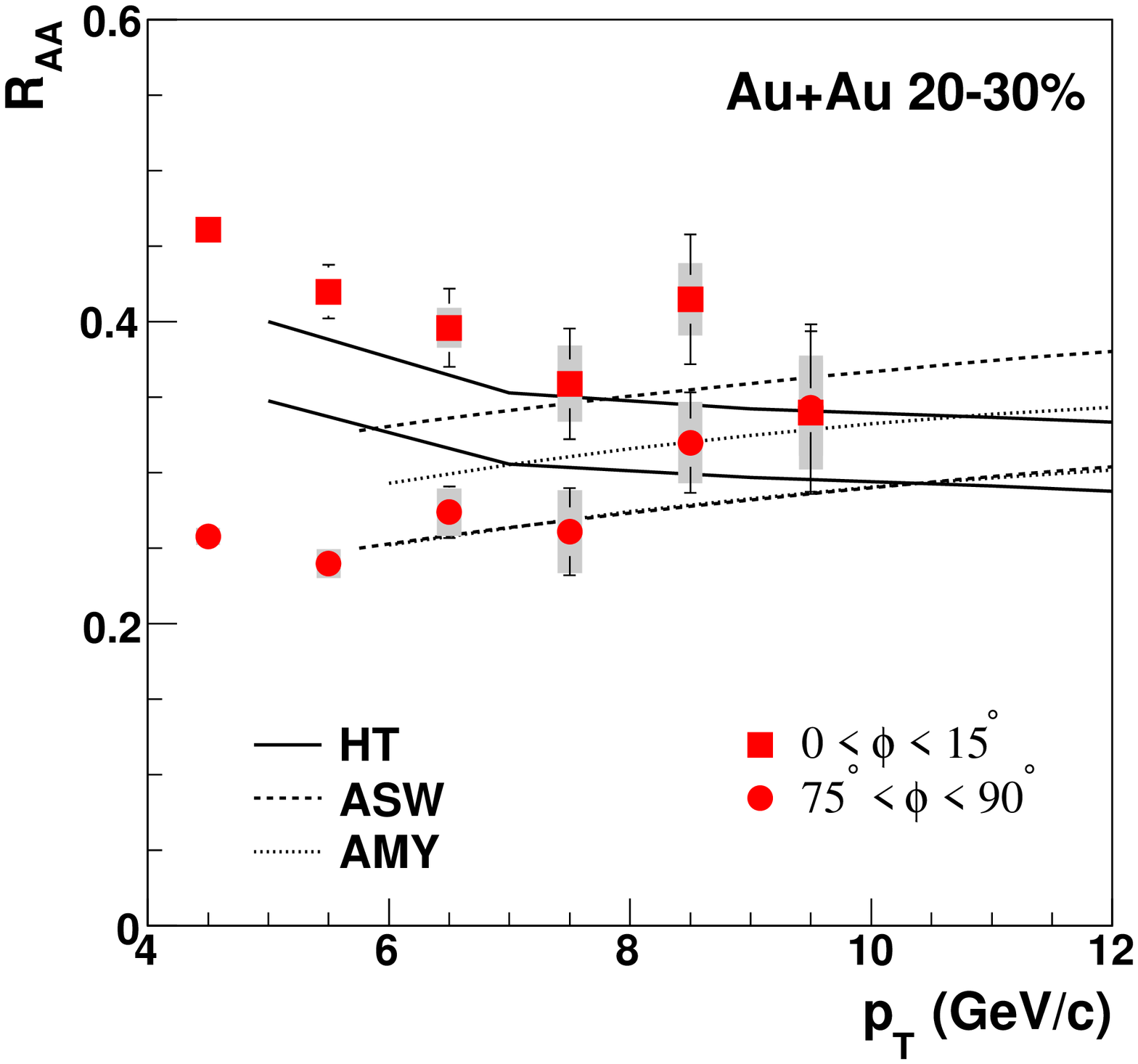,width=0.49\textwidth}
\caption{\label{fig:RAA_psi}Nuclear modification factor $R_{AA}$ for
  $\pi^{0}$ emitted at angles close to the reaction plane and
  perpendicular to the reaction plane ($75^{\circ} < \phi <
  90^{\circ}$) as measured by PHENIX in 20-30\% central Au+Au
  collisions at $\rtsnn=200$ GeV \cite{:2009iv}. The curves in the
  left panel represent calculations with a hydrodynamic medium,
  comparing radiative energy loss with three scenarios for elastic
  energy loss (see also Fig. \ref{fig:RAAIAA_renk})
  \cite{Renk:2007id}. The curves in the right panel represent
  calculations with a hydrodynamic medium at fixed impact parameter
  ($b=7.5$ fm) in the same centrality range and using three different
  energy loss formalisms \cite{Bass:2008rv}.}
\end{figure}

Experimentally, the path length of the partons through the medium can
be varied by selecting different centralities and/or using different
size nuclei. For this purpose, Cu+Cu collisions have been used at RHIC
\cite{:2008cx,Collaboration:2009ue}, but so far no detailed
comparisons with model calculations with a realistic hydrodynamic
medium have been performed.  

More differential control of the geometry is obtained by measuring
$R_{AA}$ as a function of the emission angle with respect to the
reaction plane, or by measuring elliptic flow $v_2$ at high \pt. The
most accurate measurement of this type has been performed by
PHENIX. Fig. \ref{fig:RAA_psi} shows the measured $R_{AA}$ for $\pi^0$
produced in 15 degree angular intervals around the reaction plane
(short path length, square markers) and perpendicular to the reaction
plane (long path length, circle markers) in mid-central (20-30\%
centrality) collisions are compared to two sets of model calculations
using a hydrodynamical medium . In the left panel, model calculations
for radiative energy loss (solid curves) and elastic energy loss are
shown \cite{Renk:2007id}. The three different calculations using
elastic energy all have energy loss proportional to $L$ with a gaussian
spread of different widths. The elastic models are seen to have in
general a smaller difference between $R_{AA}$ for hadrons emitted in
the reaction plane and perpendicular to the reaction plane than the
radiative calculation, as expected from the different $L$
dependence. The data are more in agreement with the radiative energy
loss curves at the highest \pt, although the differences between the
models are small. The out-of-plane measurement shows significant
deviations from the model curves at $\pt < 6$ GeV/$c$, possibly due to
contributions from hydrodynamical flow of the soft matter.

The right panel of Figure \ref{fig:RAA_psi} shows a comparison to
three different calculations with radiative energy loss (the same
calculations as Fig. \ref{fig:raa_bassetal}). It can be seen in the
figure that the ASW energy loss formalism leads to the largest
difference between in-plane and out-of-plane suppression, which is
closest to the experimental result. The statistical uncertainties in
the measurement are still sizable and will be reduced in the future
when larger data samples are analyzed.

According to model calculations, the recoil suppression $I_{AA}$ also
has a significant sensitivity to the path length dependence of energy
loss. The reason for this sensitivity is that the trigger particle
selects partons on the near side that have little or no energy loss
and therefore a short path length. As a result, the recoil parton has
a longer-than-average path length. This so-called `surface bias'
effect was
first pointed out by Dainese, Loizides and Paic in
\cite{Dainese:2004te}. Later, it was pointed out by Renk
\cite{Renk:2006nd} that there is an important interplay between the
expansion of the medium and this path-length bias effect. The longest
path-lengths arise for partons that originate close to the surface,
but then travel inwards. These partons only pass through the center of
the collision after a few fm/$c$. In the presence of rapid
longitudinal expansion, the medium will have cooled down significantly
by that time and thus despite the long path length, the average medium
density along the path is significantly reduced.

\begin{figure}
\centering
\epsfig{file=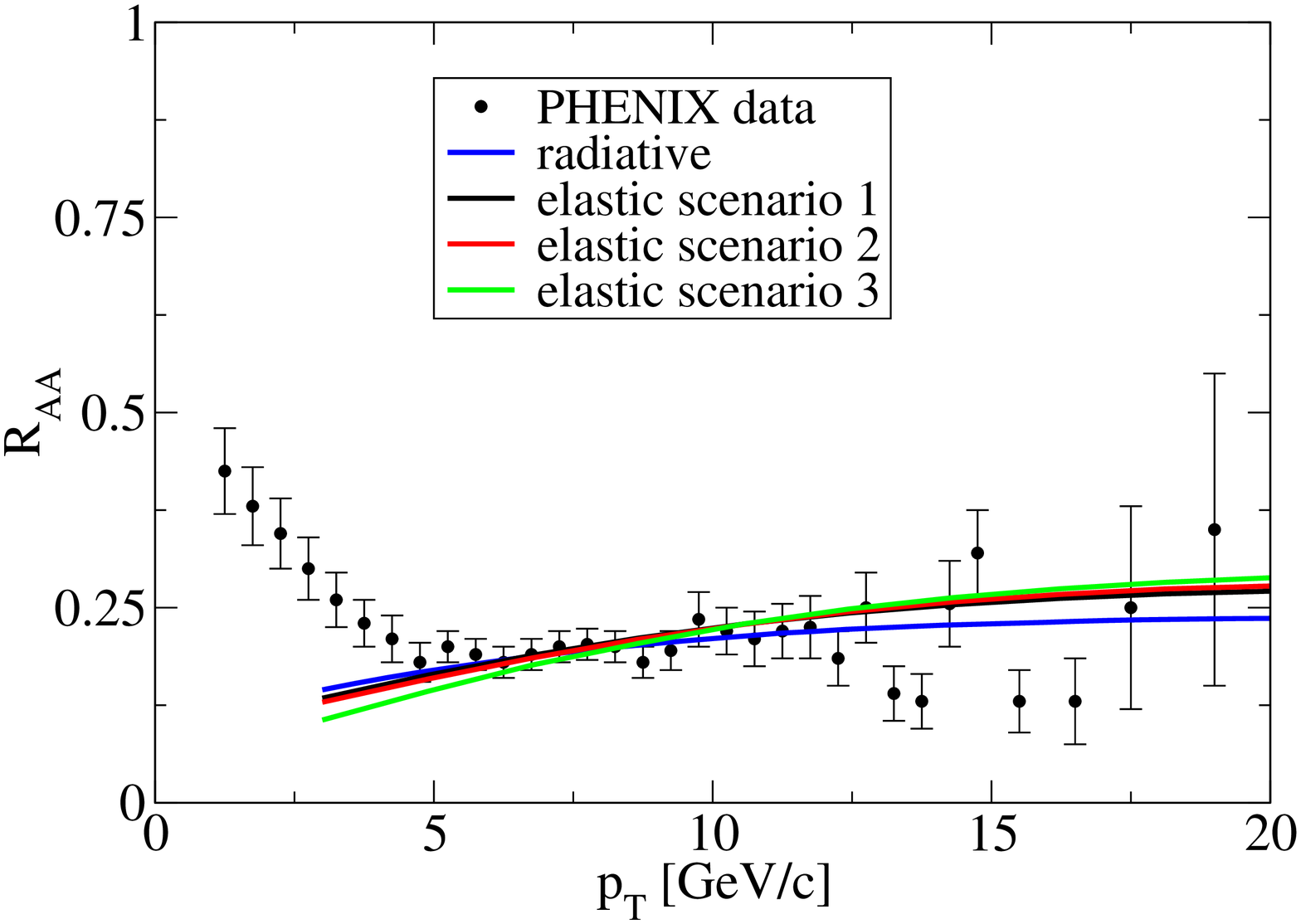,width=0.48\textwidth}\hfill%
\epsfig{file=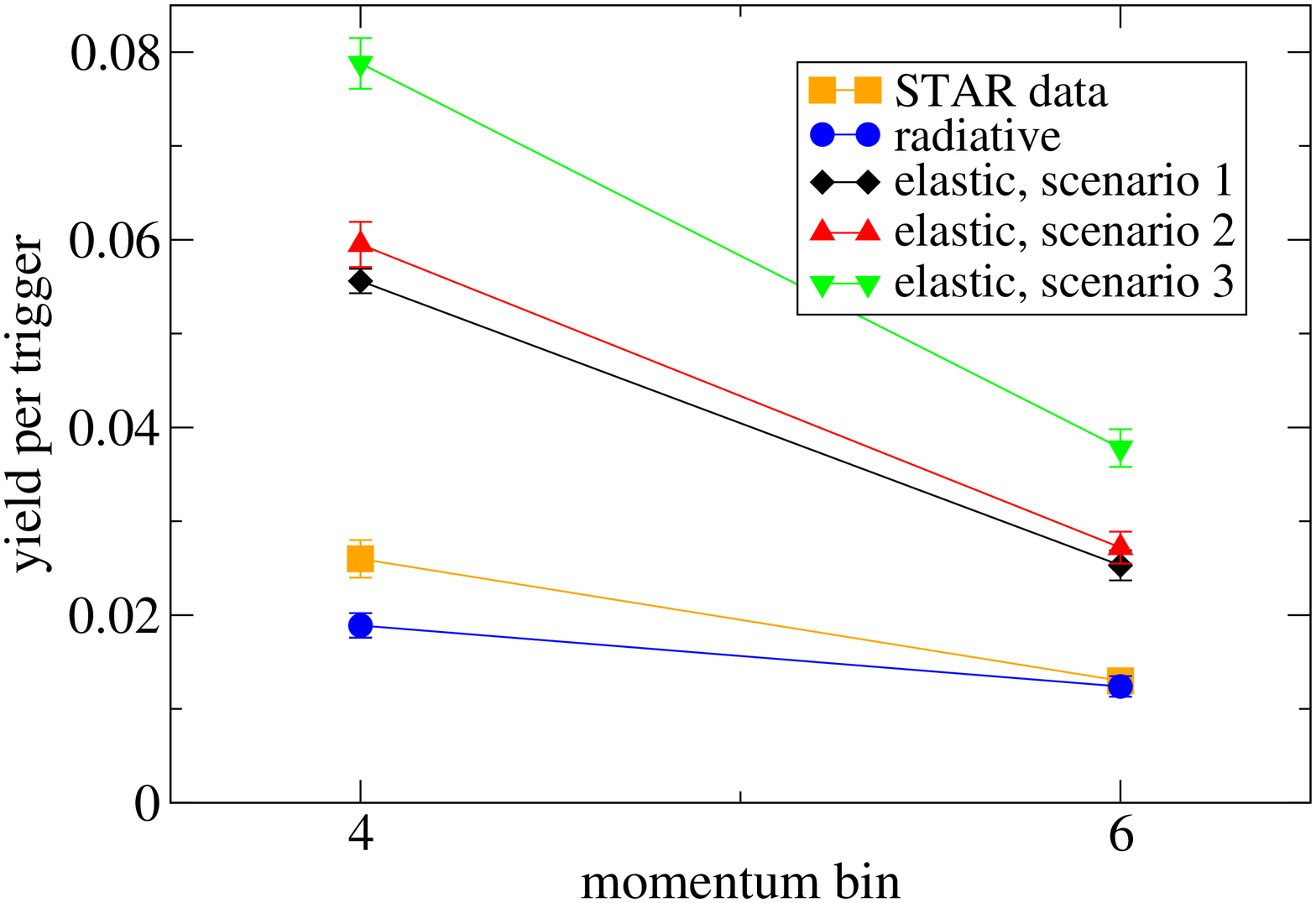,width=0.48\textwidth}
\caption{\label{fig:RAAIAA_renk}Comparison of energy loss calculations
  with the nuclear modification factor $R_{AA}$ from
  Fig. \ref{fig:raa_bassetal} (left panel) and the recoil yield
  measurement from Fig. \ref{fig:IAA} (right panel, yellow
  points). One of the calculations uses radiative energy loss and the
  other three use different schematic implementations of elastic
  energy loss. The calculations use the same geometry from a
  hydrodynamic evolution and have been tuned to reproduce the
  inclusive hadron suppression measurement $R_{AA}$. Figures reprinted
  from \cite{Renk:2007id} with permission (\copyright American
  Physical Society).}
\end{figure}
Fig. \ref{fig:RAAIAA_renk} (right panel) shows a comparison of the
measured recoil yield with high \pt{} trigger particles $8<\pt<15$
GeV/$c$ with calculations using ASW quenching weights for radiative
energy loss and the same three implementations of elastic energy loss
that were used in the left panel of Fig. \ref{fig:RAA_psi}, based on
Gaussian energy loss fluctuations with different mean and width for
the Gaussian. In each calculation, the medium density has been tuned
to reproduce the inclusive hadron suppression $R_{AA}$ (left
panel). The figure (right panel) clearly shows that all three elastic
energy loss models under predict the recoil suppression, due to the
combination of the path length bias and the linear dependence of
elastic energy loss on $L$, thus indicating that radiative energy
loss, with a quadratic dependence on $L$, is dominant for light
hadrons. Comparing the left panel of Fig. \ref{fig:RAA_psi} and the
right panel of Fig. \ref{fig:RAAIAA_renk}, we also note that the
recoil suppression $I_{AA}$ is more sensitive to the path length
dependence of energy loss than the angle dependence of $R_{AA}$.

\subsubsection{Colour factor dependence: quark versus gluon energy
  loss} Radiative energy loss is expected to be $C_A/C_F=9/4$ times
larger for high energy gluons than high-energy light quarks. I has been
suggested that this difference is experimentally accessible by
comparing nuclear modification factors for (anti-)protons and
pions. In $e^{+}e^{-}$ collisions at LEP, it has been observed that
gluon jets are more likely to produce (anti-)protons than quark jets
\cite{Abreu:2000nw}. One could therefore use anti-protons to `tag'
gluon jets and expects to see a larger suppression for (anti-)protons
than for pions. 

\mvl{There are several caveats to this simple expectation. First of
all, the argument crucially depends on the difference between
(anti-)proton production from gluon and quark fragmentation. The gluon
to proton fragmentation function is not accurately know; in fact,
recent extractions of fragmentation functions from experimental data
use the RHIC $p$-$p$ data to constrain this
\cite{Albino:2008fy,deFlorian:2007hc}. It has also been shown in more
detailed calculations using the AKK fragmentation functions, that
while the expected difference between quarks and gluon $R_{AA}$ is
almost a factor 2, the effect is diluted when comparing protons and
pions, because pions are produced by gluon as well as quark
fragmentation \cite{Renk:2007zz}.  Finally, it should be realised that
quarks showers contain gluons and vice versa, so that the distinction
between quark and gluon energy loss in a full shower evolution not be
clear. One specific implementation of this effect are `jet conversion'
processes in the medium. A calculation of this effect shows a reduced
$p/\pi$ ratio in heavy ion collisions compared to $p$-$p$ collisions
\cite{Liu:2007zz}.}
Measurements performed by STAR \cite{Abelev:2006jr},
do not show \mvl{a larger suppression of (anti-)protons compared to pions.}

\subsubsection{Heavy quark energy loss}~\label{sect:heavy_flavour_exp}

Due to their large mass ($m >> \Lambda_\mathrm{QCD}$), heavy quarks
move at speeds smaller than the speed of light through the medium. As
a result, they cannot to emit gluons at small angles, the so-called
`dead cone effect', and therefore lose less energy than light quarks
(see Section \ref{heavy_flavor:dead_cone}) \cite{Dokshitzer:2001zm}. To
verify whether radiative energy loss is dominant for heavy quarks,
measurements have been performed of heavy flavor production using
semi-leptonic decays into electrons.

\begin{figure}
\centering
\epsfig{file=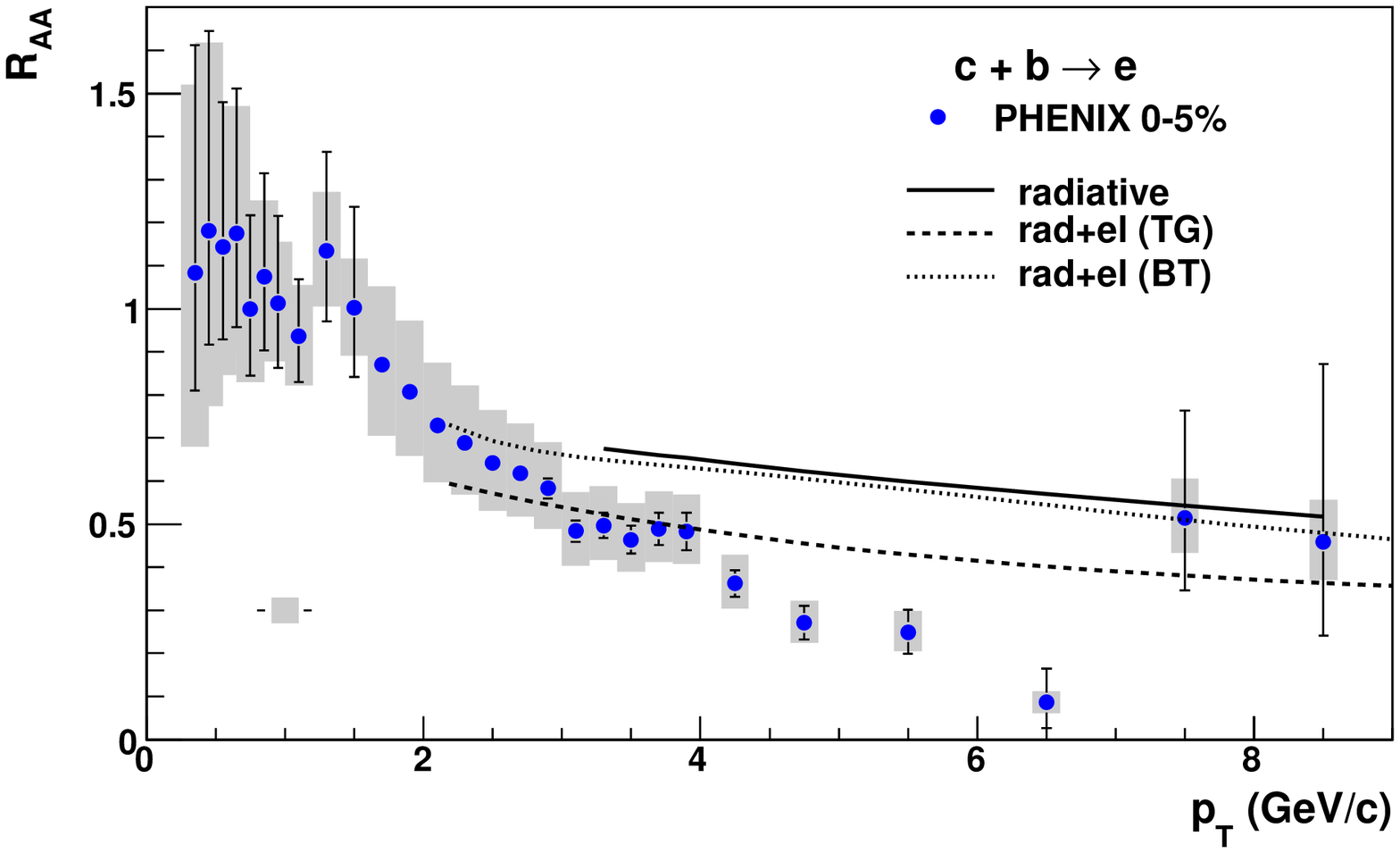,width=0.6\textwidth}
\epsfig{file=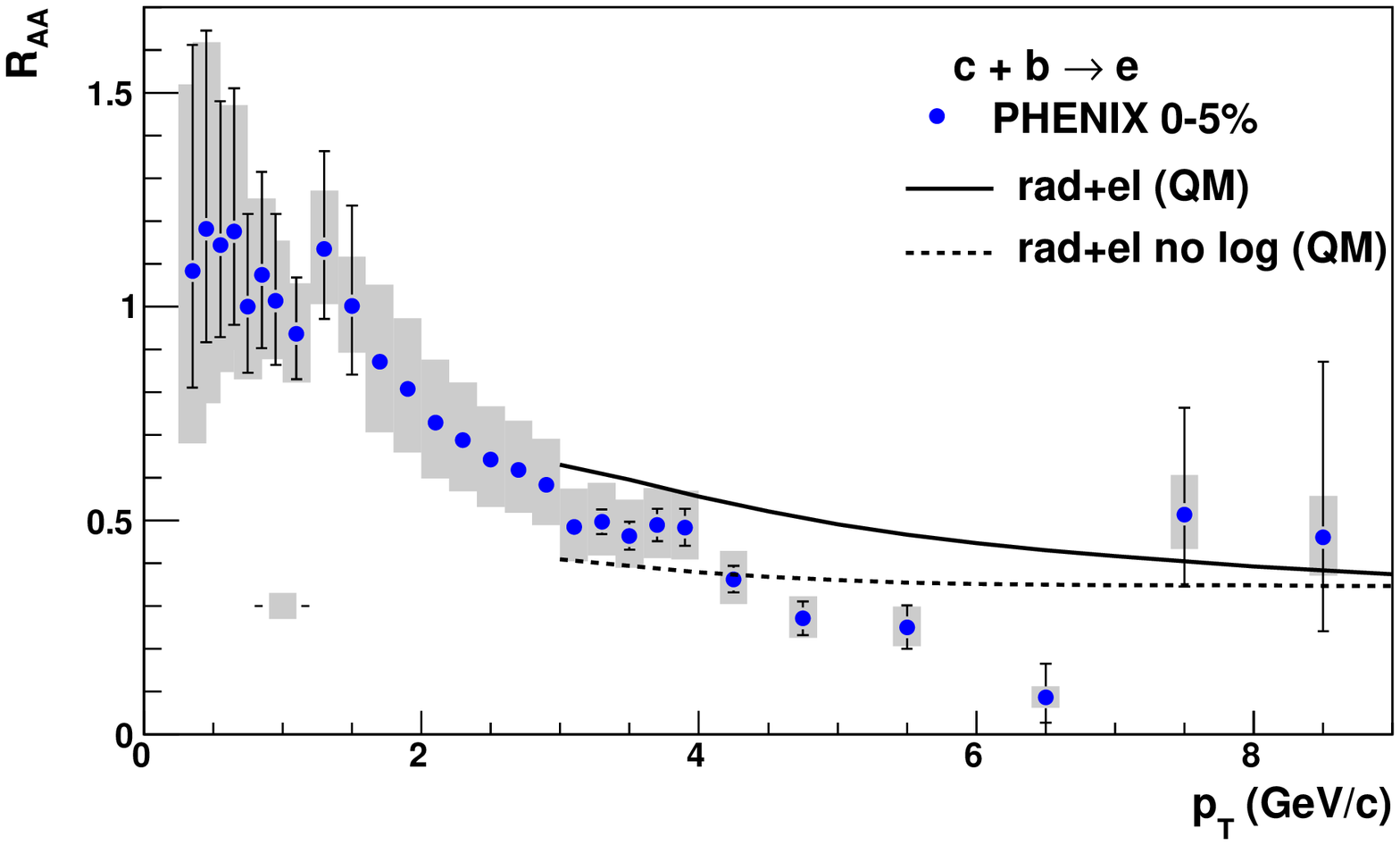,width=0.6\textwidth}
\caption{\label{fig:raa_nonphot}Nuclear modification factor $R_{AA}$
  for non-photonic electrons from heavy flavor decays as measured by
  PHENIX \cite{Adare:2006nq} in central Au+Au collisions at
  $\rtsnn=200$ GeV compared to three different model calculations. The
  uncertainty bands on the lower left indicate the overall
  normalisation uncertainty. Top panel: radiative energy loss only
  (see also Fig. \ref{fig:raaiaa_armesto}) \cite{Armesto:2009zi} and
  radiative and elastic energy loss, using two different elastic
  energy loss models \cite{Wicks:2005gt}. Bottom panel: radiative and
  elastic energy loss \cite{Qin:2009gw}, \mvlnew{using transport coefficients
  from HTL effective theory (solid line) and with an alterative
  prescription (dashed line, see text).} }
\end{figure}
The upper panel of Figure \ref{fig:raa_nonphot} shows a comparison of
predictions using only radiative energy loss by Armesto et al (solid
line in Fig. \ref{fig:raa_nonphot}), using the parameters obtained
from the light quarks measurements in Fig. \ref{fig:raaiaa_armesto},
to measurements of the nuclear modification factor of non-photonic
electrons (electrons which are not from photon conversion or light
hadron decay) from heavy flavor decays by PHENIX at RHIC
\cite{Armesto:2009zi}. The measured suppression of non-photonic
electrons is larger than the model predicts. The extracted $K$-factor
would be larger than 25, while the value for light quarks was between
2 and 5. These results indicate that the suppression of heavy quarks
is larger than expected, possibly similar to light quark
suppression. It is worth noting that the systematic uncertainties on
the measurements are sizable. These uncertainties are expected to be
quite strongly correlated point-to-point. As a result, the data are
not strictly inconsistent with the expectation from radiative energy
loss. For a more detailed discussion of the systematic uncertainties
and the implications for energy loss we refer to
\cite{Armesto:2009zi}.

Various mechanisms have been proposed to explain the observed large
suppression. For example, it has been argued that elastic energy loss
may play a significant role for heavy quarks, because the radiative
energy loss is reduced \cite{Wicks:2005gt}. A detailed evaluation of
the combined effect of radiative and elastic energy loss has been
performed using the opacity expansion for radiative energy loss and
elastic energy loss based on the HTL calculation by Thoma, Gyulassy
(TG) \cite{Thoma:1990fm} and Braaten, Thoma (BT)
\cite{Braaten:1991we}. The same mechanism should be present for light
quarks, so the model was fitted first to the light quarks and the
predicted suppression of non-photonic electrons is compared to the
measurements in the top panel of Fig. \ref{fig:raa_nonphot} (dashed
and dotted lines). The discrepancy between the model and the
experiment is smaller than when only radiative energy loss is taken
into account (compare to solid line), but the calculation still gives
less suppression than the measurement.

\amnew{In the inclusion of elastic energy loss in the calculation above 
one assumes not only that the jet is weakly coupled to the medium, 
but also that the medium may be described as a leading order HTL plasma. 
The expressions for the elastic energy loss in both the BT and TG calculations
in the upper panel of Fig.~\ref{fig:raa_nonphot} assume that the medium 
may be described as weakly interacting quasiparticles which interact directly with 
the jet leading to the elastic energy loss. In this limit, all the 
transport coefficients \mvlnew{have a characteristic 
dependence on the logarithm of the parton energy, see Eq.~(\ref{HTL_qhat}), which reduces the energy loss for lower jet energies.}
\mvlnew{Since all the relevant transport coefficients have the same logarithm,} 
there remains only one fit parameter which is 
the value of $\alpha_{S}$ in the medium. This is set by fitting to one data 
point in the light flavor sector. }

\amnew{ The ascribed limits on the viscosity to entropy ratio $\eta/s$ 
from hydrodynamical simulations~\cite{Song:2008hj} have cast strong doubts on the 
applicability of LO-HTL as a model of the dense medium created at 
RHIC. In \mvlnew{addition, }a computation of $\hat{q}$ at NLO in the HTL effective 
theory \mvlnew{was found to } introduce a large constant term in addition to the $\log[4ET/m_{D}^{2}]$ \cite{CaronHuot:2008ni}.
This has motivated the authors of Ref.~\cite{Qin:2009gw} to attempt a 
global comparison to both the light and heavy quark energy loss data assuming 
that the $\hat{q}$, $\hat{e}$ and $\hat{e}_{2}$ depend on temperature 
according to their dimensions, i.e., $\hat{q} \propto \hat{e}_{2} \propto T \hat{e} \propto T^{3}$. 
The longitudinal coefficients $\hat{e}$ and 
$\hat{e}_{2}$ are related by the fluctuation dissipation theorem and they 
also assumed $\hat{q} = 2 \hat{e}_{2}$. The one remaining overall constant 
was set by comparing light quark energy loss to data. With these simplifying 
assumptions, these authors obtained a marked improvement in comparison 
with the heavy flavor suppression. This is illustrated in the bottom panel of
Fig. \ref{fig:raa_nonphot} which 
compares non-photonic electron $R_{AA}$ with model curves using a
Woods-Saxon overlap geometry with an initial temperature of $T=400$
MeV at the hottest point. The solid line shows the result using HTL
values for the transport coefficients $\hat{q}\approx 2
\mathrm{GeV}^2$/fm at a jet energy of $20$ GeV, $\hat{e}$ and $\hat{e}_2$
are calculated using the HTL approximation and the same value of $\alpha_{s}$ as 
for $\hat{q}$ (see also 
Section \ref{heavy_flavor:other_coeffs}). The dashed line
shows the result without any logarithmic dependence of the transport coefficients on the energy of the jet.  
The same calculation also describes the measured
elliptic flow for light hadrons and non-photonic electrons.}

Other models, which include hadron
formation inside the medium followed by dissociation and recombination
\cite{Adil:2006ra} or resonant scatterings \cite{vanHees:2008gj} can also
be tuned to describe the measurement. The implications of such models for
light quark energy loss have not been studied in detail.

High-\pt{} measurements of heavy flavor production at RHIC have so far
relied on electrons from semi-leptonic decay and therefore do not
distinguish between charm and bottom. The difference in mass between
charm quarks ($m \approx 1.3$ GeV/$c^2$) and bottom quarks ($m \approx
4.2$ GeV/$c^2$) leads to a different momentum dependence of the dead-cone
effect, which will be explored by future measurements. At high \pt,
the charm quark behaves as a light quark, so that comparisons to light
hadrons are a measure of the difference between quark and gluon energy
loss. Calculations have shown that this effect leads to large effects
in the heavy-to-light ratios at RHIC and LHC \cite{Armesto:2005iq}. Future
measurements will certainly be sensitive to these effects.

\subsubsection{$\gamma$-hadron and jet based measurements at RHIC}
\label{sect:gamma_jet}
In recent years, two new types of measurements of parton energy loss
are being explored at RHIC: $\gamma$-hadron correlations and jet
reconstruction. Both measurements provide experimental control over
the initial state parton kinematics: in the case of $\gamma$-hadron
measurements, the photon and the recoil jet have equal \pt{}, while in
the case of jet reconstruction, the parton energy is measured by
summing up the energies of the fragments. 

\begin{figure}
\epsfig{file=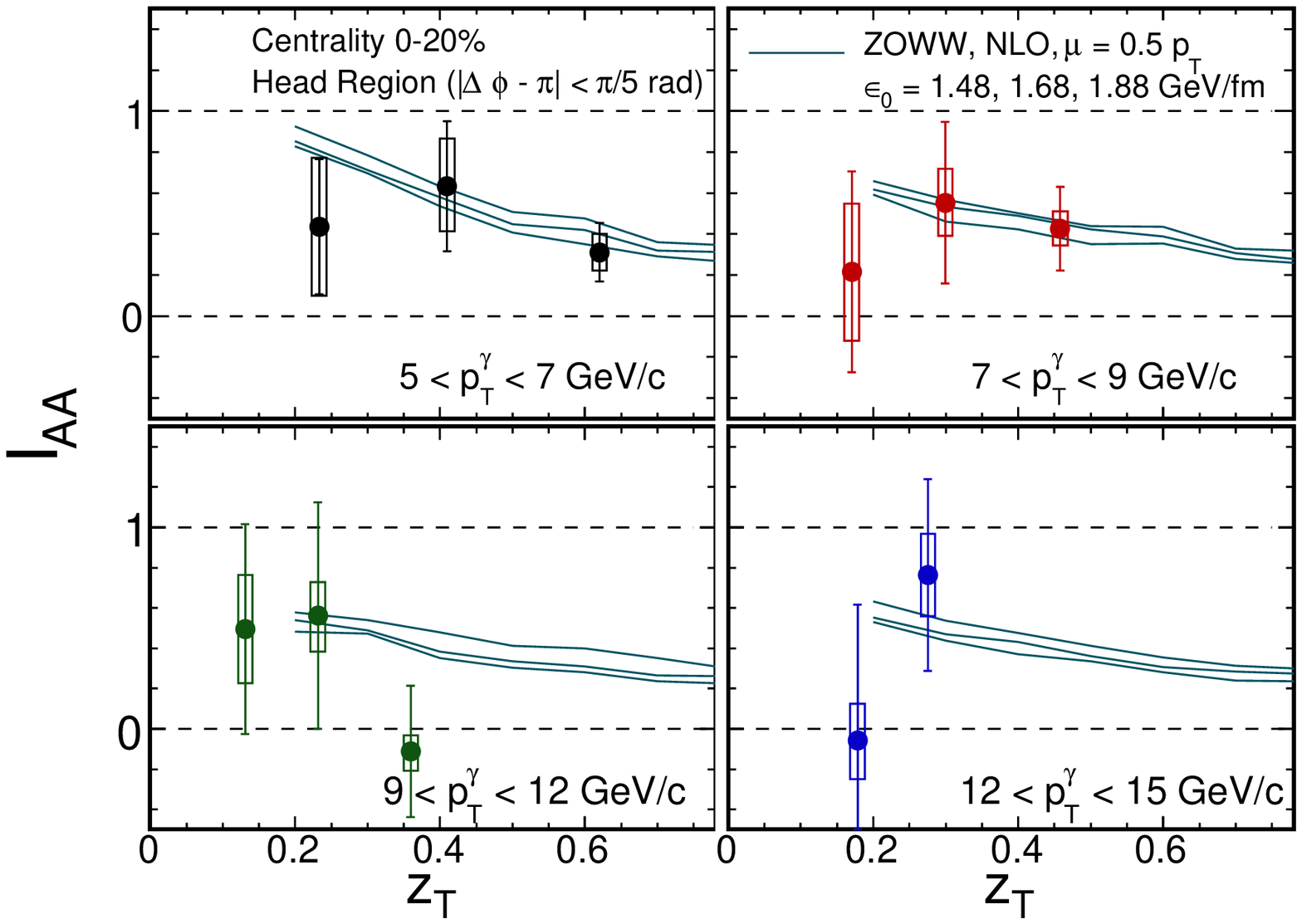,width=0.5\textwidth}\hfill%
\epsfig{file=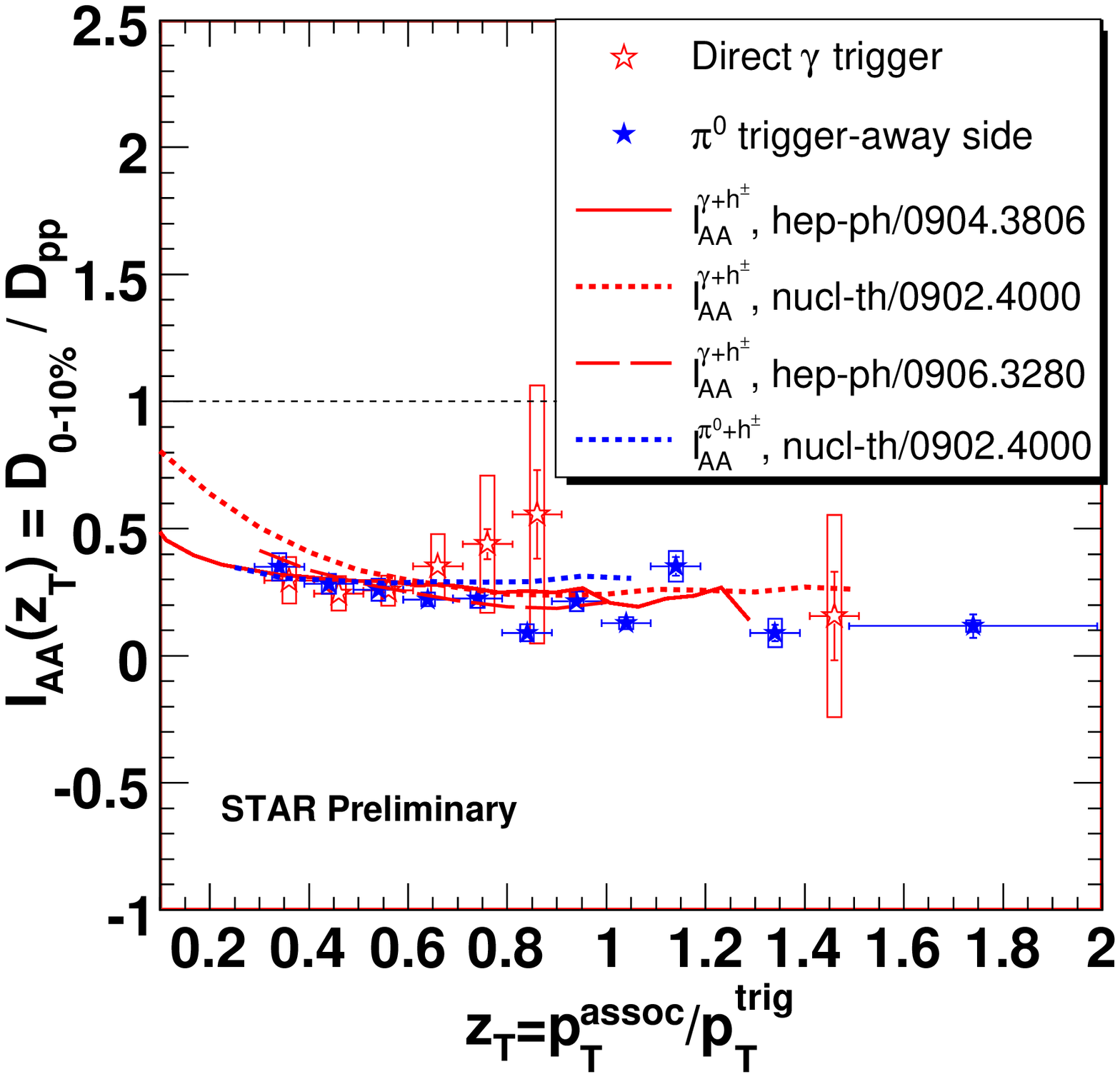,width=0.42\textwidth}
\caption{\label{fig:gamma_h}Recoil hadron suppression for direct
  photons measured by PHENIX (left panel) \cite{Adare:2009vd} and STAR
  (right panel) \cite{Hamed:2009vu} in central Au+Au collisions
  compared to $p+p$. The PHENIX measurement (left panel) is presented
  for four ranges in $\pt^{\gamma}$ between 5 and 15 GeV/$c$ and
  compared to energy loss calculations by Zhang, Owens and Wang (ZOWW)
  \cite{Zhang:2009rn}. The preliminary STAR measurement (right panel)
  has $8 < \pt^{\gamma} < 16$ GeV/$c$ and is compared to the ZOWW
  calculation (dotted curves), a calculation by Renk
  \cite{Renk:2009ur} and an AMY-based calculation that takes into
  account parton-photon conversions in the medium (dashed line)
  \cite{Qin:2009bk} and a similar measurement for $\pi^0$
  recoil. Figures reproced with permission from \cite{Adare:2009vd}
  (\copyright American Physical Society) and \cite{Hamed:2009vu}.  }
\end{figure}
The $\gamma$-hadron measurement is conceptually appealing, and has
been advocated for a long time \cite{Wang:1996pe,Arleo:2004xj}, but it
is experimentally difficult because of the low rates and the large
background from $\pi^0$ decays. First results from PHENIX
\cite{Adare:2009vd} and STAR (preliminary) \cite{Hamed:2009vu} are shown
in Fig. \ref{fig:gamma_h} and compared to model calculations. Both
experiments show a recoil suppression of 0.3, which is similar to the
value found for di-hadron suppression in Fig. \ref{fig:IAA} (charged hadrons)
and the right panel of Fig. \ref{fig:gamma_h} ($\pi^0$-charged hadron
pairs) despite the different geometrical and fragmentation bias. 

The $\gamma$-hadron measurements are compared to model calculations by
three groups. The calculation by Zhang, Owens and Wang
\cite{Zhang:2009rn}, which is shown in both panels of
Fig. \ref{fig:gamma_h} uses NLO pQCD to calculate the initial photon
rate, including fragmentation photons but uses a simplified geometry
(hard spheres) and an energy loss calculation without energy loss
fluctuations. The calculation by Renk \cite{Renk:2009ur} (solid line
in the right panel of Fig. \ref{fig:gamma_h}) uses a more realistic
geometry and ASW quenching weights for energy loss. The third
calculation in the right panel of Fig. \ref{fig:gamma_h} (dashed line)
uses a 3D hydro model for the medium and an AMY-based energy loss
formalism, which includes production of photons by interactions with
the medium \cite{Qin:2009bk}. All three model calculations are in good
agreement with the measurement, indicating that the measured recoil
suppression is consistent with the hadron suppression measurements.

For sufficiently high photon energy, one would expect to see the
$z$-dependent suppression of the fragmentation functions, similar to
what is shown in Fig. \ref{fig:ff_comp}, although the effect may be
weaker for a realistic nuclear geometry than for the fixed-length
calculations presented in Fig. \ref{fig:ff_comp}. Some of the model curves
in Fig. \ref{fig:gamma_h} indeed show a rise at small
$z_T=\pt/\pt^\gamma \sim 0.2$, where no measurements are available due
to the large combinatorial background at low $\pt$.  Future runs at
RHIC will bring larger data samples and correspondingly reduced
uncertainties.

Experimentally, jet measurements have the advantage of a large cross
section, but a good jet analysis is complicated, especially in the
presence of uncorrelated backgrounds from the underlying heavy ion
event. Jet measurements are being actively pursued at RHIC, by both
STAR \cite{Salur:2008hs,Ploskon:2009zd,Putschke:2008wn} and PHENIX
\cite{Lai:2009zq}. Initially, it was thought that it would be
necessary to reject all particles below a certain transverse momentum
threshold from the analysis to reduce the background, but experience
with data has shown that this is not necessary. This enables infrared
safe algorithms to be used, such as the recently developed SISCone
algorithm \cite{Salam:2007xv} and fast $k_T$ \cite{Cacciari:2005hq}
and anti-$k_T$ algorithms \cite{Cacciari:2008gp}.

In collisions of elementary particles, $e^+ + e^-$, $p+p$, and
$p+\bar{p}$, reconstructed jet energies provide an accurate measure of
the underlying parton momentum as evidenced by kinematic
reconstruction of $W \rightarrow q \bar{q}$
\cite{Schael:2006mz,:2008xh,Achard:2005qy} and $t \rightarrow b W
\rightarrow q \bar{q}$ decays
\cite{Abulencia:2007br,Abazov:2008ds}. The accurate mapping between
parton energy and jet energy is a phenomenological implication of QCD
factorization: an infrared and collinear safe jet definition is
insensitive to details of the soft QCD processes in jet fragmentation
and therefore measures properties of the `hard partons'. In heavy-ion
collisions, jet-fragmentation is modified by interactions with the
medium, thereby modifying the longitudinal and transverse structure of
the jet. In addition, part of the jet energy may be transported
outside the jet cone by large angle radiation, thus reducing the
reconstructed jet energy. A number of different experimental analyses
are being performed, which are sensitive to different aspects of the
medium modification of jet energies. The available preliminary results
show an increasing suppression when using a smaller jet radius in the
reconstruction algorithm, suggesting that jet broadening is an
important effect \cite{Ploskon:2009zd}. For large radius, $R=0.4$, the
jet suppression is much smaller than the inclusive hadron suppression,
indicating that a large fraction of the radiated energy is recovered
by jet reconstruction.

Attempts have also been made to
measure fragmentation functions, {\it i.e.} the longitudinal distribution
of particles in jets, with the aim to characterize the softening of the
fragmentation due to medium interactions. Such measurements, however,
are complicated by potential interplays between the softening of the
fragmentation and the jet energy reconstruction. Preliminary results
that attempt to circumvent this problem by using di-jet events, where
the jet energy is measured in one jet and the fragment distribution in
is studied in the recoil jet, show a large suppression of the overall
recoil jet rate, but no apparent modification of the fragment
distribution in the reconstructed recoil jets \cite{Bruna:2009em}.\\

\subsection{Outlook: how to reduce uncertainties}
\label{sect:outlook_precision}

In the above various measurements have been compared to theoretical
calculations to extract medium parameters. In some cases, such as the
heavy flavor measurements, the experimental uncertainties are
sizable, of the order of 20\% and seem to limit the extraction
of medium properties. As mentioned in Section \ref{sect:observables},
the first approach to improving this situation should be by defining
more selective observables using theoretical insight and
modeling. For example, one might expect a measurement of recoil
suppression $I_{AA}$ of heavy quarks to be sensitive to the path
length dependence of energy loss, like in the case of light hadrons. This
is clearly a challenging measurement, so it would be worth to first
explore the potential in model calculations. 

However, it is also worth to consider the origins of the various
experimental uncertainties to see whether improvements are
possible. In this section, we will briefly discuss the various
sources of uncertainties in the measurements and which improvements
are planned with future RHIC upgrades.

{\bf Statistical errors, luminosity}

The most straightforward source of uncertainties are statistical
uncertainties. These uncertainties are always due to counting
statistics in the experiment and can therefore be reduced by
increasing the number of analyzed events, either by improving
luminosity or by longer running times, and sometimes by selecting a
process with a larger cross section. The beam luminosity and effective
running time at RHIC have steadily increased over the years. The
largest event samples were taking in run 4 and run 7, with run 7
having about twice the integrated luminosity of run 4. It is important
to realize that not all observables can make full use of the
luminosity, because the event rates may be too high to record all
events. Both STAR and PHENIX can select events with large energy
depositions in the electromagnetic calorimeters with an on-line
trigger, so that measurements with high-\pt{} photons, $\pi^0$ and
electrons can make use of the full luminosity. Both experiments are
also continually improving the detector read-out rates to be able to
record a larger fraction of the total luminosity for other
(un-triggered) measurements. 

If a signal has a large background, the statistical uncertainty is
driven by the amount of background. This explains the relatively large
statistical uncertainties on direct photon measurements (decay photon
background) and to a lesser extent the non-photonic electron
measurements (conversion electron background).

{\bf Background uncertainties}

Apart from the statistical uncertainties on a background, there are
also systematic uncertainties. For example, in the case of the decay
photon background for direct photons, the background level is set by
the $\pi^0$ spectrum, which is known to finite precision. The
contributions of higher-mass mesons are less precisely
known. The main source of the conversion electron background for the
(non-photonic) electron spectrum are also photons from $\pi^0$ decay,
so here again the subtraction relies on the knowledge of the $\pi^0$
spectrum, in addition to a precise knowledge of the amount and
composition of the detector materials.

If the signal is smaller than the background, a relatively small
uncertainty in the background level may translate into a large
uncertainty on the signal. Larger data samples can improve the
knowledge of the background level, but only up to the level of the
intrinsic systematic uncertainties.

{\bf Intrinsic systematic uncertainties} 

Every experimental set-up has intrinsic lower limits to the
experimental uncertainties. \mvl{The two dominant effects are 
uncertainties in the detection efficiency and in the energy or momentum scale.} 
Experimental efficiency is generally
calculated using detector response simulations, which have a finite
precision. Details of what limits the uncertainty differ from detector
to detector, but it is safe to say that a precision of better than a
few per cent in the detection efficiency is difficult to
attain. In addition, energy scale calibration uncertainties may
introduce significant uncertainties in the yields at a given \pt. For
example, with a power law spectrum $1/\pt^6$, an energy scale
uncertainty of 0.5\% leads to a 3\% uncertainty on the yield.  These
types of uncertainties limit, for example,
the $\pi^0$ spectrum at intermediate \pt{} in Figs
\ref{fig:pi0_raa} and \ref{fig:raa_bassetal}.

{\bf Normalization uncertainties} 

For spectra and $R_{AA}$, normalization uncertainties are a
significant contribution, \mvl{see
Figs \ref{fig:pp_spectra}, \ref{fig:pi0_raa}, and \ref{fig:raa_nonphot}, where this type
of uncertainty is indicated by a separate error band without data
point}. There are two main contributions\mvl{: the absolute
normalization of the $p$+$p$ reference measurement and the
determination of the number of binary collisions $N_{\mathrm coll}$.}

\mvl{The normalization uncertainty of $p$+$p$ reference
measurements is due to a significant contribution of low-multiplicity
events, including double-diffractive collisions, to the total cross
section. Some of these low-multiplicity events are not detected by the
experiment. The measurement in $p$+$p$ can be given in terms of
multiplicity per event ($d^2N/dy d\pt$) or differential cross section
$d^2\sigma/dy d\pt$. The normalization uncertainty is slightly different
for these two types of measurement; in one case it enters as an
uncertainty in the total number of events, while in the other case it
enters as an uncertainty in the total cross section, which can be be
measured independently, using a Vermeer scan.} Both methods lead to a
total uncertainty on the absolute cross section of 5-10\%.

To calculate $R_{AA}$ or $R_{CP}$ in heavy ion events, the number of
binary collisions (or overlap function $T_{AA}$) is used. This
quantity is derived from Glauber models of the total cross section (cf
Section \ref{sect:glauber_geom}). \mvl{The uncertainty on this
calculation is indicated separately in Fig. \ref{fig:pi0_raa} as a
grey band on the left side of each panel.} The result is accurate for
central events, but for peripheral collisions the uncertainties can be
sizable.  This uncertainty can possibly be reduced with dedicated
studies of peripheral events, but some of the uncertainty is inherent
in the Glauber modeling.

\mvl{The normalization uncertainty is strongly reduced when concentrating on
central events, where the determination of $N_{\mathrm coll}$ is quite
reliable. The types of normalization uncertainty discussed here only
affect measurements that involve an absolute cross-section and
therefore do not affect di-hadron measurements and elliptic flow.}

\mvl{When comparing theory calculations to experimental data, all of
the above sources of uncertainties need to be taken into account. For
example, in Fig. \ref{fig:pp_spectra}, there are three types of
uncertainties indicated, statistical, systematic and an overall
normalization uncertainty. The systematic uncertainties are somewhat
correlated between points and are due to the sources listed above
under background uncertainties and intrinsic uncertainties. For the
$\pi^0$ spectrum in the left panel of Fig. \ref{fig:pp_spectra},
statistical uncertainties are dominant only for $\pt>16$ GeV/$c$. At
lower \pt, the systematic uncertainty of about 5-10\% on each data
point and the overall uncertainty of about 10\% are dominant. Reducing
the systematic uncertainties is very difficult, and often requires
building new detectors.}

\mvl{For the non-photonic electron
measurements in the right panel of
Fig. \ref{fig:pp_spectra} and in Fig. \ref{fig:raa_nonphot}, the
statistical uncertainties are dominant for $\pt > 5$ GeV/$c$, so that
longer running times or larger luminosity will still improve the
precision of the result. Note also that the statistical uncertainties
for the STAR measurement are larger, due to a smaller analyzed
(triggered) data sample. Also for this measurement, reducing the
systematic uncertainties below the current value of $15-20\%$ is
difficult to impossible.}
\mvl{The precision of the di-hadron suppression and the
$\gamma$-hadron measurements are still limited by statistical
uncertainties, so that future runs with larger luminosity will improve
the precision.}

\section{Outlook: parton energy loss at LHC}
In the near future, \am{the} LHC will start colliding protons and nuclei at
energies far greater than RHIC energies (maximum energies are
$\rtsnn=5.5$ TeV for Pb+Pb and $\sqrt{s}=14$ TeV for p+p, but the
initial runs will be at lower energy). 

First results from the LHC experiments should already provide an
important test of the 
understanding of parton energy loss developed at RHIC: if the same
models and theory can be applied to describe the LHC results, that \am{would be} 
an important sign that our basic understanding of parton energy loss
is correct. However, \am{the LHC will provide access to a far greater number 
of jet observables.}

Our current understanding of RHIC results indicate that jet energy
loss is typically a few GeV with important tails to much larger energy
losses. Since typical jet energies at RHIC are 10--20 GeV, the energy
loss is not much smaller than the jet energy, which complicates the
theoretical treatment of energy loss. At LHC, large numbers of jets
with energies above 100 GeV will be produced, while the medium density
is only expected to increase by a modest factor of
3--5~\cite{Kolb:2000sd}. This makes it possible to cleanly separate
jets from the underlying event. In addition, the regime where $\Delta
E < E$, where theory calculations are better controlled, may become
experimentally accessible. Luminosity upgrades at RHIC will also
extend the accessible \pt-range there, reaching parton energies of
40--50 GeV.  

\mvlnew{The main jet observables have already been
discussed in the context of RHIC results in Section
\ref{sect:gamma_jet}. The basic measurement is the jet cross section
which shows to what extent the radiated energy is recovered by
jet-finding algorithms. Once this is established, measurements of
longitudinal and transverse jet structure (fragmentation function and
energy profile $\Psi(R)$) are used to quantify the redistribution of
energy inside the jet. An example calculation of modified
fragmentation functions in a fixed-length medium is given in
Fig.~\ref{fig:ff_comp}. However, jet reconstruction
intrinsically involves multi-hadron observables, which are best
modeled using Monte Carlo event generators, such as discussed in the
next section.} The large jet rates at ~100 GeV at LHC will further
enable more differential studies of {\it e.g.} jet structure using
sub-jet distributions \cite{Zapp:2008gi}.

\subsection{Experimental capabilities at the LHC}
ALICE
\cite{Carminati:2004fp,Alessandro:2006yt} is the dedicated heavy-ion
detector, which has been optimized for large track densities in
heavy-ion collisions and features powerful particle
identification. ATLAS \cite{Steinberg:2007nm,Takai:2004ka} and CMS
\cite{DEnterria:2007xr} will also perform heavy-ion measurements with
their detector systems that \mvl{are} more conventional for high-\pt{} and jet
measurements, including large calorimeter coverage. For a complete
review of the potential of hard probes at the LHC, see
\cite{Accardi:2004gp,Abreu:2007kv}.

The LHC detectors are all equipped with high-precision vertex
detectors for secondary vertex reconstruction of heavy flavor
decays. This will allow to separate charm and beauty from the outset,
providing critical tests of the mass and flavour dependence
of energy loss, which will show whether radiative energy loss is
dominant or not. Since the dead-cone effect should be less pronounced
for the lighter charm quark than for the heavy bottom quark, comparing
the suppression of charm and bottom provides direct sensitivity to the
dead-cone effect. On the other hand, comparing charm fragmentation at
high $\pt$ with light parton fragmentation, which is a mix of quark
and gluon fragmentation, provides sensitivity to the difference
between quarks and gluons \cite{Armesto:2005iq}. In addition, heavy
quark fragmentation studies should be possible at LHC, but these have
not yet been explored in detail.

Finally, $\gamma$-jet and $Z^0$-jet events will be studied to provide
a well-defined calibration of the jet energy measurement. The total
rates of these events are significantly lower than inclusive jets, so
that differential measurements of jet fragmentation will most likely
be performed with single jets or di-jet events. \mvlnew{For a more
detailed discussion of the potential of such measurements, and the expected
performance of the LHC detectors, see
\cite{Collaboration:2010ze,DEnterria:2007xr}.}

\subsection{Theoretical issues: \am{From} analytical approaches to Monte-Carlo}

While the single hadron
formalism has been the central theme in this review we now report on
the emerging new directions in jet modification theory which are being
tailored to calculate multi-hadron final states.

The computation of multihadron final states for jets in vacuum is now
a rather well established procedure. The standard method uses
Monte-Carlo routine to compute the underling jet shower by sampling
the DGLAP splitting functions repeatedly in order to sequentially
decompose a single virtual parton into a shower of lower virtuality
partons, using the Sudakov prescription. We refer the reader to
Refs.~\cite{Field:1989uq,Ellis:1991qj} for details on these
algorithms. At any given step in this routine one typically has a hard
virtual parton which will survive for a time which is inversely
proportional to its virtuality and then split into two partons with
lower virtuality and so on. At low enough virtuality, this ``DGLAP''
shower routine is terminated and a phenomenological hadronization
routine is invoked to convert the parton shower into a shower of
hadrons.

In a medium\mvl{,} the showering and hadronisation process becomes far \mvl{more}
complicated, due to additional \mvl{radiations} and the possible exchange of
color with the medium \cite{Sapeta:2007ad}. In addition, hadronisation
through recombination of shower partons with partons from the medium
might play a role. At the LHC, given the
energies of the jets, one expects a large part of the jet to exit the
medium in partonic form and continue to shower. Finally one expects a
considerable part of this escaped and modified jet to hadronize
independently of the medium. While the very soft part of this hadronic
spectrum from the jet will still become indistinguishable from the
harder hadrons from the medium, the high $p_{T}$ part of the jet
should be separable from the medium.  The major challenge is to model
the partonic splittings which now contain a vacuum like part where a
higher virtuality parton splits into lower virtuality partons
interfering with an in-medium part where scatterings in the medium may
raise the virtuality of the hard parton and cause it to split into
two.

\mvlnew{One of the first Monte-Carlo event generators to include an implementation of 
parton energy loss was PYQUEN/HYDJET
\cite{Lokhtin:2005px}. The model includes elastic and radiative energy
loss. The radiative energy loss is based on the BDMPS formalism and
the radiated gluons are added to the partonic final state, which is
then fragmented using PYTHIA 6.2. HYDJET also generates a soft
underlying event using a simple parametrisation including elliptic
flow. The programs is also available in a C++ version HYDJET++
\cite{Lokhtin:2008xi}, which is still being developed.}

\mvlnew{Besides PYQUEN, }there are currently four different
\mvlnew{programs}: JEWEL \cite{Zapp:2008gi}, Q-PYTHIA
\cite{Armesto:2009fj}, YaJEM \cite{Renk:2009hv}, and MARTINI
\cite{Schenke:2009gb} in the published literature, which have been
benchmarked with single hadron data. \mvlnew{Most of these generators
are also still under active development by the authors. They all} use
a somewhat different set of approximations and calculational
techniques.  Of these Q-PYTHIA~\cite{Armesto:2009fj} is closest in
spirit to the traditional vacuum Monte-Carlo routines. Here the
authors add a medium-induced contribution to the vacuum splitting
function and use this to calculate an in-medium Sudakov form factor.
They now have a probability that the jet will emit a gluon while
propagating over a certain distance and also the probability that it
will not radiate. This is then iteratively applied to compute a shower
distribution.  In the case of JEWEL~\cite{Zapp:2008gi}, the authors
attempt an original algorithm. The very process of single gluon
radiation is calculated via a numerical algorithm. Here, the random
process of scattering is used to modify the formation time of the
radiated gluon in a stochastic process.  More scatterings tend to
increase the transverse momentum of a gluon which has not completely
formed. As the transverse momentum is increased the formation time of
the gluon shrinks till it is formed within the next time step. After
this the process begins anew. JEWEL also includes the elastic loss
that the hard jet encounters every time it scatters.  In
YaJEM~\cite{Renk:2009hv}, the author modifies the vacuum DGLAP shower
that is contained within PYTHIA. This is done by \mvlnew{increasing}
the virtuality of an intermediate state due to scattering in the
medium. This leads to a change in the radiation pattern and thus to a
different shower pattern.  In MARTINI~\cite{Schenke:2009gb},
analytical energy loss routines computed within the AMY-McGill
formalism including elastic energy loss are used to modify the shower
distribution in PYTHIA prior to hadronization. \mvlnew{MARTINI uses
tabulated hydrodynamic data for the medium density evolution, with a
choice of hydrodynamical results from different groups.}

These Monte-Carlo routines along with others being developed represent
the next wave of jet modification calculations. These will no longer
be limited to the single hadron or dihadron level but will be able to
compute a whole host of full jet observables. They will also provide
the experimental community with multiple event generator codes which
will provide necessary input and benchmarks for the experimental
analysis of multihadron final states.

\section{Conclusion}

In this review, we have discussed in some detail the theoretical
formalisms describing parton energy loss in cold and hot QCD
matter. {Conceptually, there are two main mechanisms for energy loss:
elastic and radiative. 
For radiative energy loss off high-momentum
partons in a dense medium, the formation times are large, so that the
Landau-Pemeranchuk-Migdal (LPM) effect plays an important role,
leading to energy loss scaling with the square of the path length
($L^2$)}. At the same time, a high-momentum parton is also expected to
reduce its virtuality even in the absence of a medium, leading to a
jet of hadrons. The interplay (and interference) between vacuum
radiation and medium-induced radiation is therefore also expected to
play an important role.

Due to the complex nature of the energy loss process and in-medium
{modification of the fragmentation function}, different theoretical formalisms have chosen different
approximations in the calculation. A side-by-side comparison of the
medium modifications to the fragmentation functions for a fixed-length
homogeneous medium was presented in Fig. \ref{fig:ff_comp} and shows important
quantitative differences between different formalisms. 

Unfortunately, fixed length, homogeneous hot QCD matter is not
experimentally available, but instead a rapidly expanding medium is
produced in heavy ion collisions. Comparing
Fig. \ref{fig:raa_bassetal} and \ref{fig:ff_comp}, we concluded that
the way in which the partons sample the evolving medium is an
important aspect of the modeling for heavy ion collisions, which
currently precludes the precise determination of the medium density or
transport coefficient from experiment.

There are, however, qualitative features of radiative energy loss that
can be addressed with experimental data despite the uncertainties in
some of the quantitative details of the calculations. The comparison
of measured nuclear modification factor $R_{AA}$ and the recoil
suppression $I_{AA}$ indicate that there is a leading $L^2$ dependence
to the energy loss (cf. Fig. \ref{fig:RAAIAA_renk}), which is a first
indication that the LPM interference effect indeed plays an important
role in parton energy loss.

Another prediction of radiative energy loss is that heavy quarks lose
less energy than light quarks due to the dead-cone effect. The
measured suppression of heavy quarks at RHIC is larger than expected
from pure radiative energy loss, indicating that other effects play a
role. Possibilities include the presence of elastic energy {loss} 
and  {the possible influence of the larger scale at which transport 
coefficients should be evaluated due to the larger mass of the heavy quark.} 

Future measurements at RHIC and LHC will include $\gamma$-jet
measurements and measurements with kinematically reconstructed jets{,
which will} provide more differential control of the partonic
kinematics. Such measurements will be more directly comparable to the
theoretically calculated fragmentation functions in
Fig. \ref{fig:ff_comp}.

An important ongoing theoretical development is the use of Monte Carlo
techniques, which has two important aspects. On the one hand, Monte
Carlo techniques can be used to explore the effect of certain
limitations of the analytical calculations of energy loss, most
notably the effect of kinematic limits. On the other hand, Monte Carlo
event generators can be used to calculate more exclusive (multi-hadron)
observables, and thus provide a reliable tool to explore observables
based on reconstructed jets, such as fragmentation function and jet
broadening measurements.

All in all, a quantitative understanding
of the theory and phenomenology of parton energy loss based on the
RHIC measurements is starting to emerge, including some areas of
apparent disagreement between the data and theory. As a result, all
the tools and understanding are available to start devising critical
tests of the theoretical understanding using future
measurements at RHIC and LHC. The much larger momentum range available
at LHC will likely make the regime where the energy loss is
significantly smaller than the jet energy experimentally
available. This regime is the most suited for pQCD based theoretical
treatment and therefore promises to provide the most unambiguous tests
of our understanding of fragmentation and energy loss in a hot QCD
medium.

\noindent
{\bf Acknowledgements}

The authors would like to thank William Horowitz, Guangyou Qin, Thorsten Renk, Carlos Salgado 
and Werner Vogelsang for providing numerical data from their calculations. The authors also thank 
Nestor Armesto, Ulrich Heinz, Peter Jacobs, Berndt M\"{u}ller, Jamie Nagle and Carlos Salgado 
for reading an earlier version of the manuscript and providing feedback. 
The research of A.M. is supported in part by the U. S. Department of 
Energy under grant number  DE-FG02-01ER41190. 

\bibliography{refs,eloss_review_am_mvl}

\end{document}